\documentclass[a4paper,11pt]{article}
\pdfoutput=1

\usepackage{jheppub}

\usepackage{subfig}
\usepackage{xspace}
\usepackage[countmax]{subfloat}

\newcommand{\scetii}{SCET$_{\rm II}$}
\newcommand{\sceti}{SCET$_{\rm I}$}
\newcommand{\nllp}{NLL$'$}

\newcommand{\oPpb}{ \bar{\mathbb P}_{\!\! n \! \perp} }
\newcommand{\oPp}{ {\mathbb P}_{\!\! n \! \perp} }

\def\nslash{n\hspace{-2mm}\slash}
\def\nbarslash{\bar n\hspace{-2mm}\slash}

\DeclareRobustCommand{\Sec}[1]{Sec.~\ref{#1}}

\DeclareRobustCommand{\App}[1]{App.~\ref{#1}}

\DeclareRobustCommand{\Fig}[1]{Fig.~\ref{#1}}

\DeclareRobustCommand{\Eq}[1]{Eq.~(\ref{#1})}
\DeclareRobustCommand{\Eqs}[2]{Eqs.~(\ref{#1}) and (\ref{#2})}
\DeclareRobustCommand{\Ref}[1]{Ref.~\cite{#1}}
\DeclareRobustCommand{\Refs}[1]{Refs.~\cite{#1}}

\newcommand{\tr}{\text{tr}}
\newcommand{\jet}{\text{jet}}
\newcommand{\event}{\text{event}}

\newcommand{\ang}[1]{\tau^{(#1)}}

\newcommand{\Lang}[1]{s^{(#1)}}

\newcommand{\C}[2]{C^{(#2)}_{#1}}

\preprint{MIT--CTP 4512}

\title{Jet Shapes with the Broadening Axis}

\author{Andrew J. Larkoski,}
\author{Duff Neill,}
\author{and Jesse Thaler}

\affiliation{Center for Theoretical Physics, Massachusetts Institute of Technology, Cambridge, MA 02139, USA}

\emailAdd{larkoski@mit.edu}
\emailAdd{dneill@mit.edu}
\emailAdd{jthaler@mit.edu}

\abstract{Broadening is a classic jet observable that probes the transverse momentum structure of jets.  Traditionally, broadening has been measured with respect to the thrust axis, which is aligned along the (hemisphere) jet momentum to minimize the vector sum of transverse momentum within a jet.  In this paper, we advocate measuring broadening with respect to the ``broadening axis'', which is the direction that minimizes the scalar sum of transverse momentum within a jet.  This approach eliminates many of the calculational complexities arising from recoil of the leading parton, and observables like the jet angularities become recoil-free when measured using the broadening axis.  We derive a simple  factorization theorem for broadening-axis observables which smoothly interpolates between the thrust-like and broadening-like regimes.  We argue that the same factorization theorem holds for two-point energy correlation functions as well as for jet shapes based on a ``winner-take-all axis''.  Using kinked broadening axes, we calculate event-wide angularities in $e^+e^-$ collisions with next-to-leading logarithmic resummation.  Defining jet regions using the broadening axis, we also calculate the global logarithms for angularities within a single jet.  We find good agreement comparing our calculations both to showering Monte Carlo programs and to automated resummation tools.  We give a brief historical perspective on the broadening axis and suggest ways that broadening-axis observables could be used in future jet substructure studies at the Large Hadron Collider.}

\begin{document} 
\maketitle

\section{Introduction}
\label{sec:intro}

Event shapes offer a detailed probe of the jet-like behavior of quantum chromodynamics (QCD).  Classic event shapes like thrust \cite{Farhi:1977sg} have been used to test the structure of QCD \cite{Dasgupta:2003iq,Heister:2003aj,Abdallah:2003xz,Achard:2004sv,Abbiendi:2004qz} and extract the strong coupling constant $\alpha_s$ \cite{Becher:2008cf,Davison:2008vx,Abbate:2010xh}.  Most event shapes have corresponding jet shapes, which have been used to differentiate between quark- and gluon-initiated jets \cite{Gallicchio:2011xq,Gallicchio:2012ez,Larkoski:2013eya}.  More recently, jet shapes have offered insights into jet substructure and boosted objects at the Large Hadron Collider (LHC) \cite{Almeida:2008yp,Ellis:2010rwa,Abdesselam:2010pt,Altheimer:2012mn}.  

One of the most powerful jet observables is broadening \cite{Rakow:1981qn,Ellis:1986ig,Catani:1992jc}, which is the scalar sum of transverse momentum as measured with respect to the thrust axis.\footnote{In the context of jet shapes, broadening is sometimes referred to as ``girth'' or ``width'' \cite{Gallicchio:2010dq,Gallicchio:2011xq}.  It is equivalent to angularities with $a = 1$ \cite{Berger:2003iw,Almeida:2008yp,Ellis:2010rwa}.   When we perform calculations, we will actually use a slightly different definition of broadening, namely \Eq{eq:angularity} with $\beta = 1$.}  Broadening is a recoil-sensitive observable, meaning that it responds to deflections of the leading parton away from the thrust axis.  For this reason, it is a rather complicated observable from the point of view of perturbative QCD, though event-wide broadening in $e^+ e^-$ collisions has been calculated to next-to-next-to-leading logarithmic accuracy (N$^2$LL) \cite{Becher:2012qc}, and there is an understanding of non-perturbative power corrections \cite{Dokshitzer:1998qp,Salam:2001bd}.  More recently, the technique of rapidity factorization has been used to calculate resummed broadening distributions \cite{Becher:2011pf,Chiu:2012ir}, though recoil-sensitivity leads to transverse recoil convolutions in the factorization theorem.

In this paper, we show how these recoil complications can be avoided by measuring broadening with respect to the ``broadening axis''.   The broadening axis $\hat{b}$ is defined as the direction that minimizes the scalar sum of transverse momentum within a jet (or hemisphere in the case of $e^+ e^-$ event shapes). This is in contrast to the thrust axis $\hat{t}$ which minimizes the vector sum and is therefore aligned along the jet (or hemisphere) momentum.   For the case of a single jet, the broadening and thrust axes minimize the following quantities:\footnote{It is straightforward to show that minimizing \Eq{eq:introthrustaxisdef} is the same as minimizing $\left|\sum_{i \in J} \hat{t} \times \vec{p}_{i} \right|$ and that $\hat{t} \propto \sum_{i \in J} \vec{p}_{i}$ at the minimum.}
\begin{alignat}{2}
\text{Broadening axis}~\hat{b}: \quad & \min_{\hat{b}} \sum_{i \in \jet} | \hat{b} \times \vec{p}_{i}| \quad &&\approx \min_{\hat{b}} \sum_{i \in  \jet} E_i \, \theta_{i,\hat{b}}, \label{eq:introbroadaxisdef} \\
\text{Thrust axis}~\hat{t}: \quad & \min_{\hat{t}} \sum_{i \in  \jet} 2(|\vec{p}_{i}| - \hat{t} \cdot \vec{p}_{i})  \quad &&\approx \min_{\hat{t}}  \sum_{i \in  \jet} E_i \, \theta^2_{i,\hat{t}}, \label{eq:introthrustaxisdef}
\end{alignat}
where the sum runs over particles $i$ with momentum $\vec{p}_{i}$ in a given jet.  The approximations hold in the small angle limit, with particle energies $E_i$ and angles $\theta_{i,\hat{n}}$ with respect to the axis $\hat{n}$.  To our knowledge, the first paper that defined something like the broadening axis was \Ref{Georgi:1977sf} (the ``spherocity axis'', see \Sec{sec:history}), though this definition was lost to history.  The broadening axis was reintroduced in the context of the jet substructure observable $N$-subjettiness \cite{Brandt:1978zm,Stewart:2010tn,Thaler:2010tr} as the ``$\beta = 1$ minimization'' axis \cite{Thaler:2011gf}, and has been subsequently utilized in at least two jet substructure studies \cite{Larkoski:2013eya,Larkoski:2013paa}.\footnote{\Ref{Brandt:1978zm} introduced an observable identical to $N$-jettiness for $e^+e^-$ events shortly after thrust was defined.  In their language, 3-jettiness is ``triplicity''.}

As we will see, almost any jet or event shape that is measured with respect to the broadening axis will be recoil-free, including broadening itself.  As a concrete example, we calculate the jet angularities \cite{Berger:2003iw,Almeida:2008yp,Ellis:2010rwa} with respect to the broadening axis $\hat{b}$.  For a single jet, we define the angularities about an arbitrary axis $\hat{n}$, forming a light-cone vector $n=(1,\hat{n})$, as\footnote{This definition of the angularities differs from any of the previous definitions in \Refs{Berger:2003iw,Almeida:2008yp,Ellis:2010rwa}, both because of the form of the angular measure and because of an overall factor of $2^{\beta -1}$.  We prefer this angular measure because it is monotonic on $\theta_i\in[0,\pi/2]$ for all $\beta>0$.  The angular measures used in previous definitions, e.g. \Eq{eq:compare_new_old_ang}, are only monotonic in $\theta_i$ for $\beta\geq1$, which has the unfortunate consequence of making the same value of the angularity sensitive to both two and three jet configurations for $\beta < 1$.  We choose our factor of 2 convention such that the angularities have a simple form in the collinear limit.}
\begin{align}
\label{eq:angularity}
\ang{\beta} &=\frac{1}{E_\text{jet}}\sum_{i\in \text{jet}}E_i\left(2\frac{n\cdot p_i}{E_i}\right)^{\frac{\beta}{2}}\nonumber \\
&= \frac{1}{E_ \jet} \sum_{i\in  \jet} E_i \left(2\sin \frac{\theta_{i, \hat{n}}}{2}\right)^\beta \nonumber \\
&\approx \frac{1}{E_ \jet}\sum_{i\in  \jet} E_i \theta_{i, \hat{n}}^\beta  \ ,
\end{align}
where $p^\mu_i$ is the four-vector of a massless particle $i$, $E_i$ is its energy, and $\theta_{i, \hat{n}}$ is the angle between particle $i$ and axis $\hat{n}$.\footnote{In this paper, we will always treat hadrons as being effectively massless.  Hadron masses are relevant for non-perturbative power corrections \cite{Salam:2001bd,Mateu:2012nk}, which are beyond the scope of this paper.}  For massless hadrons, the limit $\beta = 1$ corresponds to broadening and $\beta = 2$ to thrust.  True to their names, the broadening axis minimizes $\ang{1}$ and the thrust axis minimizes $\ang{2}$.  In terms of the exponent $a$ in the original angularities paper, we are using the notation $\beta = 2-a$, and $\beta>0$ ($a < 2$) is needed for infrared and collinear (IRC) safety.  Crucially, in order for $\ang{\beta}$ to be recoil-free, we have to identify the jet region using a recoil-free jet algorithm such that the jet center coincides with the hard-collinear radiation and not the jet momentum axis.

Remarkably, we find a simple factorization theorem for $\ang{\beta}$ about the broadening axis that has the same form for the entire range $0 < \beta < \infty$, including broadening itself with $\beta = 1$.  This factorization theorem is derived using soft-collinear effective theory (SCET) \cite{Bauer:2000ew,Bauer:2000yr,Bauer:2001ct,Bauer:2001yt,Bauer:2002nz}.  Our result is in contrast to the usual case of $\ang{\beta}$ about the thrust axis, where there is a factorization theorem for $1 < \beta < \infty$ that expands any recoil contributions as power suppressed \cite{Berger:2003iw,Hornig:2009vb}, and a different factorization theorem that accounts for recoil at $\beta = 1$ \cite{Becher:2011pf,Chiu:2012ir}.  Because broadening-axis observables do not require transverse recoil convolutions, our factorization theorem is $\beta$-independent, which is quite surprising given that the relative scaling of the collinear and soft modes depends on $\beta$.  As a proof of concept, we resum the event-wide broadening-axis $\ang{\beta}$ distributions to next-to-leading logarithmic (NLL) accuracy for any value of $\beta$.  We also apply our analysis to jet-based broadening-axis $\ang{\beta}$ distributions to next-to-leading logarithmic (NLL) accuracy, ignoring non-global logarithms \cite{Dasgupta:2001sh} at present.

The behavior of our factorization theorem is particularly interesting in the vicinity of $\beta = 1$ (corresponding to broadening itself).  Strictly speaking, rapidity factorization \cite{Chiu:2012ir} is needed at exactly $\beta = 1$, with or without recoil.  However, ordinary soft/collinear factorization is valid for $\beta = 1 \pm \epsilon$ for non-zero $\epsilon$, and we can obtain the $\beta = 1$ distributions from the limits $\beta \to 1^+$ or $\beta \to 1^-$.  By measuring angularities with respect to the broadening axis, we therefore achieve a smooth interpolation between the thrust-like ($\beta = 2$) and broadening-like ($\beta = 1$) regimes.

Furthermore, we argue that our factorization theorem also holds for the two-point energy correlation function \cite{Banfi:2004yd,Jankowiak:2011qa,Larkoski:2012eh,Larkoski:2013eya} as well as for angularities measured with respect to the ``winner-take-all axis'' (defined in \Sec{subsec:WTA}, see also \Ref{Bertolini:2013iqa}).  This is true in a very strong sense:  not only is the form of the factorization theorem identical, but the anomalous dimensions of all corresponding functions in each observable are identical for a given $\beta$.  Only the finite terms of the functions can differ.  Even though the observables are clearly distinct, the fact that even one of the soft or collinear limits agrees is sufficient to prove that they have the same large logarithmic behavior.

The outline of the remainder of this paper is as follows.  In \Sec{sec:axisdef} we define the broadening axis, highlight the important recoil properties of broadening-axis observables, and show how to define jets using the broadening axis.  We then introduce a factorization theorem for general angularities about the broadening axis in \Sec{sec:factor}, calculating event-wide angularities in $e^+ e^-$ collisions to \nllp\ accuracy.  We briefly discuss how to apply this factorization theorem to jet-based angularities in \Sec{sec:shapefactor}.  In \Sec{sec:anom_dim_beta_1_limit}, we discuss the broadening limit ($\beta = 1$) and sketch how to obtain rapidity factorization from ordinary soft/collinear factorization.  We compare our distributions to showering Monte Carlo programs and automated resummation tools in \Sec{sec:mc}.  We give a historical perspective on the broadening axis in \Sec{sec:history}, and conclude in \Sec{sec:conclude} with some potential applications for broadening-axis observables at the LHC.  Various calculational details are left to the appendices.

\section{Jet Observables with the Broadening Axis}
\label{sec:axisdef}

\subsection{Defining the Broadening Axis}
\label{sec:broad_def}

To simplify the notation in parts of this paper, we will sometimes display expressions that are only valid in the small-angle limit.  The difference between, e.g., $\sin \theta$ and $\theta$ does not matter at NLL order, though it is relevant for fixed-order corrections.  Similarly, the dynamics of the broadening axis is dominated by the collinear radiation, where $\sin \theta \simeq \theta$ is a good approximation.  For our actual SCET calculations, we use the full expressions in \Eq{eq:angularity}, but we use the small angle limit when focusing on the dynamics.  

For a single jet, the broadening axis is defined as the axis that minimizes the broadening jet shape.  Repeating the equations in the introduction for convenience:
\begin{align}
\text{Angularities}~\ang{\beta}: \quad & \ang{\beta} = \frac{1}{E_\jet} \sum_{i\in  \jet} E_i \left(2 \sin \frac{\theta_{i,\hat{n}}}{2}\right)^\beta,\\
\text{Broadening axis}~\hat{b}: \quad & \hat{b} = \hat{n}~~\text{ with}~~\min_{\hat{n}} \ang{1}. \label{eq:broadaxisdef}
\end{align}

\begin{figure}
\begin{center}
\includegraphics[scale= 1.0]{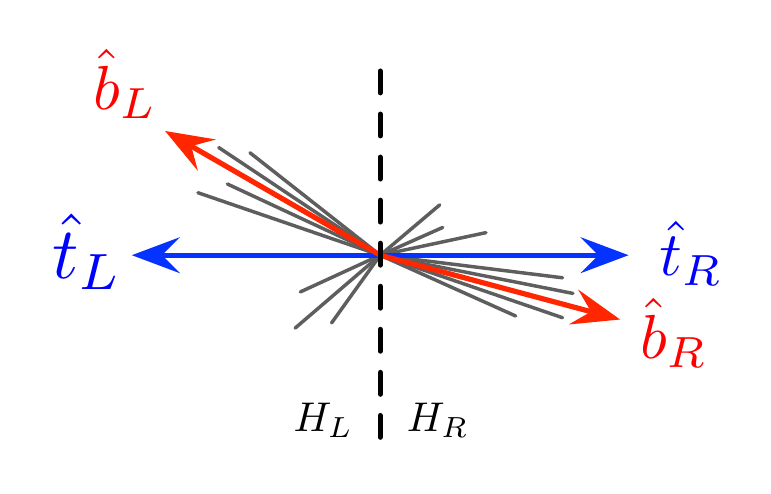}
\end{center}
\caption{Kinked broadening axes $\hat{b}_L$ and $\hat{b}_R$.  While we will use the thrust axes $\hat{t}_L$ and $\hat{t}_R$ to partition the event into left ($H_L$) and right ($H_R$) hemispheres, we measure the angularities with respect to the broadening axis in each hemisphere.}
\label{fig:kinkbroad}
\end{figure}

For $e^+ e^-$ event shapes, the situation is more subtle.  As discussed more in \Sec{sec:history}, there are various pathologies associated with applying \Eq{eq:broadaxisdef} to an entire $e^+ e^-$ event (where the resulting axis is called the spherocity axis \cite{Georgi:1977sf}).  The hard-collinear dijets need not be back-to-back due to wide-angle emissions of soft radiation, so to have a recoil-free observable, it is necessary to define two independent broadening axes.  In order to more easily compare results between broadening and thrust in \Sec{sec:mc}, we will partition the event into a left hemisphere $H_L$ and a right hemisphere $H_R$ using the thrust axis, such that the event-wide angularities with respect to two arbitrary axes $\hat{n}_L$ and $\hat{n}_R$ are
\begin{equation}
\label{eq:eventwideangularity}
\ang{\beta}_{\rm event} = \frac{1}{E_{\event}}  \left( \sum_{i \in H_L} E_i \left( 2 \sin \frac{\theta_{i, \hat{n}_L}}{2}\right)^\beta  + \sum_{i \in H_R} E_i \left(2\sin\frac{\theta_{i, \hat{n}_R}}{2}\right)^\beta \right) \equiv \ang{\beta}_L + \ang{\beta}_R.
\end{equation}
When clear from context, we will often drop the ``event'' subscript.  We can then find the broadening axes $\hat{b}_L$ and $\hat{b}_R$ separately in each hemisphere using \Eq{eq:broadaxisdef}.\footnote{\label{footnote:2jettinessregions}It is a bit unsatisfying that \Eq{eq:eventwideangularity} depends on both the broadening and thrust axes.  An alternative approach is to minimize 2-jettiness with a suitably chosen measure \cite{Stewart:2010tn,Thaler:2011gf}:
\begin{align}
\text{2-jettiness}~\ang{\beta}_2: \quad & \ang{\beta}_2 = \frac{1}{E_{\event}}\sum_{i \in \text{event}} E_i \min\left \{ \left( 2 \sin \frac{\theta_{i, \hat{n}_L}}{2}\right)^\beta , \left( 2 \sin \frac{\theta_{i, \hat{n}_R}}{2}\right)^\beta \right\},\\
\text{Broadening axes}~\hat{b}_L,\hat{b}_R: \quad & \hat{b}_i = \hat{n}_i~~\text{ with}~~\min_{\hat{n}_L,\hat{n}_R} \ang{1}_2.  \label{eq:kinkbroadaxisdef}
\end{align}
The minimum inside the sum in $\ang{1}_2$ effectively partitions the event into two regions, and $\hat{b}_i$ is the broadening axis for the $i$-th region.  Note that the two axes $\hat{b}_L$ and $\hat{b}_R$ still partition the event into hemispheres separated by a plane, though the $\hat{b}_i$ are not perpendicular to that plane.  One can think of these broadening hemispheres as being a rotation of the usual thrust hemispheres. For the event shape, these distinctions do not matter, since the total angularity is unchanged regardless what hemisphere a boundary parton is assigned.  Because the distributions for $\ang{\beta}_2$ and $\ang{\beta}_{\rm event}$ are the same to leading power, we will stick with the thrust partitioning for convenience.}
In general $\hat{b}_L$ and $\hat{b}_R$ will not be back-to-back, so we will refer to them as kinked broadening axes, as shown in \Fig{fig:kinkbroad}. This is in contrast to thrust, where $\hat{t}_L = - \hat{t}_R$ in the $e^+ e^-$ center-of-mass frame by momentum conservation.

\subsection{Recoil Properties of the Broadening Axis}
\label{sec:recbroad}

Traditionally, the jet shape $\ang{\beta}$ and event shape $\tau_{\event}^{(\beta)}$ have been measured with respect to thrust axes.  However, this induces problematic dependence on the soft radiation, since the thrust axis itself depends on the pattern of soft radiation.  This effect is known as ``recoil'' \cite{Catani:1992jc,Dokshitzer:1998kz,Banfi:2004yd,Larkoski:2013eya}, since the soft radiation causes the direction of collinear particles to recoil away from the thrust axis.  This recoil sensitivity can be formally ignored for jets whose soft wide-angle radiation has momentum transverse to the jet axis that is parametrically smaller than that of the hard-collinear radiation \cite{Bauer:2008dt}.  This can be selected for with a thrust value $\ang{2}$ that is sufficiently small.  However, to achieve robust insensitivity to the dynamics of soft radiation, one wants to study recoil-free observables.

As we will now explain, the broadening axis $\hat{b}$ is one example of a recoil-free axis, in the sense that the broadening axis is insensitive to the pattern of soft radiation.  To illustrate this, first consider a jet populated by two particles, $1$ and $2$, with energy fractions
\begin{equation}
z_i \equiv \frac{E_i}{E_\jet} \ .
\end{equation}
In the small angle limit, for a generic axis $\hat{n}$, the broadening takes the form
\begin{align}\label{two_particle_broadening}
\ang{1}&=z_1\theta_{1,\hat{n}}+z_2\theta_{2,\hat{n}},
\end{align}
and the broadening axis is obtained by minimizing $\ang{1}$ with respect to the choice of $\hat{n}$.  Without loss of generality we can assume $z_1<z_2$, and we can take the two particles to lie in the same plane as the candidate axis, since moving out of the plane increases the angle to both particles. Rewriting \Eq{two_particle_broadening} in terms of the relative angle $\theta_{12} \equiv \theta_{2,\hat{n}}+\theta_{1,\hat{n}}$ of the two particles, 
\begin{align}\label{two_particle_broadening_rewrite}
\ang{1}&=(z_2-z_1)\theta_{2,\hat{n}}+z_1\theta_{12}\,.
\end{align}
This is minimized for $\theta_{2,\hat{n}}=0$, so that the broadening axis coincides with the momentum of particle $2$. Note that it does not matter how much smaller $z_1$ is compared to $z_2$; the broadening axis always tracks the most energetic particle and is insensitive to the softer particle.   See \App{sec:3 particle broadening axis} for an analysis for a jet with three constituents.

Now consider a single energetic particle $J$ accompanied by arbitrary soft radiation $S$.  We assume that the sum of the soft radiation's energy is less than half of the total energy of the jet,\footnote{In terms of factorization in SCET, this formally means we are not specifying whether we are in \sceti-~or \scetii-like theories.  Taking the sum of soft energy to be less than half the jet energy is a rather mild requirement, and holds for all observables that we are aware of where there is some notion of soft factorization.} as with the two particle case, and the soft radiation is generically at wide angle relative to the energetic particle:
\begin{equation}
\sum_{i\in S}z_i <\frac{1}{2}<z_J, \qquad \theta_{iJ} \sim {\cal O}(1),\,\,i\in S.
\end{equation}
For an arbitrary axis $\hat{n}$, the broadening takes the form:
\begin{align}
\label{broadening_in_simple_jet}
\ang{1} & =z_J \theta_{J,\hat{n}}+\sum_{i\in S}z_i\theta_{i,\hat{n}},\\
& =z_J\theta_{J,\hat{n}}+z_S\theta_{S,\hat{n}}, \label{effective_two_particle_jet}
\end{align} 
where in the second line we have effectively treated the soft radiation as a single particle, defining $z_S \equiv \sum_{i\in S}z_i$ and $\theta_{S,\hat{n}} \equiv \frac{1}{z_S}\sum_{i\in S}z_i\theta_{i,\hat{n}}$.  This is similar to \Eq{two_particle_broadening} but with the difference that the ``two'' particles are in no sense co-planar as the soft ``particle'' is a sum of wide angle radiation. This implies that $\theta_{S,\hat{n}}\sim {\cal O}(1)$ for any choice of $\hat{n}$, such that the sum is minimized for $\theta_{J,\hat{n}}=0$. Again, the broadening axis tracks the hardest particle, even after considering the net effect of recoil from the soft radiation. 

To determine the broadening axis for a generic jet, we allow our energetic particle $J$ to undergo collinear splittings.  The broadening axis will move in a nonlinear fashion with subsequent splittings, and in general, there is no closed form expression for $\hat{b}$.  However, if $\theta_C$ is the largest angle involved in the collinear splittings (subject to the condition that the splitting is genuinely collinear, $\theta_C\ll 1$), then the broadening axis remains confined in a disc of radius $\theta_C$ about the initial energetic particle. 
The broadening axis never leaves this disc, since doing so would only decrease the broadening contribution from {some} of the soft radiation, while increasing the contribution from collinear radiation and that of other soft particles.\footnote{Specifically, the variation of the broadening axis under collinear splittings induces changes in the soft contribution:
\begin{align}
\sum_{i\in S}z_i\theta_{\hat{b}i}\rightarrow \sum_{i\in S}z_i(\theta_{\hat{b}i}+{\cal O}(\theta_C))\approx\sum_{i\in S}z_i\theta_{\hat{b}i}\,,
\end{align}
where again $\theta_C\ll \mathcal{O}(1)$ for a collinear splitting.  Any collinear splitting that produces a soft particle can be included in the initial haze of soft radiation.}

Thus, the broadening axis is immune to recoil from soft radiation because it tracks the hard, collinear radiation, and it recoils coherently with the collinear radiation under soft emission. While the broadening axis does depend on the precise configuration of the collinear radiation, this is of less concern since that radiation is naturally clustered together.  Indeed, for the purposes of factorization, we could use any axis that depends on the dynamics of the collinear radiation alone, since that axis (by definition) would be recoil free.  We will see an explicit example of this with the ``winner-take-all axis'' defined in \Sec{subsec:WTA}.

This behavior of the broadening axis is in contrast to that of the thrust axis.  The thrust axis lies along the total jet momentum, and is therefore conserved in the evolution of the jet, so the energetic core of a jet can be displaced from its thrust axis.  In the case of the setup in \Eq{broadening_in_simple_jet} with one energetic particle in a haze of soft radiation, the thrust axis is generically displaced from the energetic particle $J$ by an angle set by $z_S$.  Thus, whether the thrust axis can be considered recoil-free depends sensitively on the precise scaling of the soft energy fraction $z_S$ versus the typical collinear splitting angles $\theta_C$.  In contrast, the broadening axis is recoil free as long as the soft radiation is soft (i.e.~$z_S < 1/2$). 

\subsection{Comparison to Other Recoil-Free Observables}
\label{sec:comparison}

The broadening-axis angularities $\ang{\beta}$ are a recoil-free observable for any value of $\beta$.  To our knowledge, the only other jet-like observables in the QCD literature with this property are the energy correlation functions \cite{Banfi:2004yd,Jankowiak:2011qa,Larkoski:2013eya}.  The (normalized) two-point energy correlation function for a jet is defined as\footnote{
To agree with the recoil-free angularities in the soft limit, we have changed the angular factor from $\theta_{ij}^\beta$ as defined in \Ref{Larkoski:2013eya} to that given in \Eq{eq:C1}.
}
\begin{equation}
\label{eq:C1}
\C{1}{\beta} = \frac{1}{E_\jet^2}\sum\limits_{i<j\in \jet} E_i E_j\left(2\sin\frac{\theta_{ij}}{2}\right)^\beta \ ,
\end{equation}
where the sum runs over all distinct pairs of particles in the jet.  Because the angles $\theta_{ij}$ are measured between pairs of particles, this observable is manifestly insensitive to recoil, since it does not depend on the overall pattern of soft radiation.

In fact, for the same value of $\beta$, $\C{1}{\beta}$ and broadening-axis $\ang{\beta}$ are identical observables in the soft-collinear limit \cite{Larkoski:2013eya}.  In \Sec{subsec:anomdimrelatsion}, we will show that this implies the much stronger result that the logarithmic resummation of these two observables are identical to all orders, though the specific functions appearing in the factorization theorem will be different.  We will verify that the NLL resummation of $\ang{\beta}$ in this paper matches the NLL resummation of $\C{1}{\beta}$ presented in \Ref{Banfi:2004yd} in \Sec{subsec:NLL}.  

\subsection{Recoil-Free Jet Algorithms and the ``Winner-Take-All'' Axis}
\label{subsec:WTA}

The key feature of the broadening axis is that it recoils coherently with the collinear radiation.  In order to have a fully recoil-free observable, though, the particles that enter into the observable should be selected in a recoil-free fashion as well.  This is automatic for the event shape $\ang{\beta}_{\event}$ since all particles enter the observable,\footnote{Strictly speaking, the partitioning of the event into left and right hemispheres is recoil sensitive, since it depends on the recoil-sensitive thrust axis.  However, the effect of this partitioning is power suppressed.  For a truly recoil-free partitioning, see footnote \ref{footnote:2jettinessregions}.} but not for the jet shape $\ang{\beta}$ which depends on which particles are clustered into the jet.

A full study of the recoil properties of jet algorithms is beyond the scope of this work, but suffice it to say that all recursive jet algorithms  using standard ``$E$-scheme'' recombination \cite{Blazey:2000qt}\footnote{Not to be confused with the $E$-scheme for treating hadron masses \cite{Salam:2001bd,Mateu:2012nk}.} are recoil-sensitive, including anti-$k_T$ \cite{Cacciari:2008gp}.  In the $E$-scheme, one simply adds the four-vectors in a pair-wise recombination ($p_r = p_1 + p_2$), which ensures that the jet axis (i.e.~the center of the jet) and the jet momentum are aligned at each stage of the recursion, and therefore the final jet is centered on the thrust axis.  Similarly, iterative cone algorithms \cite{Salam:2007xv} search for stable cones where the jet axis and jet momentum align.  For these recoil-sensitive jet algorithms, the jet axis depends on both the collinear and soft radiation.

In this paper, we will use two different recoil-free jet algorithms.  The first jet algorithm is based on $N$-jettiness minimization using the $\beta = 1$ measure \cite{Thaler:2011gf}.  To define a single jet with radius $R$, we augment 1-jettiness with an out-of-jet measure:
\begin{align}
\text{1-jettiness}~\ang{\beta}_1: \quad & \ang{\beta}_{1} = \frac{1}{E_\text{event}} \sum_{i \in \text{event}} E_i \min \left\{ \left(2 \sin \frac{\theta_{i,\hat{n}}}{2}\right)^\beta,  \left(2 \sin \frac{R}{2}\right)^\beta \right\}, \\
\text{Jet axis}~\hat{b}: \quad & \hat{b} = \hat{n}~~\text{ with}~~\min_{\hat{n}} \ang{1}_1.
\end{align}
After minimizing to find the jet axis, the particles that contribute to the first minimum term in $\ang{\beta}_1$ are inside the jet (i.e.~particles with $\theta_{i, \hat{b}} < R$) and the rest are outside the jet.  Note that the resulting jet axis is the same as the broadening axis for the found jet (hence the notation $\hat{b}$), so by construction the jet region is defined in a recoil-free fashion.  For a multi-jet final state, 1-jettiness will typically identify the hardest jet in the event, since $\ang{\beta}_1$ penalizes unclustered radiation proportional to the unclustered energy.

The second jet algorithm is based on recursive jet clustering algorithms with an alternative recombination scheme.\footnote{We thank Gavin Salam for discussions on this point.  The winner-take-all scheme was recently used in \Ref{Bertolini:2013iqa} for determining a jet axis that was guaranteed to align along the direction of one of the input particles.}  Consider the ``winner-take-all'' recombination scheme, where we define the four-vector from pair-wise recombination to be massless, i.e.~$p_r = (E_r, E_r \hat{n}_r)$, with momentum pointing in the direction of the harder particle:
\begin{align}
E_r &= E_1 +E_2, \\
\hat{n}_r &=
\begin{cases} \hat{n}_1 & \text{if $E_1 > E_2$}, \\
\hat{n}_2 & \text{if $E_2 > E_1$},
\end{cases}
\end{align}
where $\hat{n}_i = \vec{p}_i/|\vec{p}_i|$ are unit-normalized.  This recombination scheme is (perhaps surprisingly) IRC safe, just like other weighted schemes like the $p_t^2$-scheme \cite{Catani:1993hr,Butterworth:2002xg}, and it can be applied to any of the generalized $k_T$ algorithms including anti-$k_T$.  Because the jet axis always aligns with the harder particle in a pair-wise recombination, soft radiation cannot change the jet axis, so the resulting jet axis is recoil-free.  Note that the jet axis is only needed to determine the particles clustered into a given jet, but the actual jet four-vector can be defined by adding the jet's constituents (just as in the $E$-scheme, though here the jet momentum and jet axis will be offset because of recoil).  Because finding the winner-take-all axis is computationally much faster than minimizing $\ang{\beta}$, we expect it will become the default way to define a recoil-free axis.  We leave a more in depth study of the winner-take-all axis for future work.

In \Sec{subsec:pythia}, we will see that this winner-take-all axis yields nearly identical results to the broadening axis. This is a consequence of both being dominated by collinear dynamics. The difference between angularities measured with respect to the broadening axis versus the winner-take-all axis must be proportional to the typical collinear splitting angle. Since soft radiation cannot resolve such splittings, the two observables share the same soft function to leading power.  Then to all orders in perturbation theory, the anomalous dimensions of all functions for either axis are identical; see \Sec{subsec:anomdimrelatsion}.  Indeed, to \nllp\ order, the two cross sections are identical.

\section{Resummation of Event-Wide Angularities}
\label{sec:factor}

To illustrate the recoil insensitivity of the broadening axis, we now present a factorization theorem for the event-wide angularities in \Eq{eq:eventwideangularity} for all $\beta>0$, measured with respect to kinked broadening axes.  We use this factorization theorem to calculate the broadening-axis angularities to \nllp\ order in \Sec{subsec:NLLp_result}, including matching to the full $\mathcal{O}(\alpha_s)$ fixed-order result.

\subsection{Relevant Collinear and Soft Modes }

We are interested in the process $e^+e^-\rightarrow \text{hadrons}$ with kinked broadening axes $\hat{b}_{L,R}$ and measured hemisphere angularities $\ang{\beta}_{L,R}$.  By requiring $\ang{\beta}_{L,R}\ll 1$, we can select for events with energy clustered about the two axes, which defines a two-jet state.  These jets are dominated by collinear and soft radiation, and to find the specific modes that contribute to the observable, we set our power counting parameter by $\ang{\beta}_{L,R} \sim \lambda$.  We are interested in describing the double-differential cross section in $\ang{\beta}_{L}$ and $\ang{\beta}_{R}$ to leading power in $\lambda$.  

Because we are measuring $\ang{\beta}_{L,R}$ with respect to kinked broadening axes, there is an important subtlety regarding the choice of coordinate system.  Consider measuring just the left hemisphere contribution $\ang{\beta}_{L}$ about a light-cone vector $n_L=(1,\hat{n}_L)$ aligned along the left broadening axis $\hat{b}_{L}$.  As shown already in \Fig{fig:kinkbroad}, $\bar{n}_L = (1, -\hat{n}_L)$ is \emph{not} aligned along the right broadening axis $\hat{b}_{R}$, and we will return to that subtlety in a moment. Looking only in the left hemisphere and using the notation of \Sec{sec:recbroad}, the power counting $\ang{\beta}_{L} \sim \lambda$ implies the following scalings for the energy fractions and splitting angles:
\begin{align}
\text{Collinear modes:} &\quad z_C \sim 1, \quad \theta_C\sim\lambda^{\frac{1}{\beta}}, \\
\text{Soft modes:} &\quad z_C \sim \lambda, \quad \theta_C\sim 1.
\end{align}
Thus, in light-cone coordinates $p=(\bar{n}_L\cdot p, n_L \cdot p,\vec{p}_\perp)$, the relevant collinear and soft modes (in the left hemisphere) have the scaling
\begin{align}
\label{mode_power_counting}
p_{n_L}&\sim Q(1,\lambda^{\frac{2}{\beta}},\lambda^{\frac{1}{\beta}}), \nonumber\\
p_s&\sim Q(\lambda,\lambda,\lambda),
\end{align}
where $Q$ is the collision energy of the event.  Running the same argument for the right hemisphere about a light cone vector $n_R=(1,\hat{n}_R)$, we get the same soft modes, but find additional contributions to $\ang{\beta}_{R}$ from right-collinear modes $p_{n_R}$.  Therefore, in the small $\lambda$ limit, the relevant modes of the theory are
\begin{equation}
p_{n_L}~(\text{left-collinear}), \qquad p_{n_R}~(\text{right-collinear}), \qquad p_s~(\text{soft}).
\end{equation}

Having identified the relevant modes, we are free to use reparametrization invariance (RPI) of the effective theory \cite{Chay:2002vy,Manohar:2002fd} to perform small changes in the jet axis directions.  In particular, instead of kinked broadening axes, we can invoke RPI to use back-to-back thrust axes to characterize the modes and operators of the effective theory.  Physically, one can think about performing a small $\mathcal{O}(\lambda)$ boost on the system to make $\hat{b}_{L}$ and $\hat{b}_{R}$ back-to-back.  Practically, RPI will allow us to recycle known factorization theorems for thrust-axis observables to generate a factorization theorem for broadening-axis observables.  It is well-known that thrust-axis observables are described by collinear, anti-collinear, and soft modes, though for general angular exponent $\beta$, we have to use the scaling from \Eq{mode_power_counting}:
\begin{align}\label{alt_mode_power_counting}
p_{n}&\sim Q(1,\lambda^{\frac{2}{\beta}},\lambda^{\frac{1}{\beta}}), \nonumber\\
p_{\bar{n}}&\sim Q(\lambda^{\frac{2}{\beta}},1,\lambda^{\frac{1}{\beta}}),\nonumber\\
p_s&\sim Q(\lambda,\lambda,\lambda).
\end{align}
As we will see in \Sec{sec:eventfactor}, unlike traditional factorization theorems, there will be a formal difference between the directions used for the measurement (i.e.\ $\hat{b}_{L,R}$) and the directions used to define the fields (i.e. $n, \bar{n}$).

\begin{figure}
\begin{center}
\subfloat[]{
\includegraphics[width=6.1cm]{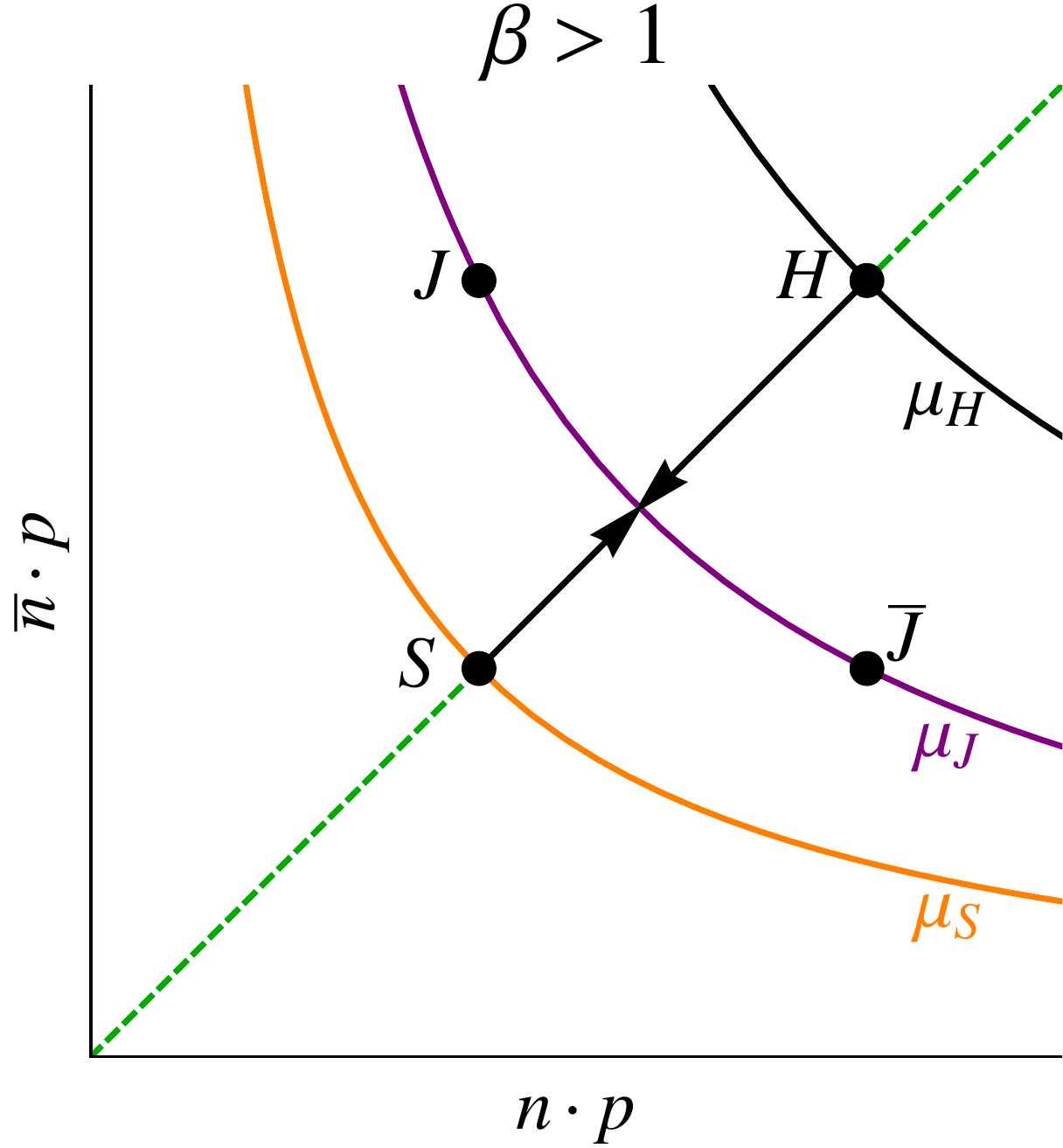}
}
$\qquad\qquad$
\subfloat[]{
\includegraphics[width=6.1cm]{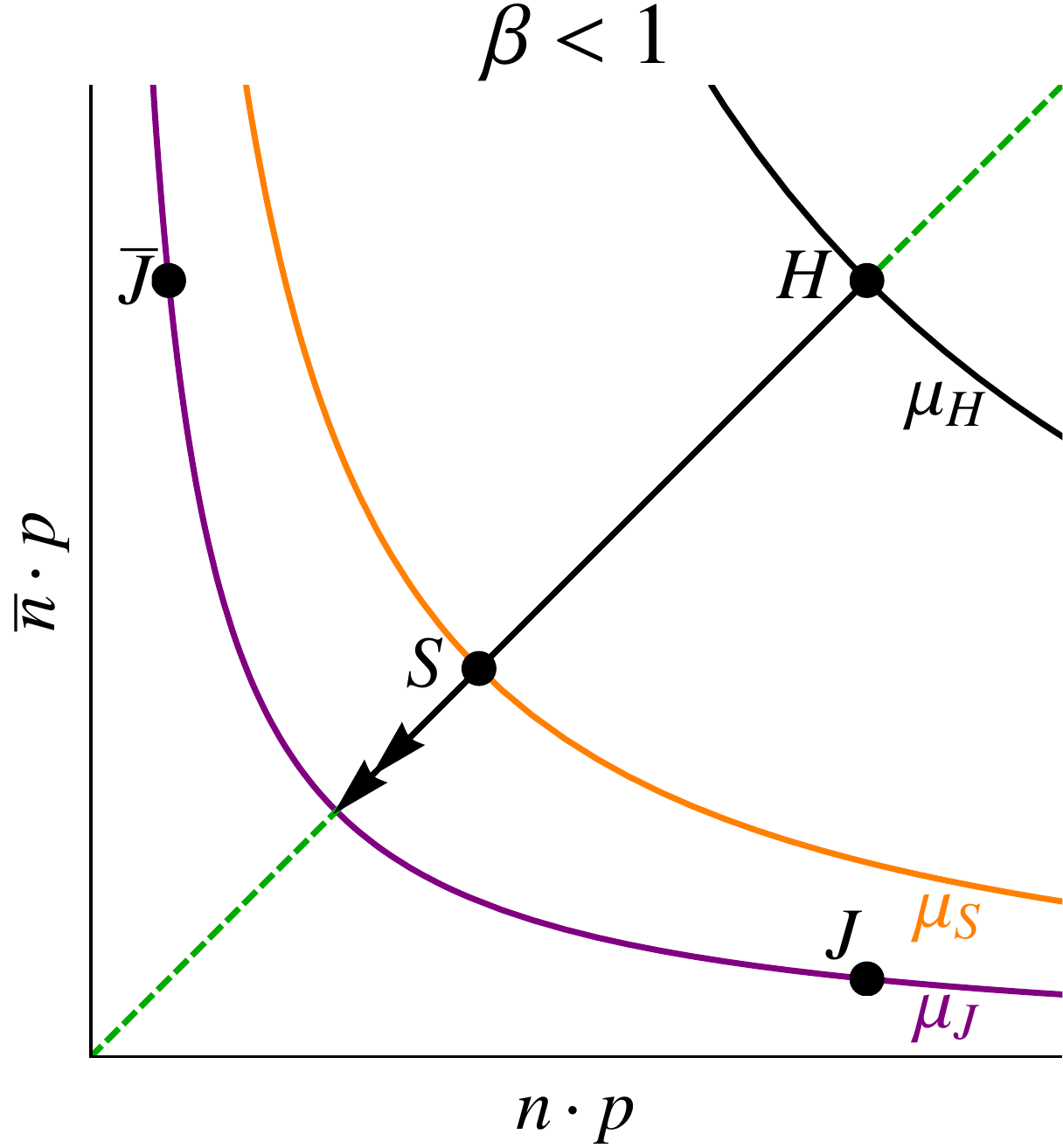}
}
\end{center}
\caption{Hierarchy of modes in the effective theory for $\beta > 1$ (left) and $\beta < 1$ (right).  Shown is the light-cone momentum plane $(n\cdot p, \bar{n}\cdot p)$ and the dots correspond to the hard ($H$), left-collinear ($J$), right-collinear ($\bar{J}$), and soft ($S$) modes.  The curves indicate the natural invariant mass scale for the hard ($\mu_H$), jet ($\mu_J$), and soft ($\mu_S$) functions.   For $\beta > 1$, the soft modes have smaller invariant mass than the collinear modes, but this hierarchy is inverted for $\beta < 1$.  The arrows indicate the direction of renormalization group evolution if modes are factorized at the jet scale (as we will do in this paper).
}
\label{fig:invariant_mass}
\end{figure}

From the assigned scaling in \Eq{alt_mode_power_counting}, we can derive the natural hard, jet, and soft scales by the invariant mass of the corresponding modes:
\begin{align}
\label{eq:natural_scales}
\mu_H & = Q, \nonumber \\
\mu_J&=\theta_CQ\sim (\ang{\beta}_{L,R})^{1/\beta}Q \ ,\nonumber\\
\mu_S&=z_S Q\sim \ang{\beta}_{L,R} Q \ .
\end{align}
Depending on the choice of the angular exponent $\beta$, the jet scale can lie either above ($\beta > 1$) or below ($\beta < 1$) the soft scale.   For the special case $\beta = 1$ (discussed more in \Sec{sec:anom_dim_beta_1_limit}), the jet and soft scales coincide.   The relative positions of all modes in the theory are illustrated in \Fig{fig:invariant_mass}.  From the point of view of the factorization theorem, the relative hierarchy between $\mu_J$ and $\mu_S$ will be largely irreverent, though we will have to choose appropriate scales in \App{sec:scalechoice}.\footnote{The relative hierarchy is interesting from the point of view of non-perturbative physics.  For $\beta<1$, the leading non-perturbative power correction will be controlled by the collinear modes, not the soft modes as it is for $\beta > 1$ observables like thrust \cite{Lee:2006nr}.  Non-perturbative corrections to the jet function are not well-understood, so we will not discuss this issue further in this paper.}  

\subsection{Factorization Theorem for Broadening-Axis Angularities}
\label{sec:eventfactor}

Because we have used RPI to define the relevant modes in terms of \Eq{alt_mode_power_counting}, we can recycle the factorization theorem in \Ref{Chiu:2012ir}, which is applicable to any $e^+ e^-$ event shape that selects two-jet structures.  We will then use the recoil-free nature of the broadening axis to derive a simplified broadening-axis factorization theorem.

To lowest order in $\lambda$, the double-differential cross section in $\ang{\beta}_{L}$ and $\ang{\beta}_{R}$ is 
\begin{align}
\label{eq:factorized-X-sec-start}
\frac{d^2\sigma}{d\ang{\beta}_Ld\ang{\beta}_R}
& =  \sigma_0 \, H(Q^2,\mu) \, \int de_n \, de_{\bar n}   \, \int d^2 \vec{k}_{1\!\perp} \, d^2 \vec{k}_{2\!\perp} \nonumber \\
& \quad \quad \quad {\cal J}_{n}(Q,e_n,\vec k_{1\perp}) \,  {\cal J}_{\bar n}(Q,e_{\bar n},\vec k_{2\perp}) \, {\cal S}(\ang{\beta}_L-e_{n},\ang{\beta}_R-e_{\bar{n}},\vec k_{1\perp},\vec k_{2\perp}).
\end{align} 
Here, $H(Q^2,\mu)$ is the hard function  
and $\sigma_0$ is the Born-level cross section for the process $e^+e^- \to $ dijets.
Note that the above factorization theorem allows for recoil between the soft and collinear modes. This recoil is captured by the $2$-dimensional transverse momentum convolution between the jet and soft functions (i.e.~the $\vec{k}_{i\perp}$ integrals).  The jet and soft functions themselves are defined as:
\begin{align}
{\cal J}_{n}(Q,e_n,\vec k_{1\perp})  &= \frac{(2\pi)^3}{N_c}\tr   \langle 0 \vert    \frac{\nbarslash}{2}    \chi_{n}   \delta( \bar n \cdot \hat P - Q )  \delta(e_n -\hat e_{\hat b_L})     \delta^{(2)}(\hat P_{\!\!\perp} - k_{1\!\perp})     \bar{\chi}_{n}      \vert 0\rangle\ ,\nonumber \\
{\cal J}_{\bar{n}}(Q,e_{\bar{n}},\vec k_{2\perp})&=  \frac{(2\pi)^3}{N_c}   \langle 0 \vert     \bar{\chi}_{\bar n}   \delta( n \cdot \hat P - Q )  \delta(e_{\bar n} -\hat e_{\hat b_R})     \delta^{(2)}(\hat P_{\!\!\perp} - k_{2\!\perp})   \frac{\nslash}{2}  \chi_{\bar n}     \vert 0\rangle\ ,\nonumber \\
{\cal S}(e_L,e_R,\vec k_{1\perp},\vec k_{2\perp})& = \frac{1}{N_c}\tr  \langle 0 \vert       S^{\dagger}_{\bar n} S_{n}   \delta^{(2)}(\oPp + k_{1\!\perp})   \delta^{(2)}(\oPpb + k_{2\!\perp})   \nonumber \\
&\qquad\qquad\qquad\qquad\times\delta(e_L-\hat e_{\hat b_L})\delta(e_R-\hat e_{\hat b_R})  
 S^{\dagger}_{n}  S_{\bar n}    \vert 0\rangle \ .
\end{align}
Here, we are using the notation of \Ref{Chiu:2012ir}, where $\chi_{\bar n}$ is a collinear field operator,\footnote{In position space, this is  defined as $\chi_{\bar n}  =W_{\bar{n}}^{\dagger}\xi_n(0)$ with $\xi_n$ being the standard quark field operator and $W$ a collinear Wilson line defined as $W_{n}={\mathcal P}\text{ exp}\left(ig\int_{0}^{\infty}d\lambda\, \bar{n}\cdot A_c(\bar{n}\lambda)\right) $.} and $S_{n}$ is a soft wilson line operator defined as
\begin{align} 
S_{n}&={\mathcal P}\text{ exp}\left(ig\int_{0}^{\infty}d\lambda\, n\cdot A_s(n\lambda)\right) \ .
\end{align}
The directions of the jets are set by the momentum-constraining delta functions inserted between the two collinear field operators. Within these delta functions, the operator $\hat{P}$ returns the momentum of the collinear state.  The soft recoil contribution is accounted for by the operators $\oPp$ and $\oPpb$ in the soft function, which measure the total transverse momentum generated by the soft radiation in the left and right hemispheres respectively; this sets the amount of recoil injected into the jet function by the $\vec{k}_{i\perp}$ convolutions. As discussed above \Eq{alt_mode_power_counting}, we have defined the field operators in terms of the thrust axis, even though the measurement operators $\hat e_{\hat b_{L,R}}$ are defined with respect to the broadening axis.  

We can specialize \Eq{eq:factorized-X-sec-start} to broadening-axis observables by making use of the fact that the broadening axis is recoil-free.  As argued in \Sec{sec:recbroad}, the collinear modes recoil coherently with the broadening axis, so any injected transverse momentum from the soft radiation merely translates the broadening axis with the collinear radiation.  Thus, the jet function is independent of the injected transverse momentum $k_{i\!\perp}$ to leading power:
\begin{align}
\label{recoil_power_corrections_for_jet_function}  
{\cal J}_{n}(Q,e_n,\vec k_{\perp})&={\cal J}_{n}(Q,e_n,0)+{\cal O}\left(\lambda^{2}\right) \ .
\end{align}
Using the analysis of \Sec{sec:recbroad}, it is straightforward to check explicitly that this form holds at tree-level and one-loop.  Because \Eq{recoil_power_corrections_for_jet_function} is true for any $\beta>0$, recoil is a power-suppressed correction for all of the broadening-axis angularities.  

It is instructive to compare this behavior to the thrust-axis angularities, where only for $\beta > 1$ is recoil power-suppressed \cite{Bauer:2008dt}.  For the thrust-axis angularities, one can perform a multipole expansion \cite{Grinstein:1997gv,Bauer:2000yr,Beneke:2002ni} of the jet function in $\vec{k}_\perp$ to account for the displacement of the collinear direction due to soft recoil.  This displacement is power suppressed for $\beta>1$ since the $p_\perp$ of a soft mode is a factor of $\lambda^{(\beta - 1)/\beta}$ smaller than the $p_\perp$ of a collinear mode, so the effect of recoil can be formally ignored to leading power.  By contrast, the broadening axis lies in the collinear direction for all $\beta>0$, independent of the soft emissions in the jet, so the multipole expansion is \emph{irrelevant}.  There are still power corrections to \Eq{recoil_power_corrections_for_jet_function} for the broadening axis, but they are generated by hard perturbative emissions, the finite size of the collinear region, and similar effects.

Using \Eq{recoil_power_corrections_for_jet_function}, we can trivially perform the transverse momentum integrals in \Eq{eq:factorized-X-sec-start} to achieve the simplified factorization theorem for broadening-axis observables: 
\begin{align}
\label{eq:factorized-X-sec-fin}
\frac{1}{\sigma_0}\frac{d^2\sigma}{d\ang{\beta}_Ld\ang{\beta}_R}
& =   H(Q^2,\mu)  \int de_n \, de_{\bar n}\,   {\cal J}_{n}(Q,e_n)   {\cal J}_{\bar n}(Q,e_{\bar n}) \, {\cal S}(\ang{\beta}_L-e_n,\ang{\beta}_R-e_{\bar{n}}),
\end{align}
where the jet and soft functions are defined as
\begin{align}\label{eq:jet-and-soft-functions-final-form}
{\cal J}_{n}(Q,e_n)  &= \frac{(2\pi)^3}{N_c}\tr   \langle 0 \vert    \frac{\nbarslash}{2}    \chi_{n}   \delta( \bar n \cdot \hat P - Q )  \delta(e_n -\hat e_{\hat b_L})  \delta^{(2)}(\hat P_{\perp})   \bar{\chi}_{n}      \vert 0\rangle\ ,\nonumber \\
{\cal J}_{\bar{n}}(Q,e_{\bar{n}})&=  \frac{(2\pi)^3}{N_c}   \langle 0 \vert     \bar{\chi}_{\bar n}   \delta( n \cdot \hat P - Q )  \delta(e_{\bar n} -\hat e_{\hat b_R})  \delta^{(2)}(\hat P_{\perp})\frac{\nslash}{2}  \chi_{\bar n}     \vert 0\rangle\ ,\nonumber \\
 {\cal S}(e_L,e_R)& = \frac{1}{N_c}\tr  \langle 0 \vert       S^{\dagger}_{\bar n}  S_{n}    \delta(e_L-\hat e_{\hat b_L})\delta(e_R-\hat e_{\hat b_R})  
  S^{\dagger}_{n}  S_{\bar n}    \vert 0\rangle \ .
\end{align}
This factorization theorem will be the basis for our subsequent analysis.  The event-wide observable from \Eq{eq:eventwideangularity} is obtained by summing over the left and right hemispheres
\begin{equation}
\frac{d\sigma}{d\ang{\beta}_{\rm event}} = \int d \ang{\beta}_L \, d \ang{\beta}_R \, \frac{d^2\sigma}{d\ang{\beta}_Ld\ang{\beta}_R} \, \delta(\ang{\beta}_{\rm event} - \ang{\beta}_L - \ang{\beta}_R),
\end{equation}
and the procedure for calculating the jet-based observables is given in \Sec{sec:shapefactor}.

The key ingredient needed to derive this factorization theorem was \Eq{recoil_power_corrections_for_jet_function}, so one can really think of \Eq{recoil_power_corrections_for_jet_function} as \emph{defining} what it means to be a recoil-free observable.  Indeed, the energy correlation functions (\Sec{sec:comparison}) and angularities with respect to the winner-take-all axis (\Sec{subsec:WTA}) satisfy the same property, so the factorization theorem above applies equally well to those observables.  Of course, even though the jet function is insensitive to soft recoil effects, the jet function does depend on the relative angles between collinear radiation.  Consequently,  different recoil-free observables will have different jet (and soft) functions, even though they share the same factorization theorem.  As we discuss in detail in \Sec{subsec:anomdimrelatsion}, recoil-free observables with the same behavior in the soft limit will necessarily have the same jet and soft anomalous dimensions to all orders, a fact which is generically not true for recoil-sensitive observables.

\subsection{Resummation}
\label{sec:event}

The recoil insensitivity of the broadening-axis angularities greatly simplifies the resummation of the cross section, since the form of the factorization theorem in \Eq{eq:factorized-X-sec-fin} is independent of the angular exponent $\beta$.  As with traditional thrust-axis angularities, rapidity divergences arise when $\beta=1$ (i.e. the broadening limit), and we will treat that case separately in \Sec{sec:anom_dim_beta_1_limit}.  That said, we will find in \Sec{sec:anom_dim_beta_1_limit} that the renormalization group (RG) evolution discussed here will smoothly merge onto the rapidity RG of \Ref{Chiu:2012ir} as $\beta \to 1$.

Conveniently, the form of the broadening-axis factorization theorem in \Eq{eq:factorized-X-sec-fin} is identical in form to the familiar thrust factorization theorem \cite{Schwartz:2007ib,Fleming:2007qr}.   Correspondingly, the resummation of large logarithms in the angularities cross section using RG evolution can be performed in exactly the same way as for thrust, which we review here.  We start by writing the factorization theorem in \Eq{eq:factorized-X-sec-fin} schematically as
\begin{align}
\frac{1}{\sigma_0}\frac{d^2\sigma}{d\ang{\beta}_L\, d\ang{\beta}_R}&=H \times J_n\left(\ang{\beta}_L\right)\otimes J_{\bar{n}}\left(\ang{\beta}_R\right)\otimes S\left(\ang{\beta}_L,\ang{\beta}_R\right),
\end{align}
where $\times$ denotes ordinary multiplication, and $\otimes$ denotes a convolution in the observables $\ang{\beta}_L,\ang{\beta}_R$. We can remove the convolutions by transforming to Laplace space, so the factorization theorem becomes
\begin{equation}
\frac{1}{\sigma_0}\frac{d^2\tilde{\sigma}}{d\Lang{\beta}_L\, d\Lang{\beta}_R}=H\tilde{J}_n\left(\Lang{\beta}_L\right)\tilde{J}_{\bar{n}}\left(\Lang{\beta}_R\right)\tilde{S}\left(\Lang{\beta}_L,\Lang{\beta}_R\right) \ ,
\end{equation}
where the Laplace transform of a function $g(\ang{\beta})$ is defined as
\begin{equation}
\tilde{g}(\Lang{\beta})=\int_0^\infty d\ang{\beta} e^{-\Lang{\beta}\ang{\beta}}g(\ang{\beta}) \ .
\end{equation}

Being a physical quantity, the cross section itself is finite. In each sector, though, divergences appear in loop calculations which are removed by corresponding $Z$-factors, which carry all the divergent terms in a given function $F$:
\begin{align}
\tilde{F}^B(\Lang{\beta})&=Z_F\left(\Lang{\beta},\frac{\mu}{Q}\right)\tilde{F}^R\left(\Lang{\beta},\frac{\mu}{Q}\right) \ .
\end{align}
The superscript $B$ ($R$) indicate the bare (finite renormalized) functions. The product of these $Z$-factors from all sectors is $1$ due to the finiteness of the cross section.  That is,
\begin{align}
\label{eq:Zfactorequals1}
1=Z_H\left(\frac{\mu}{Q}\right)Z_n\left(\Lang{\beta}_L,\frac{\mu}{Q}\right)Z_{\bar{n}}\left(\Lang{\beta}_R,\frac{\mu}{Q}\right)Z_S\left(\Lang{\beta}_L,\Lang{\beta}_R,\frac{\mu}{Q}\right) \ .
\end{align}
Applying this to the cross section, we have:
\begin{align}
\frac{1}{\sigma_0}\frac{d^2\tilde{\sigma}}{d\Lang{\beta}_L\, d\Lang{\beta}_R}&=H^B\tilde{J}^B_n\left(\Lang{\beta}_L\right)\tilde{J}^B_{\bar{n}}\left(\Lang{\beta}_R\right)\tilde{S}^B\left(\Lang{\beta}_L,\Lang{\beta}_R\right)\nonumber \\
\label{Renorm_xsec}&=H^{R}\left(\frac{\mu}{Q}\right)\tilde{J}^R_n\left(\Lang{\beta}_L,\frac{\mu}{Q}\right)\tilde{J}^R_{\bar{n}}\left(\Lang{\beta}_R,\frac{\mu}{Q}\right)\tilde{S}^R\left(\Lang{\beta}_L,\Lang{\beta}_R,\frac{\mu}{Q}\right)
\end{align}
Thus, in removing these divergences, we have introduced into each function the (same) factorization scale $\mu$. The physical cross section is independent of this factorization scale for exactly the reason it contains no divergences. 

To minimize large logarithms in the cross section, we want to evolve each function from its initial factorization scale $\mu$ to its ``natural'' scale using its RG equation.  The anomalous dimensions can be calculated from the $Z$-factors as
\begin{align}
\tilde{\gamma}_F\left(\Lang{\beta},\frac{\mu}{Q}\right)&=-\mu\frac{d}{d\mu}\log\,Z_F\left(\Lang{\beta},\frac{\mu}{Q}\right) \ ,
\end{align}
and the renormalized function $F^R$ satisfies the RG equation
\begin{align}
\mu\frac{d}{d\mu}F^R\left(\Lang{\beta},\frac{\mu}{Q}\right) &=\tilde{\gamma}_F\left(\Lang{\beta},\frac{\mu}{Q}\right)F^R\left(\Lang{\beta},\frac{\mu}{Q}\right) \ .
\end{align}
The presence of the eikonal lines in the functions \Eq{eq:jet-and-soft-functions-final-form} meeting at a cusp implies that the anomalous dimensions also depend on the observable through $\Lang{\beta}$.  In Laplace space, solving the RG equation is quite simple,
\begin{align}\label{eq:basic_RG_solution}
\log\frac{F^R\left(\Lang{\beta},\frac{\mu_F}{Q}\right)}{F^R\left(\Lang{\beta},\frac{\mu_I}{Q}\right)}&=\int_{\mu_I}^{\mu_F}\frac{d\mu}{\mu}\tilde{\gamma}_F\left(\Lang{\beta},\frac{\mu}{Q}\right) \ .
\end{align}
By solving the RG equations, the scales in each function can be set to independent values and large logarithms of the ratio of those scales are resummed.  Note that from the one-loop results in \App{sec:cal_one_loop_details} for the jet and soft functions, the scale choice discussed already in \Eq{eq:natural_scales} minimizes the $\mu$-dependent logarithms.  By \Eq{eq:Zfactorequals1}, we have the following consistency condition among the anomalous dimensions
\begin{equation}
\label{eq:anomdimconsistency}
0 = \gamma_H+\gamma_J+\gamma_{\bar{J}}+\gamma_S \ ,
\end{equation}
where $\gamma_J$ ($\gamma_{\bar{J}}$) is the anomalous dimension of the left (right) hemisphere jet function.

For a given function $F=H,J,S$, it is convenient to break up the anomalous dimension $\tilde{\gamma}_F$ into a cusp term $\Gamma_F$ and non-cusp term $\gamma_F$.
The evolution in $\mu$ of the function $F$ can then be written as
\begin{align}
\label{eq:anom_dim_def}
\mu\frac{d}{d\mu}\log F(\mu)&=\gamma_F\left[\alpha_s(\mu)\right]+\Gamma_F\left[\alpha_s(\mu)\right]\log(\mu) \ .
\end{align}
The logarithm in the cusp-term will generically depend on the canonical scale of the specific function, which we have suppressed here. We can express the solution to the RG equation, \Eq{eq:basic_RG_solution}, entirely in terms of the running coupling using the definition of the $\beta$-function:
\begin{align}
\log\frac{F\left(\mu_f\right)}{F\left(\mu_i\right)}=\int_{\alpha_s(\mu_i)}^{\alpha_s(\mu_f)}\frac{d\alpha}{\beta(\alpha)}\left[\gamma_F[\alpha]+\Gamma_F[\alpha]\int_{\alpha_s(\mu_i)}^{\alpha}\frac{d\alpha'}{\beta(\alpha')}\right] \ ,
\end{align} 
where we have used
\begin{align}
\frac{d\alpha_s}{\beta(\alpha_s)}=d\log\mu \ .
\end{align}

\subsection{Resummed Cross Section to NLL and \nllp }
\label{subsec:NLLp_result}

Here, we present the complete resummed cross section at NLL and \nllp\ order.  At NLL order, the two-loop cusp and one-loop non-cusp anomalous dimension are included in the resummation, but only the tree-level expressions for the hard, jet, and soft functions are necessary.  At \nllp~order, one further includes the ${\cal O}(\alpha_s)$ corrections to the hard, jet, and soft functions, which allows both for scale profiling in the resummation and for matching to fixed-order corrections.  As we will show in \Sec{subsec:NLLprime}, increasing the accuracy of the cross section significantly reduces the dependence on the renormalization scale $\mu$.

To \nllp~order, the cross section for the hemisphere recoil-free angularities can be written as
\begin{align}
\label{eq:NLLPorderFinal}
\frac{1}{\sigma_0}\frac{d^2\sigma}{d\ang{\beta}_L\, d\ang{\beta}_R}&= e^{K_H(\mu,\mu_H)}\left(\frac{\mu_H}{Q}\right)^{\omega_H(\mu,\mu_H)}\nonumber\\
&\!\!\!\!\!\times\int \frac{d\Lang{\beta}_L}{2\pi i}\frac{d\Lang{\beta}_R}{2\pi i}\, e^{\ang{\beta}_L\Lang{\beta}_L}\, e^{\ang{\beta}_R\Lang{\beta}_R}
e^{K_S(\mu,\mu_S)}\left(\frac{\mu_S}{\Lang{\beta}_LQ}\right)^{\omega_S(\mu,\mu_S)}\left(\frac{\mu_S}{\Lang{\beta}_RQ}\right)^{\omega_S(\mu,\mu_S)}\nonumber\\
&\qquad\times e^{K_J(\mu,\mu_J)}\left(\frac{\mu_J}{\Lang{\beta}_LQ}\right)^{\omega_J(\mu,\mu_J)}e^{K_{\bar{J}}(\mu,\mu_{\bar{J}})}\left(\frac{\mu_{\bar{J}}}{\Lang{\beta}_RQ}\right)^{\omega_{\bar{J}}(\mu,\mu_{\bar{J}})}
\nonumber \\
&\qquad\times\left\{1+f_{H}(Q,\mu_H)+f_{J}(\Lang{\beta}_L,\mu_J)+f_{\bar{J}}(\Lang{\beta}_R,\mu_J)+f_{S}(\Lang{\beta}_L,\Lang{\beta}_R,\mu_S)\right\} \ .
\end{align}
The integrals over $\Lang{\beta}_L,\Lang{\beta}_R$ represent the inverse Laplace transform.  The functions $K_F(\mu,\mu_F)$ and $\omega_F(\mu,\mu_F)$ are defined through the anomalous dimensions as
\begin{align}
K_F(\mu,\mu_F) & =  \int_{\alpha_s(\mu_F)}^{\alpha_s(\mu)}\frac{d\alpha}{\beta(\alpha)}\left[\gamma_F[\alpha]+\Gamma_F[\alpha]\int_{\alpha_s(\mu_F)}^{\alpha}\frac{d\alpha'}{\beta(\alpha')}\right],   \nonumber \\
\omega_F(\mu,\mu_F)&=  \frac{1}{j_F} \int_{\alpha_s(\mu_F)}^{\alpha_s(\mu)}\frac{d\alpha}{\beta(\alpha)}\Gamma_F[\alpha]   \ ,
\end{align}
where $j_F$ is a number that depends on the scale choice $\mu_F$ of the function $F$.  For example, for the canonical scale choice for the jet and soft functions in \Eq{eq:natural_scales}, $j_J = \beta $ and $j_S = 1$.  The $f$-functions correspond to the ${\cal O}(\alpha_s)$ finite pieces of the jet and soft functions which is necessary for \nllp~accuracy.  The anomalous dimensions and the finite pieces of the jet and soft functions to one loop are presented in \App{sec:cal_one_loop_details}.

For accuracy just to NLL order, \Eq{eq:NLLPorderFinal} still applies, but the $f$-functions can simply be set to 0.  

At \nllp\ (but not NLL) order, we can match to the ${\cal O}(\alpha_s)$ fixed-order cross section by simply adding the non-singular terms of the cross section:  
\begin{equation}
\label{eq:fixedorderpiece}
\frac{d^2\sigma}{d\ang{\beta}_L\, d\ang{\beta}_R} = \frac{d^2\sigma^{\text{NLL}'}}{d\ang{\beta}_L\, d\ang{\beta}_R} + \frac{d^2\sigma^{\text{non-sing}}}{d\ang{\beta}_L\, d\ang{\beta}_R},
\end{equation}
where the non-singular correction factor $\sigma^{\text{non~sing.}}$ is chosen such that $\sigma$ gives the complete ${\cal O}(\alpha_s)$ cross section when resummation is turned off (i.e~$\mu_H = \mu_J = \mu_S = Q$). This additive matching is possible since the fixed-order singular contributions to the cross section are capture by the $f$ functions in \Eq{eq:NLLPorderFinal}, so even when the resummation is turned off, one gets the correct singular cross section out of the \nllp\ result.   At NLL, these singular pieces are only encoded by the resummation, so that if the resummation is turned off, one is left with nothing.  Schemes do exist to match at NLL (see e.g. \Ref{Catani:1992ua}), but we use \nllp\ resummation to exploit the full power of the RG approach.  To achieve a smooth transition between the resummation and fixed-order regimes, we use profile scales \cite{Abbate:2010xh} which interpolate between the canonical scales in \Eq{eq:natural_scales} and a common scale choice $\mu_i = Q$. A comparison of the relative strengths of the singular and non-singular contributions to the fixed order can be found in \Fig{fig:non-sing}, and the details of the profiling is given in \Sec{sec:scalechoice}.  

The total event-wide recoil-free angularity $\ang{\beta}$ cross section is found by integrating over one of the the hemisphere angularities subject to the constraint that $\ang{\beta} = \ang{\beta}_L+\ang{\beta}_R$.  Setting the renormalization scale to the scale of the jet function, $\mu=\mu_J$ as illustrated in \Fig{fig:invariant_mass}, we find
\begin{align}\label{eq:totangexp}
\frac{1}{\sigma_0}\frac{d\sigma}{d\ang{\beta}}&= e^{K_H(\mu_J,\mu_H)}\left(\frac{\mu_H}{Q}\right)^{\omega_H(\mu_J,\mu_H)}\nonumber\\
&\qquad\times\int \frac{d\Lang{\beta}}{2\pi i}\, e^{\ang{\beta}\Lang{\beta}}
e^{K_S(\mu_J,\mu_S)}\left(\frac{\mu_S}{\Lang{\beta} Q}\right)^{\omega_S(\mu_J,\mu_S)}\nonumber\\
&\qquad\qquad\times\left\{1+f_{H}(Q,\mu_H)+2f_{J}(\Lang{\beta},\mu_J)+f_{S}(\Lang{\beta},\mu_S)\right\} \ .
\end{align}

\subsection{Equality of Anomalous Dimensions}
\label{subsec:anomdimrelatsion}

As already mentioned, different recoil-free observables will in general have different jet and soft functions.  Surprisingly, though, any two factorizable recoil-free observables which share the same soft, small angle behavior (and the same hard function) will exhibit the same anomalous dimensions.  In particular, broadening-axis angularities $\ang{\beta}$, winner-take-all-axis angularities $\ang{\beta}$, and the energy correlation function $\C{1}{\beta}$ all have the same anomalous dimensions.  We are not aware of a similar generic statement being true for recoil-sensitive observables.

\begin{figure}
\begin{center}
\includegraphics[width=14cm]{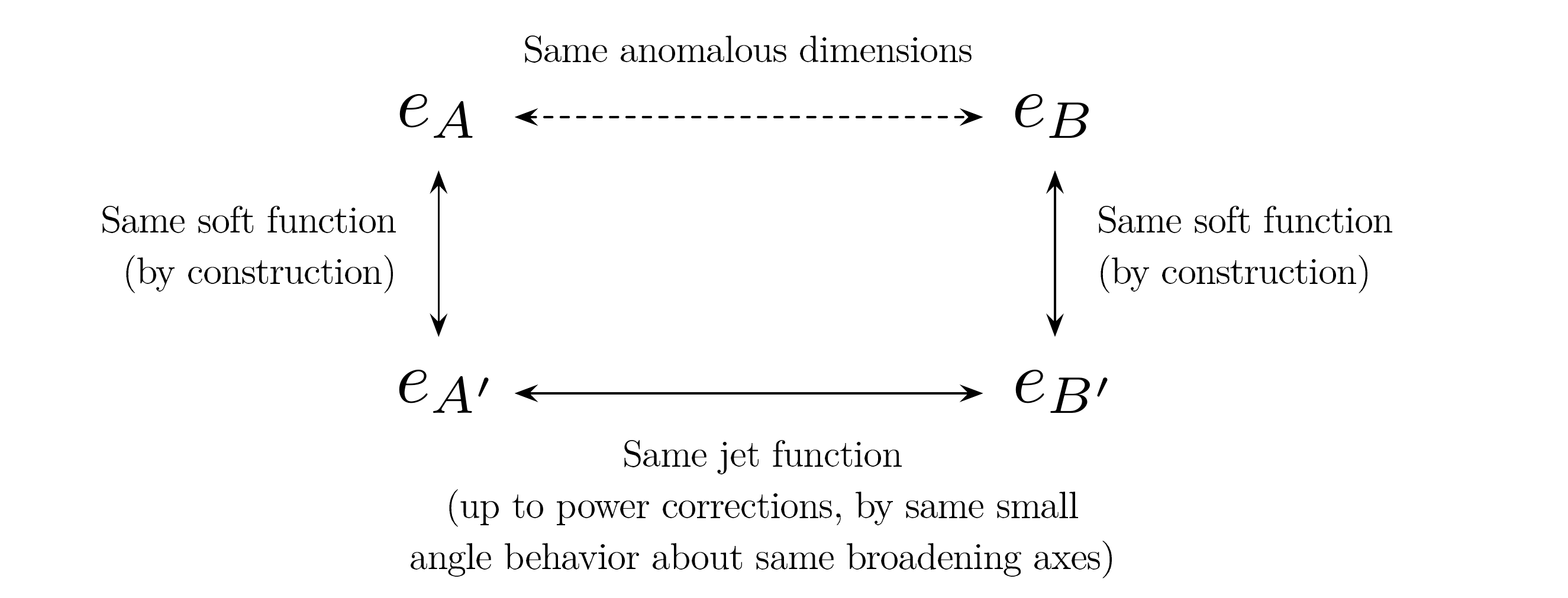}
\end{center}
\caption{Relationship between the anomalous dimensions of observables $e_A$ and $e_B$ that share the same soft, small angle behavior (and the same hard function).  The observable $e_{A'}$ ($e_{B'}$) is constructed to have the same soft behavior as $e_A$ ($e_B$), albeit with angles measured with respect to broadening axes.  But this then implies that $e_{A'}$ and $e_{B'}$ have the same jet function (up to power corrections).  By consistency of the anomalous dimensions, this means that $e_A$ and $e_B$ have the same jet and soft anomalous dimensions.}
\label{fig:anom_relation}
\end{figure}

To see why this is the case, we will follow the logic depicted in \Fig{fig:anom_relation}.  Consider two recoil-free observables $e_A$ and $e_B$ with the same soft, small angle behavior.  These observables may or may not have explicit dependence on a recoil-free axis.  Because it is recoil-free, in the soft limit, $e_A$ only depends on the pattern of soft radiation about one or more collinear directions. We can then build an alternative observable $e_{A'}$ which uses one or more broadening axes to define the collinear directions but extrapolates the soft behavior of $e_A$ to all energy scales.  Because by construction $e_A$ and $e_{A'}$ have the same soft function, they must have the same soft anomalous dimension: $\gamma_{S}^A = \gamma^{A'}_S$.  Because the total cross section is independent of renormalization scales, they must have the same anomalous dimensions for the jet function $\gamma^A_J = \gamma^{A'}_J$ as well, assuming that the hard functions are identical.  We can perform the same manipulations to generate $e_{B'}$ from $e_B$.

Now consider the collinear behavior of $e_{A'}$ and $e_{B'}$.  Since $e_{A'}$ and $e_{B'}$ were built by extrapolating the soft (arbitrary angle) behavior of $e_A$ and $e_B$ to all energies, and since $e_A$ and $e_B$ have the same soft, small angle behavior by assumption, $e_{A'}$ and $e_{B'}$ will have the same small angle behavior for all energy scales.  Because $e_{A'}$ and $e_{B'}$ measure the same (small) angles with respect to the same broadening axes, they must have the same jet function (up to possible power corrections), and therefore the same jet function anomalous dimension: $\gamma^{A'}_J = \gamma^{B'}_J$.  By construction, it then follows that
\begin{equation}
\gamma^A_J = \gamma^{A'}_J = \gamma^{B'}_J = \gamma^{B}_J\ ,
\end{equation}
and by consistency of the anomalous dimensions, we also have $\gamma^A_S = \gamma^B_S$.  Therefore, the anomalous dimensions of recoil-free observables $e_A$ and $e_B$ are identical if they have the same soft, small-angle behavior.

As a concrete example, compare the broadening-axis angularities with the angular measure defined in \Eq{eq:angularity} to the traditional angular measure \cite{Berger:2003iw,Ellis:2010rwa}:
\begin{equation}
\label{eq:compare_new_old_ang}
\ang{\beta}=\frac{1}{E_\text{jet}}\sum_{i\in \text{jet}}E_i\left(2\sin\frac{\theta_{i\hat{b}}}{2}\right)^{\beta}, \qquad \ang{\beta}_{\rm trad}=\frac{1}{E_\text{jet}}\sum_{i\in \text{jet}} E_i \sin \theta_{i\hat{b}}\left( \tan \frac{\theta_{i\hat{b}}}{2}  \right)^{\beta-1} \ ,
\end{equation}
where $\hat{b}$ is the broadening axis.  Up to a factor of $2^{\beta-1}$, they have the same collinear limit
\begin{equation}
\label{eq:collinear-limit for new ang.}
\lim_{\theta\to 0} \ang{\beta}  = \lim_{\theta\to 0} 2^{\beta-1} \ang{\beta}_{\rm trad} = \frac{1}{E_\text{jet}}\sum_{i\in \text{jet}}E_i\theta_{i\hat{b}}^\beta \ .
\end{equation}
and therefore have the same jet function (and same jet anomalous dimension).  The soft functions differ, but by the above argument, the soft anomalous dimensions must be the same.

Similarly, the broadening-axis angularities agree with the energy correlation functions $\C{1}{\beta}$ from \Eq{eq:C1} in the soft limit:
\begin{align}\label{eq:soft-limit for new ang.}
\lim_{E\to 0}\frac{1}{E_\text{jet}^2}\sum_{i<j\in \text{jet}}E_iE_j\left(2\sin\frac{\theta_{ij}}{2}\right)^{\frac{\beta}{2}}&=
\frac{1}{E_\text{jet}}\sum_{i\in \text{jet}}E_i\left(2\sin\frac{\theta_{i\hat{b}}}{2}\right)^{\frac{\beta}{2}} \ ,
\end{align}
where $\hat{b}$ is the axis of the hardest particle in the jet, which coincides with the broadening axis in the soft limit.  This means that they share the same soft function (and the same soft anomalous dimension).  The jet functions differ, but by the above argument, the jet anomalous dimensions must be the same.  The same logic holds for the broadening-axis angularities versus winner-take-all-axis angularities, which have distinct behavior in the collinear limit but identical behavior in the soft limit.  Among other things, this means that the fact that the winner-take-all axis always lies along the direction of an input particle is irrelevant to logarithmic accuracy.   

Thus, for the same value of $\beta$, all of the recoil-free observables considered in this paper have the same anomalous dimensions, and thus the same large logarithmic behavior.  We will see numerical evidence for this in \Sec{sec:mc}.

\section{Resummation of Jet-Based Angularities}
\label{sec:shapefactor}

The analysis of the previous section for event-wide angularities can be repeated for jet-based angularities.  This requires the introduction of a jet algorithm to identify a region of radius $R_0$ in the event about a single collinear direction.  The factorization of jet-based angularities measured with respect to the thrust axis was explicitly derived in \Ref{Ellis:2010rwa} for cone \cite{Salam:2007xv} and $k_T$-type \cite{Catani:1993hr,Ellis:1993tq,Dokshitzer:1997in,Wobisch:1998wt,Wobisch:2000dk,Cacciari:2008gp} jet algorithms.\footnote{Because of recoil and rapidity divergences, the analysis of \Ref{Ellis:2010rwa} only formally holds for angularities for which $\beta = 2-a > 1$.}  The global component of the cross section factorizes when the jets in the event are well-separated in angle compared to the jet radius $R_0$.  Factorization-violating non-global \cite{Dasgupta:2001sh} or clustering \cite{Banfi:2005gj} logarithms are beyond the scope of this paper, but we do not expect the broadening axis to present any additional complications with respect to those issues.  

We now sketch a derivation of the factorization theorem for broadening-axis angularities measured on a single jet.  The key point is that in the absence of recoil, it is possible to extend the factorization proof for thrust-axis angularities given in \Ref{Ellis:2010rwa} to the broadening-axis case.  What \Ref{Ellis:2010rwa} showed is that even in the presence of a jet algorithm, the jet-based angularities factorize into contributions from soft and collinear modes within the jet (where the soft function itself depends on the light-like directions of unmeasured jets).  The proof of \Ref{Ellis:2010rwa} is only valid for $\beta>1$, though, since only then is the jet function recoil-free (i.e.\ independent of the injected transverse momentum to leading power).  However, apart from recoil, nothing in the proof of \Ref{Ellis:2010rwa} precluded an extension to $\beta < 1$, since the same power counting and identification of modes holds equally well for all $\beta > 0$.  For the broadening-axis angularities, \Eq{recoil_power_corrections_for_jet_function} guarantees that there are no recoil effects for all $\beta > 0$.  Thus, in the absence of recoil, we can simply recycle the $\beta > 1$ factorization theorem from \Ref{Ellis:2010rwa} and apply it to the jet-based broadening-axis angularities for all $\beta > 0$.

The only subtlety in the above argument is that for $\beta < 1$, one has to use a recoil-free jet algorithm (see \Sec{subsec:WTA}), such that the jet region is determined solely by the location of the collinear modes and is independent of soft recoil.  Having dealt with that subtlety, then the jet-based factorization theorem is analogous to the event-wide factorization theorem from \Sec{sec:eventfactor} in the sense that one can smoothly continue the result for $\beta > 1$ thrust-axis angularities to the $\beta>0$ broadening-axis angularities.

In general, we would have to use the full machinery of \Ref{Ellis:2010rwa} to calculate the (recoil-free) jet-based angularities, including a careful treatment of the color structure of unmeasured jets.  Since we will only ever work to \nllp\ accuracy, though, there is a simple procedure to determine the global structure of the cross section for the jet-based broadening-axis angularity from the corresponding event-wide result.  Recall that our factorization theorem for the event-wide angularities in \Eq{eq:factorized-X-sec-fin} divided the event into two hemispheres, with each hemisphere containing a single collinear direction.  As we define a jet by a single collinear direction, we can integrate over the unmeasured hemisphere to compute the cross section of the angularity in the measured hemisphere:
\begin{equation}
\frac{1}{\sigma_0}\frac{d\sigma}{d\ang{\beta}_{\rm jet}} = \int d\ang{\beta}_L\, \left. \frac{1}{\sigma_0}\frac{d^2\sigma}{d\ang{\beta}_Ld\ang{\beta}_R} \right|_{\ang{\beta}_R = \ang{\beta}_{\rm jet}} \ .
\end{equation}
At this point, we have made no approximations, but we are only describing hemisphere jets.  To \nllp\ accuracy, the jet radius can be incorporated by boosting the hemisphere into a region of radius $R_0$ about the collinear direction.  As shown in \Ref{Ellis:2010rwa}, this can be accomplished by appropriately rescaling the canonical soft scale by the jet radius,
\begin{equation}
\label{eq:softrescalingtojet}
\mu_S \to \frac{\ang{\beta}Q_J}{2^{\beta-1}\sin^{\beta - 1} \frac{R_0}{2}} \ ,
\end{equation}
where $Q_J$ is the energy of the jet and the jet scale is unchanged.  
This holds for both cone and $k_T$-type algorithms to \nllp, but higher order corrections will in general differ between the two algorithms.  We will only consider the recoil-free jet algorithms presented in \Sec{subsec:WTA} in the rest of this paper for which these results hold at \nllp\ accuracy.

\section{The Broadening Limit ($\beta=1$)}
\label{sec:anom_dim_beta_1_limit}

Thus far, we have avoided the special case of $\beta = 1$, which corresponds to broadening itself.  At $\beta=1$, rapidity divergences arise that are unregulated by dimensional regularization schemes.  These divergences require an additional regularization procedure, and an additional rapidity RG evolution to resum all large logarithms of the matrix elements \cite{Chiu:2012ir}.  However, the recoil-free factorization theorem in  \Eq{eq:factorized-X-sec-fin} has a continuous behavior as $\beta \to 1$, which suggests that one can achieve the exponentiation of the rapidity divergences without direct use of the rapidity RG.  Here, we will show that this is indeed the case by sketching the mapping between the traditional $\mu$ RG and the rapidity RG.

To see explicitly that the $\beta = 1$ limit is singular, consider the RG equations in Laplace space for $e^+e^-\to q\bar{q}$ from \Sec{sec:event}.  From \Eqs{eq:renorm_jet_function}{eq:renorm_soft_function}, the one-loop RG equations for the jet and soft functions are 
\begin{align}
\mu\frac{d}{d\mu}\log\tilde{J}\left(\Lang{\beta},\frac{\mu}{Q}\right)&=\frac{\alpha_s(\mu)}{\pi}C_F\left[\frac{3}{2}-\frac{2}{1-\beta}\,\log\left(2^{\beta-1}\Lang{\beta} e^{\gamma_E}\frac{\mu^\beta}{Q^\beta}\right)\right] \ ,\nonumber \\
\mu\frac{d}{d\mu}\log\tilde{S}\left(\Lang{\beta},\frac{\mu}{Q}\right)&=\frac{4C_F}{1-\beta}\frac{\alpha_s(\mu)}{\pi}\,\log\left(2^{\beta-1}\Lang{\beta} e^{\gamma_E}\frac{\mu}{Q}\right) \ ,
\end{align}
which are singular at $\beta = 1$.  However, the resummed cross section is continuous through $\beta= 1$, suggesting that there should be a way to recover  (non-singular) RG evolution from the (singular) $\mu$ RG equations.

To do this, consider the $\mu$ evolution in the neighborhood of the canonical scales.  From \Eq{eq:natural_scales}, we can write the jet scale $\mu_J$ in terms of the soft scale $\mu_S$ as 
\begin{equation}\label{jet_scale}
\mu_J=Q\left(\frac{\nu}{\mu_S}\right)^{\frac{1-\beta}{\beta}}\left(\frac{\mu_S}{Q}\right)^{\frac{1}{\beta}} \ ,
\end{equation}
where $\frac{\nu}{\mu_S}$ is an ${\cal O}(1)$ constant.   The scale $\nu$ is introduced to keep track of the arbitrariness in the scale setting, which will allow us to connect to the rapidity RG parameter $\nu$ introduced in \Ref{Chiu:2012ir}.   For convenience, we choose the factorization scale $\mu$ in \Eq{Renorm_xsec} to be the jet scale $\mu_J$, such that the evolution of the jet function trivial.  The soft  function must be evolved from $\mu_S$ to $\mu_J$ in order to resum logarithms.  In principle, the hard function also needs to be evolved from $\mu_H$ to $\mu_J$, but we will ignore that evolution here since are no divergences in the hard anomalous dimension at $\beta = 1$.

In the limit $\beta\to1$, the soft function RG solution from \Eq{eq:basic_RG_solution} is
\begin{equation}
\label{us_beta_one_limit}
\lim_{\beta\to 1}\log\frac{S(\Lang{\beta},\mu_J)}{S(\Lang{\beta},\mu_S)}=\lim_{\beta\to 1}\frac{4 C_F}{1-\beta}\int_{\mu_S}^{\mu_J}\frac{d\mu}{\mu}\frac{\alpha_s(\mu)}{\pi}\log\left(\Lang{\beta} e^{\gamma_E}\frac{\mu}{Q}\right)  \ .
\end{equation}
As $\beta\to 1$, the singularity in the anomalous dimension is regulated by the decreasing range of integration, since from \Eq{jet_scale}:
\begin{equation}
\mu_J = \mu_S \left(1 + (1- \beta) \log \frac{\nu}{Q} +\dotsc\right).
\end{equation}
At $\beta = 1$, we have 
\begin{equation}
\label{eq:betaonelimitSRG}
\lim_{\beta\to 1}\log\frac{S(\Lang{\beta},\mu_J)}{S(\Lang{\beta},\mu_S)}\equiv \log\frac{S(\Lang{\beta},\mu_J)}{S(\Lang{\beta},\mu_J,\nu)}=4\frac{\alpha_s(\mu_J)}{\pi}C_F\log\left(s_1 e^{\gamma_E}\frac{\mu_J}{Q}\right)\log\frac{\nu}{Q} \ .
\end{equation}
At the end of the evolution, $\mu_S=\mu_J$, but the number of renormalization scales remains constant for all $\beta$: when $\beta\neq 1$, $\mu_S$ and $\mu_J$ are the renormalization scales, and when $\beta = 1$, $\mu_J$ remains but the soft scale transmutes into the rapidity scale $\nu$.  Importantly, this result is precisely of the form dictated by the rapidity RG, with $\nu$ being the effective rapidity scale.  In particular, we show in \App{app:betaonecheck} that we can interpret \Eq{eq:betaonelimitSRG} as the rapidity evolution of the soft function:\footnote{This rapidity scale evolution is distinct from that presented in \Ref{Chiu:2012ir} because there broadening was a recoil-sensitive observable.  The effect of recoil modifies the non-cusp component of the rapidity anomalous dimensions from that for recoil-free broadening presented here.}
\begin{equation}
\nu \frac{d }{d \nu}\log S = -4\frac{\alpha_s(\mu)}{\pi}C_F\log\left(s_1 e^{\gamma_E}\frac{\mu}{Q}\right) \ .
\end{equation}

Our analysis in this section has been confined to the renormalization at one-loop for which only the cusp anomalous dimension is singular as $\beta\to 1$.  At higher loop order, one expects that the non-cusp anomalous dimension will also have a pole at $\beta =1$, which would subsequently contribute to the soft function evolution studied in this section.  Nevertheless, because the physical cross section is continuous through $\beta  = 1$, we expect that, to all orders, the terms in the anomalous dimensions that are singular at $\beta = 1$ should map directly onto the rapidity anomalous dimensions.  This also assumes that the RG evolution of the hard, jet and soft functions resums all (global) logarithms, but this should be guaranteed from the factorization theorem.

\section{Numerical Results}
\label{sec:mc}

Having established the resummation of broadening-axis angularities, we now show numerical results from our analytic calculation and compare them to results obtained from a showering Monte Carlo program and an automated resummation tool.

\subsection{Effect of Axes Choice}
\label{subsec:pythia}

To begin, it is instructive to see how the angularities depend on the various possible axes choices.  These effects are easiest to demonstrate using a showering Monte Carlo program.  Our event sample is $e^+ e^- \to u \bar{u}, d \bar{d}$ at a center-of-mass energy of 1 TeV, generated using \textsc{Pythia 8.165} \cite{Sjostrand:2006za,Sjostrand:2007gs}.   In order to isolate the final state parton evolution, we turn off initial state radiation and hadronization effects, though we maintain \textsc{Pythia}'s matrix element matching for the first emission \cite{Bengtsson:1986et}.  To later compare to our \nllp\ calculations, we take $\alpha_s(m_Z) = 0.118$ (different from \textsc{Pythia}'s default of 0.1383) and include two-loop running (different from the default of one-loop running).

\begin{figure}
\begin{center}
\subfloat[]{\label{subfig:mc_axes_broad}
\includegraphics[scale= 0.55]{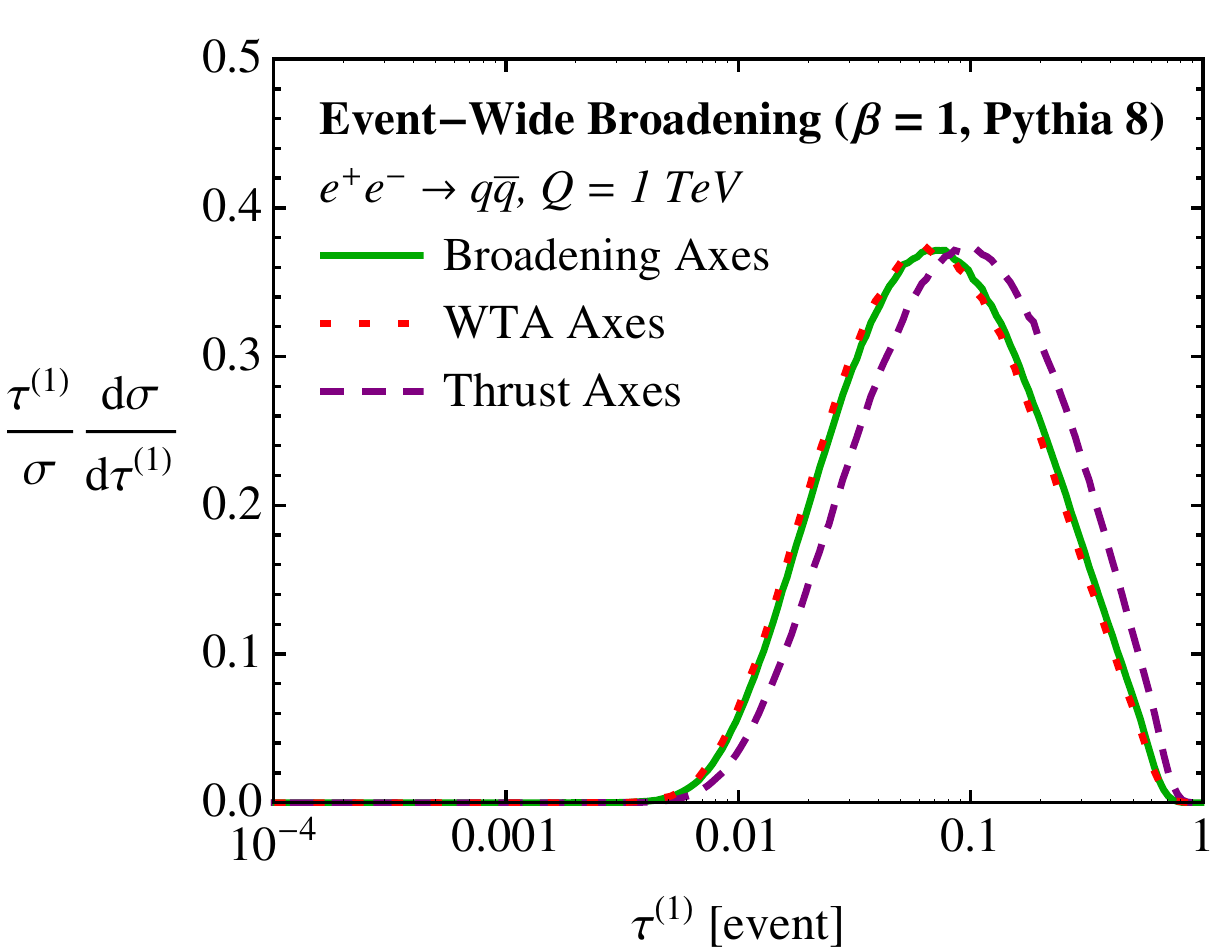}
}
$\quad$
\subfloat[]{\label{subfig:mc_axes_thrust}
\includegraphics[scale= 0.55]{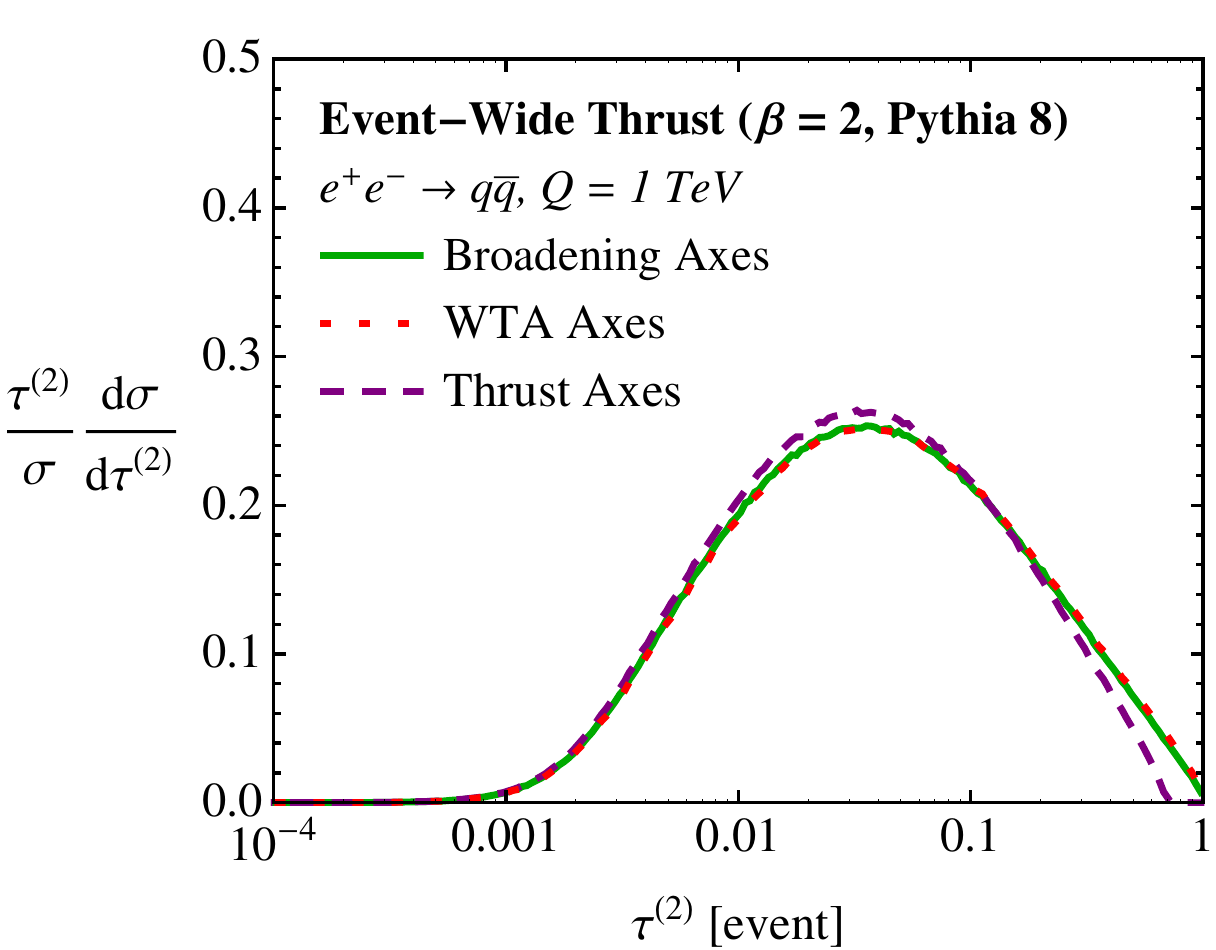}
}
\end{center}
\caption{Event-wide broadening ($\ang{1}$, left) and thrust ($\ang{2}$, right) in \textsc{Pythia 8.165}.  We test three different axes choices:  broadening axes, winner-take-all axes, and thrust axes.  The effect of recoil is seen clearly in the left figure, where the recoil-sensitive thrust axes give a larger value of $\ang{1}$ compared to the recoil-free axes.  In the right figure, all the curves are quite similar in the Sudakov peak region, since the effect of recoil is power-suppressed for $\ang{2}$.}
\label{fig:mc_axes}
\end{figure}

\begin{figure}
\begin{center}
\subfloat[]{
\includegraphics[scale= 0.55]{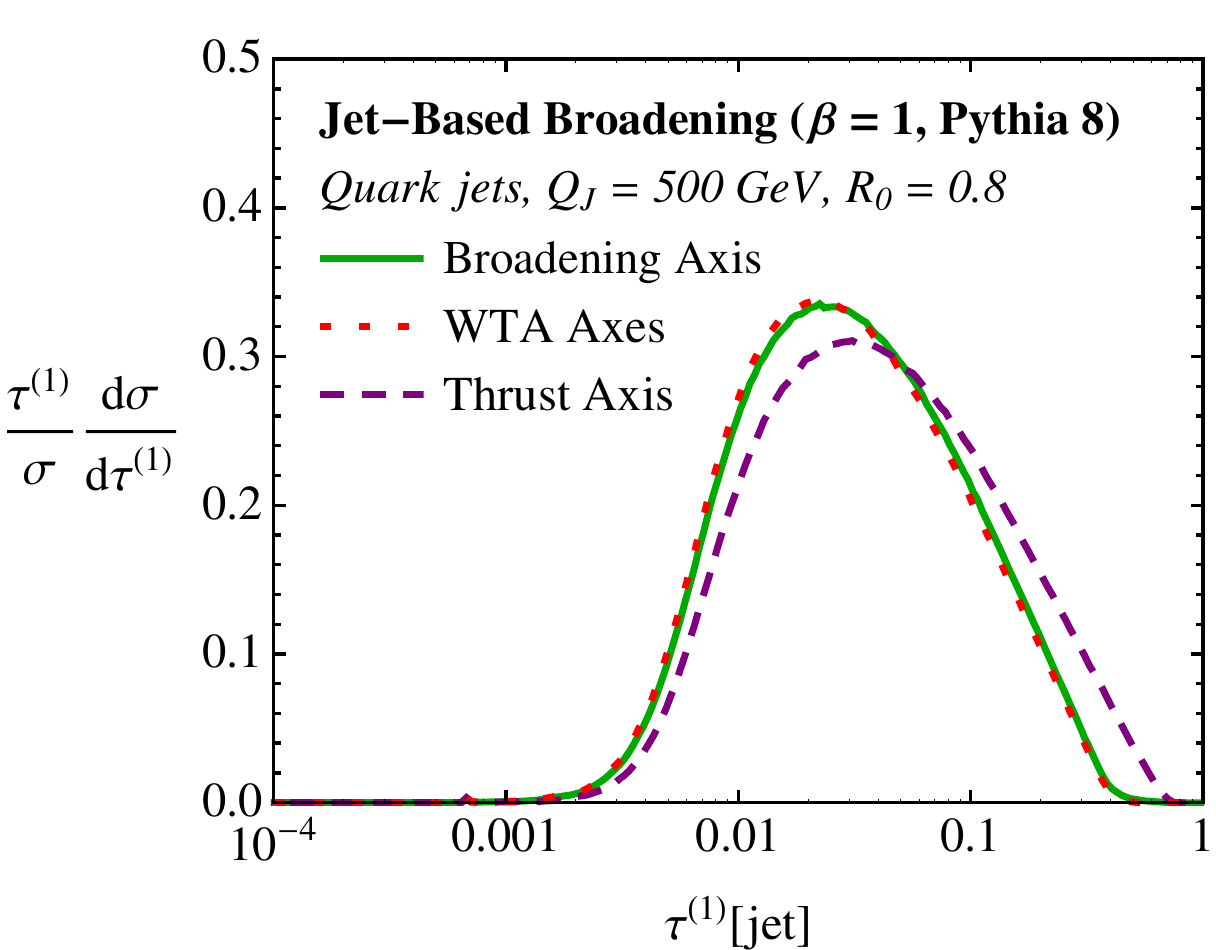}
}
$\quad$
\subfloat[]{
\includegraphics[scale= 0.55]{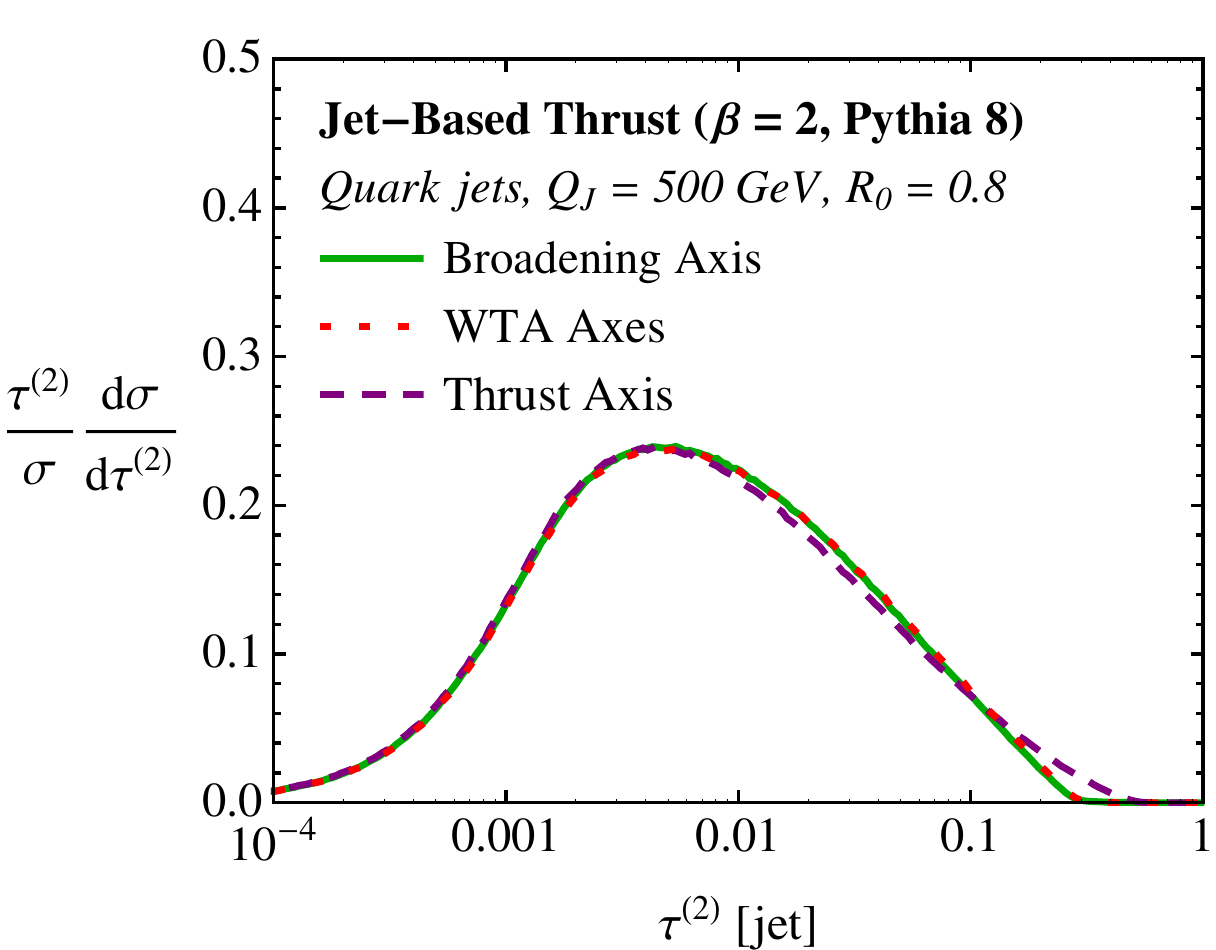}
}
\end{center}
\caption{Same as \Fig{fig:mc_axes}, but for the jet-based observable with $R= 0.8$.  As expected, the qualitative effect of recoil is the same as for the event-wide observables.}
\label{fig:mc_jetbroad}
\end{figure}

In \Fig{fig:mc_axes}, we show the normalized distributions for event-wide broadening and event-wide thrust for three different axes choices:  kinked broadening axes, kinked winner-take-all axes, and  back-to-back thrust axes.  To find the winner-take-all axis in each hemisphere, we use the Cambridge-Aachen (CA) clustering metric relevant for $e^+ e^-$ collisions \cite{Dokshitzer:1997in,Wobisch:1998wt,Wobisch:2000dk} and use the winner-take-all recombination scheme described in \Sec{subsec:WTA}.  We have plotted the results with a logarithmic abscissa, such that the Sudakov peak looks like a downward parabola.  For broadening in \Fig{subfig:mc_axes_broad}, the broadening-axis distribution is peaked at lower values than the thrust-axis one, as expected since the broadening axis minimizes broadening.  Amazingly, the broadening-axis and the winner-take-all-axis give nearly identical results.  For thrust in \Fig{subfig:mc_axes_thrust}, all of the distributions look quite similar, since the effect of recoil is formally power-suppressed at $\beta  = 2$.  That said, there are differences at large values of  $\ang{2}$ where the observables have different non-singular corrections.

In \Fig{fig:mc_jetbroad}, we take the same \textsc{Pythia} event sample but now calculate angularities on a single jet with radius $R = 0.8$ (measured by solid angle).  We fill the histogram with two jets per event (roughly corresponding to one from each hemisphere).  To find the jets for the broadening axis, we use $1$-jettiness as a jet algorithm with $\beta = 1$, using the hemisphere broadening axes as seeds for one-pass minimization \cite{Thaler:2011gf}.  For the thrust axis, we perform the same procedure, but using $\beta = 2$.  For the winner-take-all axis, we find the two hardest jet axes from CA clustering with winner-take-all recombination and $R = 0.8$, and to avoid boundary effects, we draw cones of radius $R$ around those axes.  The jet angularities exhibit roughly the same qualitative features as for the event-wide angularities, showing again that the recoil-free observables have very similar behavior.

\subsection{Broadening at NLL}
\label{subsec:NLL}

To better understand the difference between recoil-free and recoil-sensitive broadening, we can compare the two distributions to NLL order using our analytic calculation.  To see the effect of different color factors, we show distributions for two different processes at a center-of-mass energy of 1 TeV:
\begin{align}
\text{Quarks:} \quad e^+ e^- &\to q \bar{q}, \\
\text{Gluons:} \quad e^+ e^- & \to g g,
\end{align}
where the second process occurs through an off-shell Higgs boson.  We will consider both event-wide and jet-based broadening, but we will not account for non-global logarithms \cite{Dasgupta:2001sh} in the jet-based version.

For the (recoil-free) broadening-axis broadening, our NLL result is described by the anomalous dimensions given in \Eqs{eq:br_anom}{eq:br_anom_rap}, with the jet and soft functions set to their tree-level expressions (see \Eq{eq:NLLPorderFinal}).  For the (recoil-sensitive) thrust-axis broadening, we use the results of \Ref{Chiu:2012ir} which includes an additional transverse momentum convolution.  In both cases, we use the natural hard, jet, and soft scales given in \Eq{eq:natural_scales}.  Note that these NLL results are not directly comparable to the \textsc{Pythia} distributions in \Sec{subsec:pythia}, since they lack the $\mathcal{O}(\alpha_s)$ singular corrections which show up only at \nllp\ order.

\begin{figure}
\begin{center}
\subfloat[]{
\includegraphics[scale= 0.55]{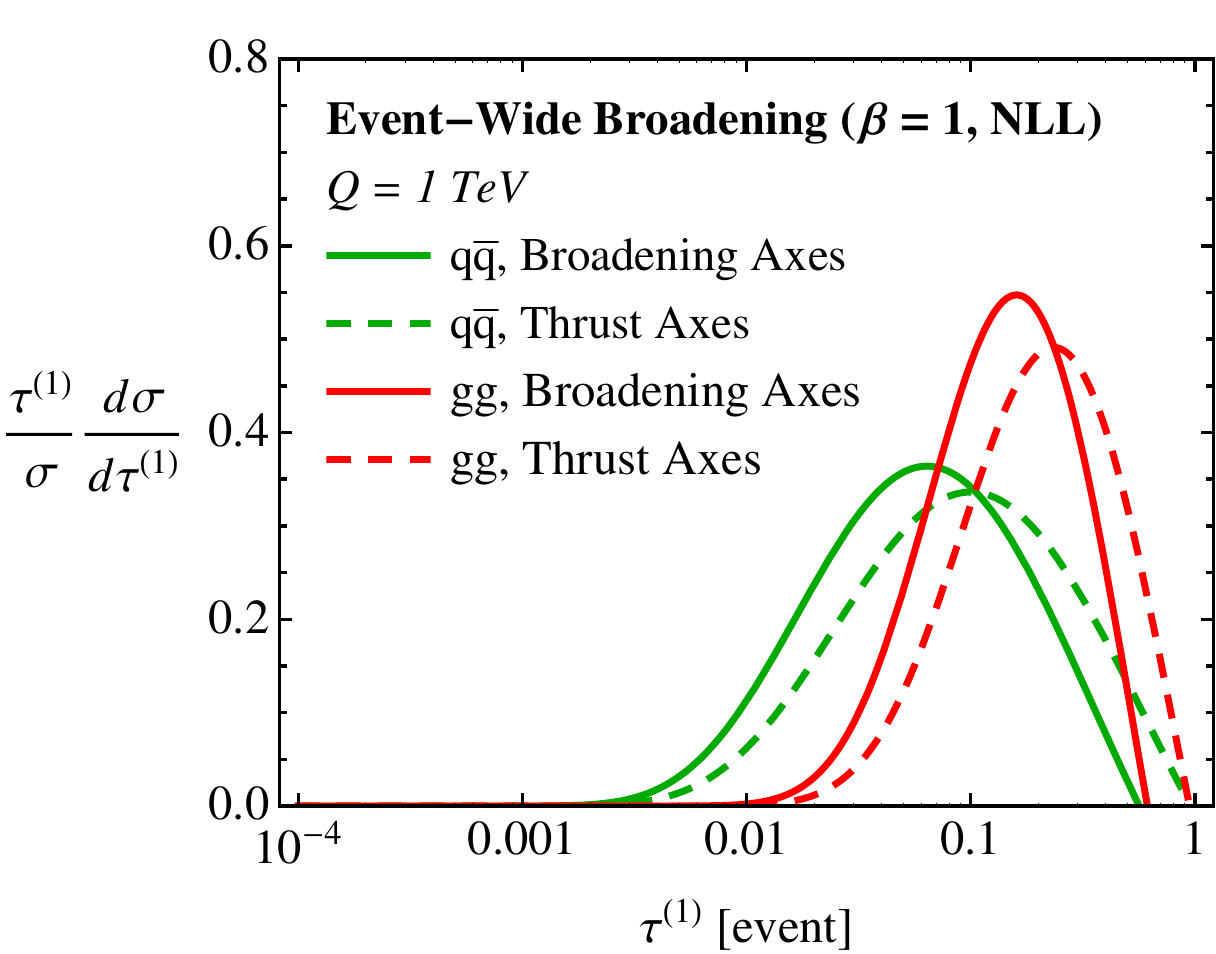}
}
$\qquad$
\subfloat[]{
\includegraphics[scale= 0.55]{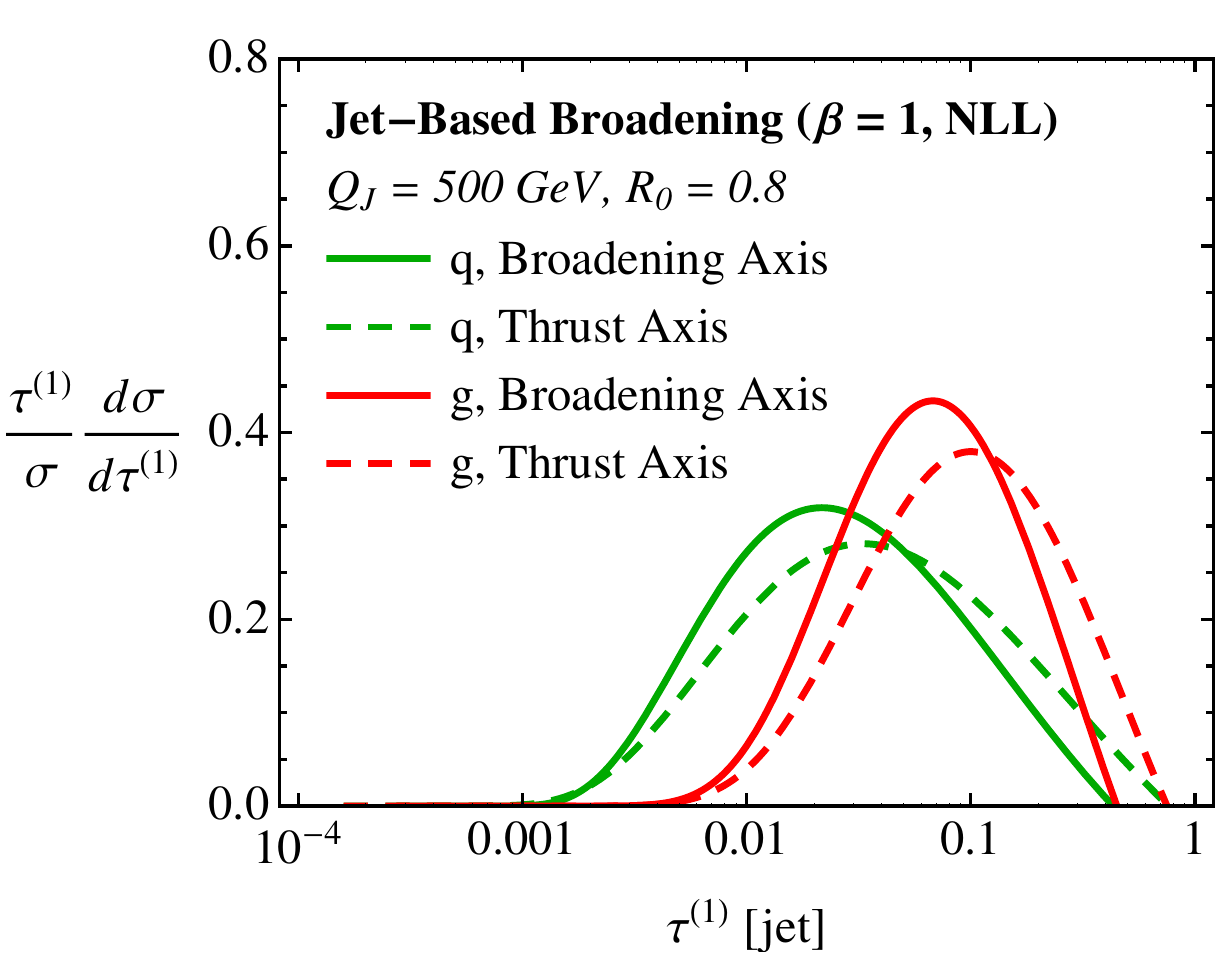}
}
\end{center}
\caption{Event-wide (left) and jet-based (right) broadening calculated with NLL resummation.  At this order, the distributions are functionally identical between the SCET and CAESAR calculational methods.  Shown are $e^+ e^- \to q \bar{q}$ (quarks) and $e^+ e^- \to gg$ (gluons) distributions, measured with respect to either the broadening axes or the thrust axes.  The effect of recoil is again seen in the relative shift between the two axes choices.}
\label{fig:caesar_broadthrust}
\end{figure}

\begin{figure}
\begin{center}
\subfloat[]{
\includegraphics[width=6.7cm]{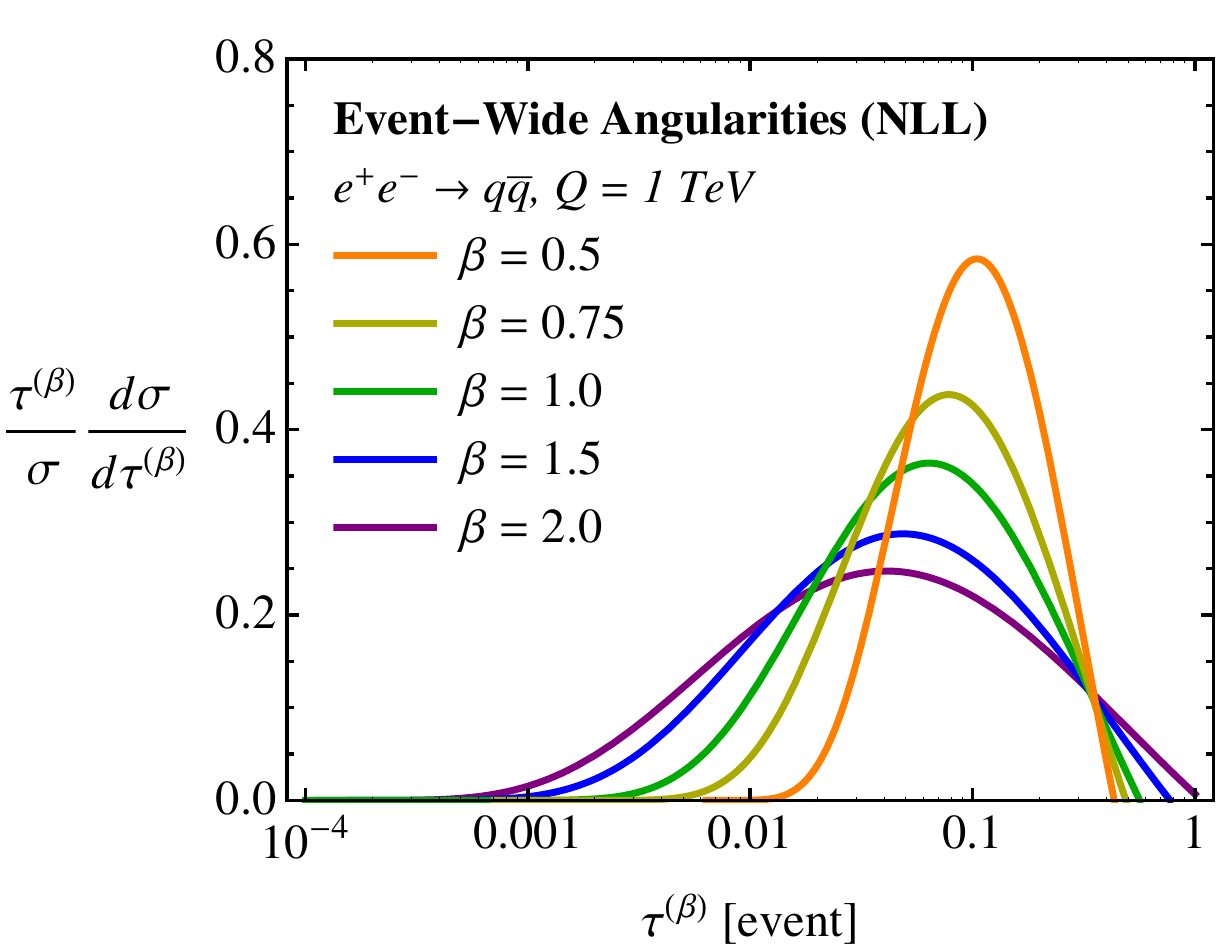}
}
$\qquad$
\subfloat[]{
\includegraphics[width=6.7cm]{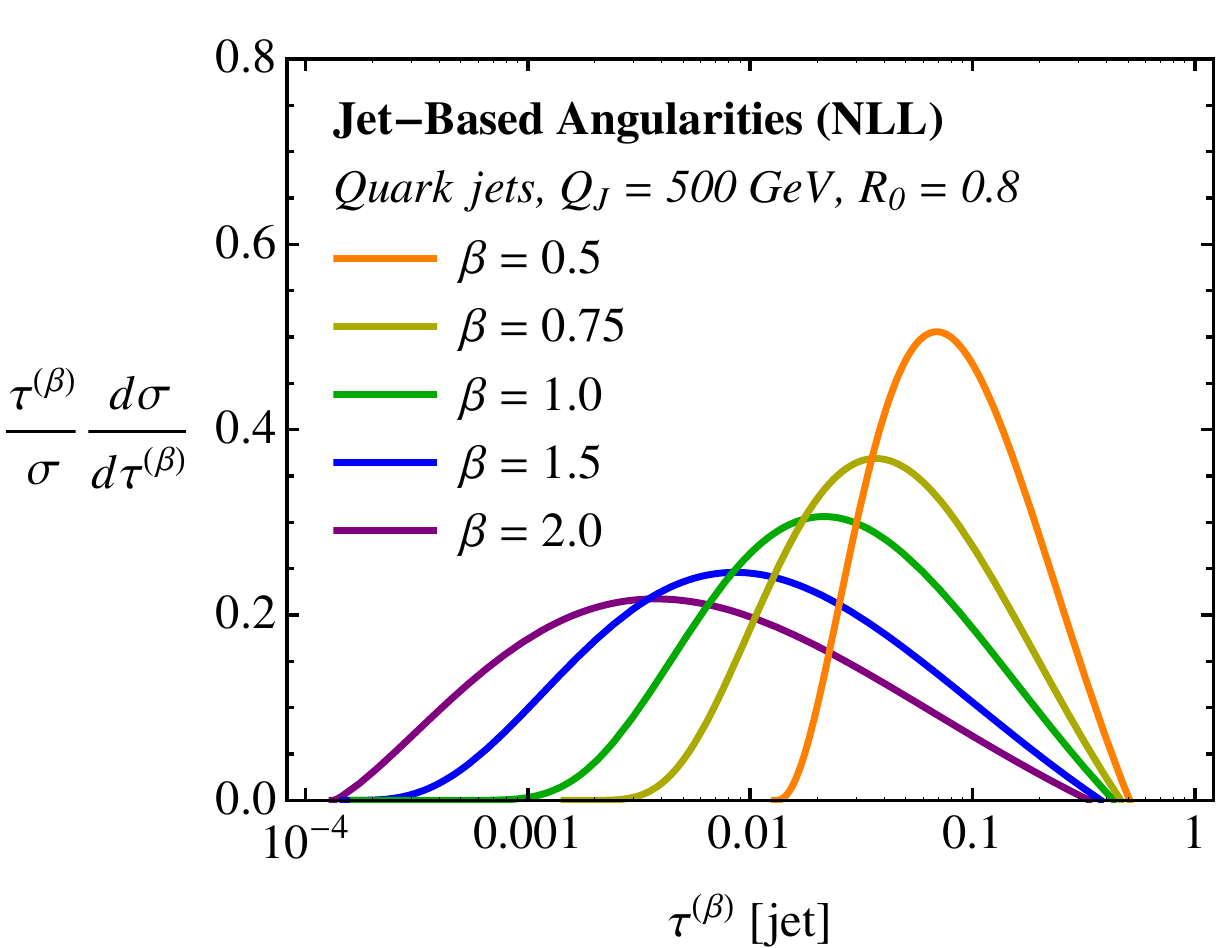}
}
\end{center}
\caption{Event-wide (left) and jet-based (right) angularities at NLL for $\beta = \{0.5,0.75,1.0,1.5,2.0\}$.  Here we use the $e^+ e^- \to q \bar{q}$ event sample and the recoil-free broadening axes.}
\label{fig:caesar_betasweep}
\end{figure}

As a cross check of our SCET factorization theorem, we can also compare our results to the CAESAR approach for NLL resummation \cite{Banfi:2004yd}.\footnote{Though CAESAR is available as an automated tool, here we are simply using the formulas given in \Ref{Banfi:2004yd}.}  Since recoil-free broadening and $\C{1}{1}$ have the same distribution at NLL order, we can use the CAESER result for $\C{1}{1}$ derived in \Ref{Banfi:2004yd} (where $\C{1}{\beta}$ is called $FC_x$ with $x = 2-\beta$).  For the thrust-axis broadening, we use the expression for the resummed cross section first computed in \Ref{Dokshitzer:1998kz}.   Interestingly, the CAESAR and SCET results are functionally \emph{identical}, as long as we use the scale choice in \Eq{eq:natural_scales}.\footnote{Strictly speaking, there is a small difference because we use two-loop $\alpha_s$ running throughout our computation, whereas the CAESAR formulae truncate the expansion to only include effects that are formally NLL order.}  Therefore, we will label the plots below as corresponding generically to ``NLL'', since the two calculational methods agree.

In \Fig{fig:caesar_broadthrust}, we show broadening for the two different event samples (quarks vs.\ gluons), the two different axes choices (broadening axes vs.\ thrust axes), and the two different particle selections (event-wide vs.\ jet-based).  As the color factor is changed from $C_F = 4/3$ (quarks) to $C_A = 3$ (gluons), the broadening distribution predictably moves to larger values.  Going from thrust axes to broadening axes decreases the value of broadening, in agreement with the behavior seen in \Fig{fig:mc_axes}.  The jet-based version is qualitatively similar to the event-wide version, but the distributions are pushed to lower values because of the soft rescaling in \Eq{eq:softrescalingtojet}.

As a further cross check of our factorization theorem at NLL, we can compare the recoil-free angularities between SCET and CAESAR for different values of $\beta$.  In the SCET calculation, we have already emphasized that the same factorization theorem can be used for all values of $\beta$.  Similarly, to get different values of $\beta$ in CAESAR, one needs only adjust the $b_\ell = \beta - 1$ parameter in the CAESAR cross section.    Again, we find identical functional forms between the SCET and CAESAR results to NLL order.  In \Fig{fig:caesar_betasweep}, we show the NLL recoil-free angularities for $\beta = \{0.5,0.75,1.0,1.5,2.0\}$, for the $e^+ e^- \to q \bar{q}$ sample.

\subsection{Effects at \nllp}
\label{subsec:NLLprime}

Our SCET result for the differential cross section of the recoil-free angularities is systematically improvable and as discussed in \Sec{subsec:NLLp_result}, the first realistic distributions are obtained at \nllp~order.  \nllp~order is defined as resummation to NLL accuracy and the inclusion of the ${\cal O}(\alpha_s)$ non-singular terms of the jet and soft functions.  At \nllp~order, it is important to properly choose renormalization scales, and we distinguish between the canonical scale choices from \Eq{eq:natural_scales} (\nllp$^{c}$) versus the more accurate profile method \cite{Abbate:2010xh} discussed in \App{sec:scalechoice} (which we indicate by just \nllp).  For the distributions of the recoil-free angularities in $e^+e^-\to q\bar{q}$ events, we will also consider the full ${\cal O}(\alpha_s)$ fixed-order corrections, which we will refer to as NLL$'+$NLO accuracy.  Here, we study the progression to higher accuracy and, in particular, emphasize the significant decrease in dependence on renormalization scale at \nllp~as compared to NLL order.

\begin{figure}
\begin{center}
\subfloat[]{
\includegraphics[height=5.05cm]{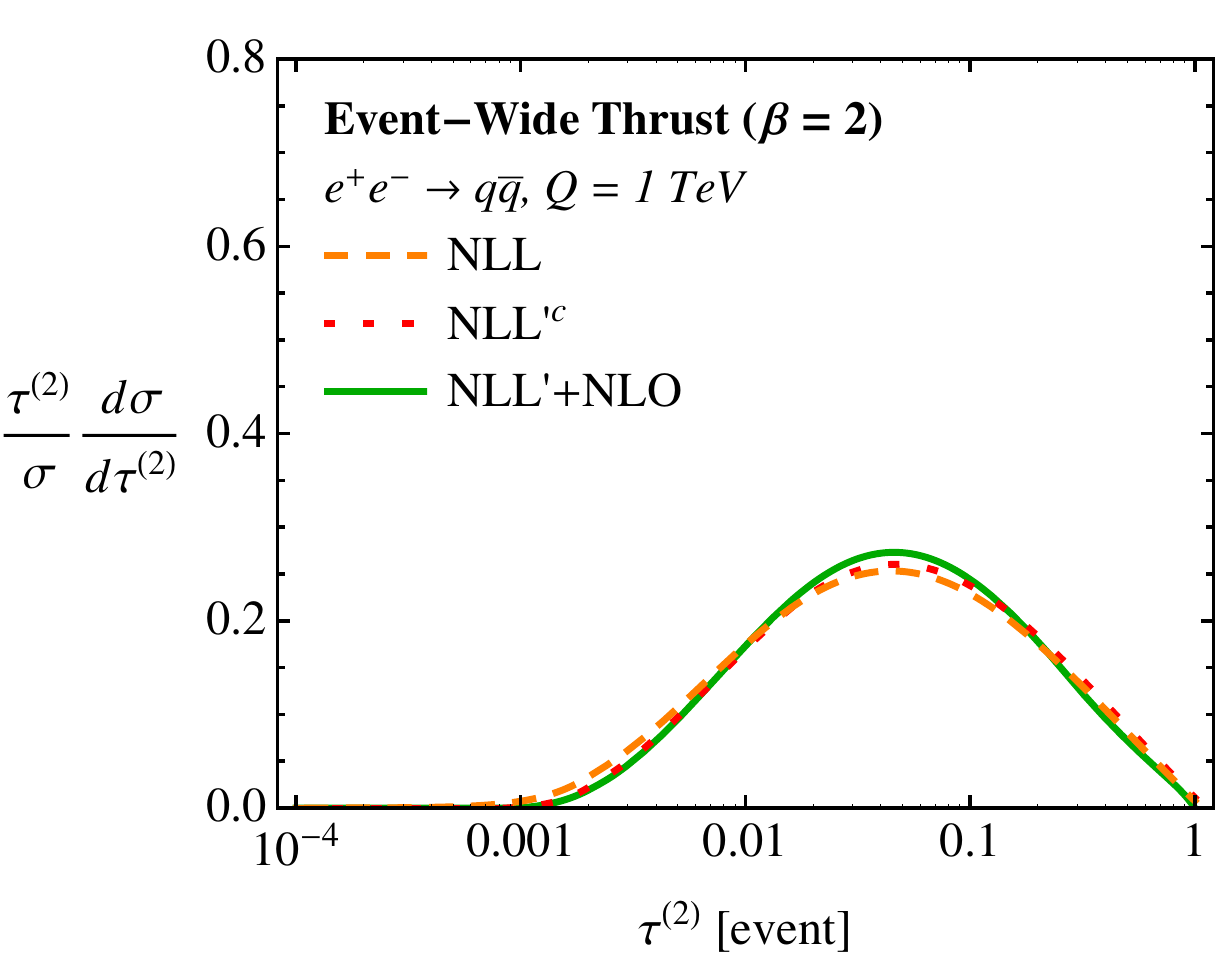}
}
$\quad$
\subfloat[]{
\includegraphics[height=5.05cm]{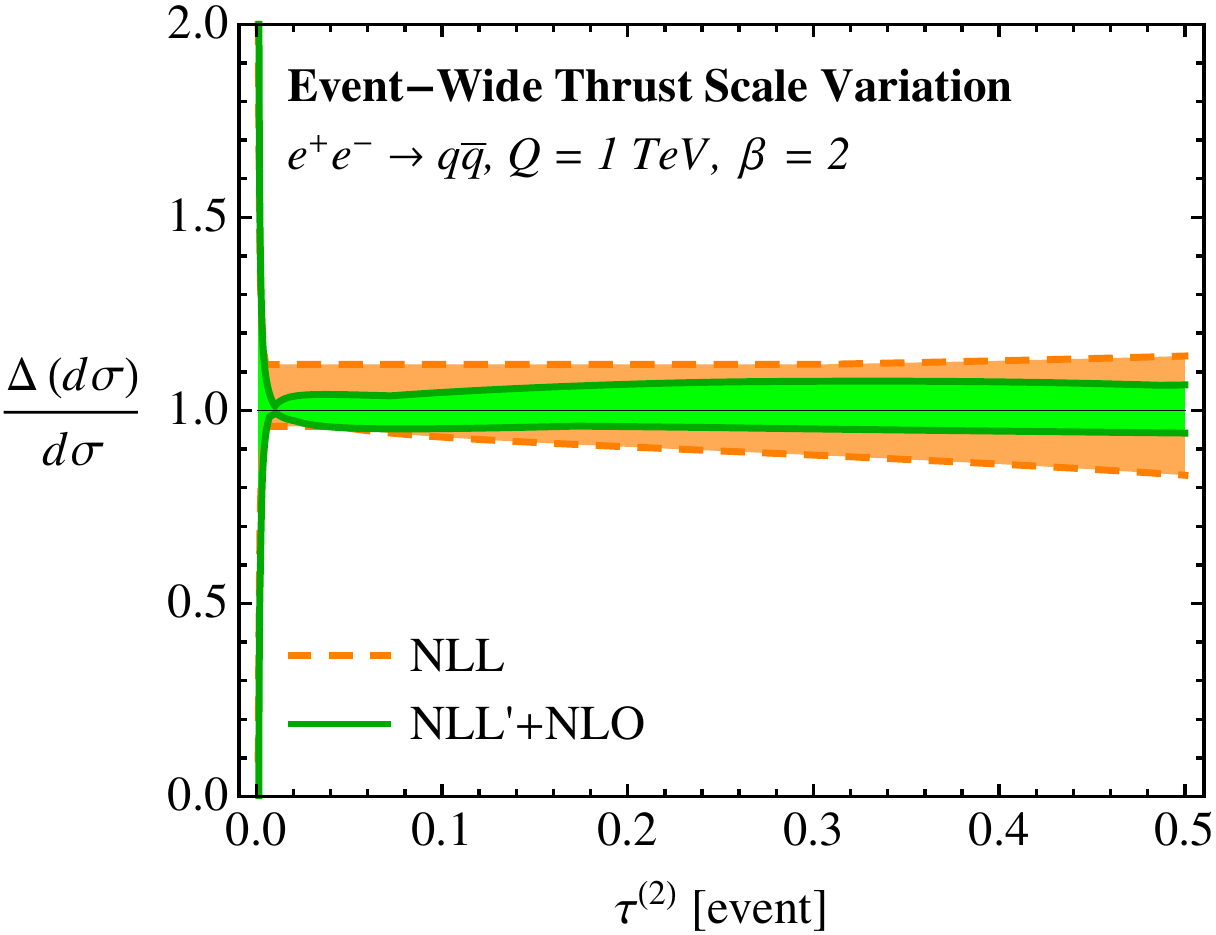}
} \\
\subfloat[]{
  \includegraphics[height=5.05cm]{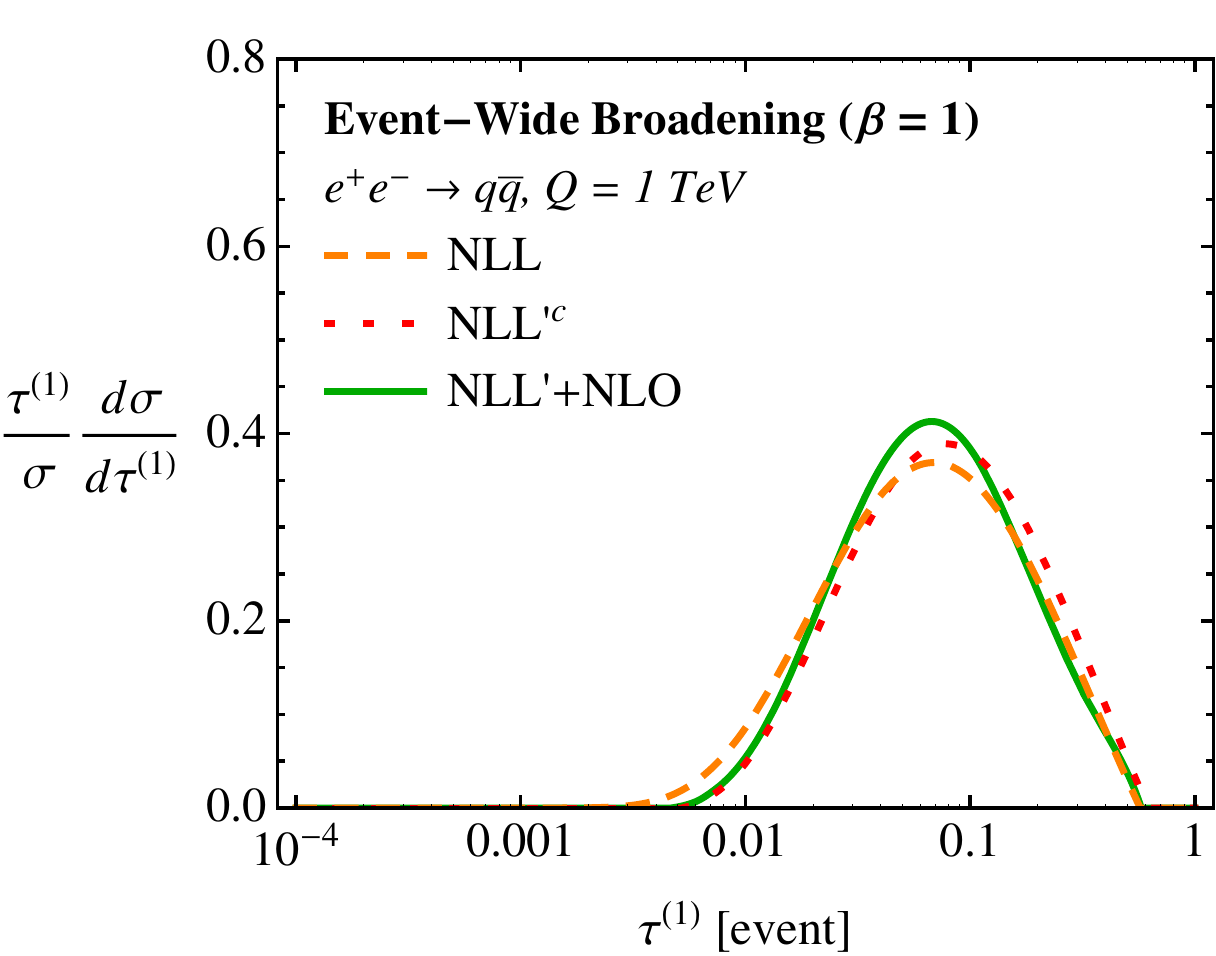}
}
$\quad$
\subfloat[]{
\includegraphics[height=5.05cm]{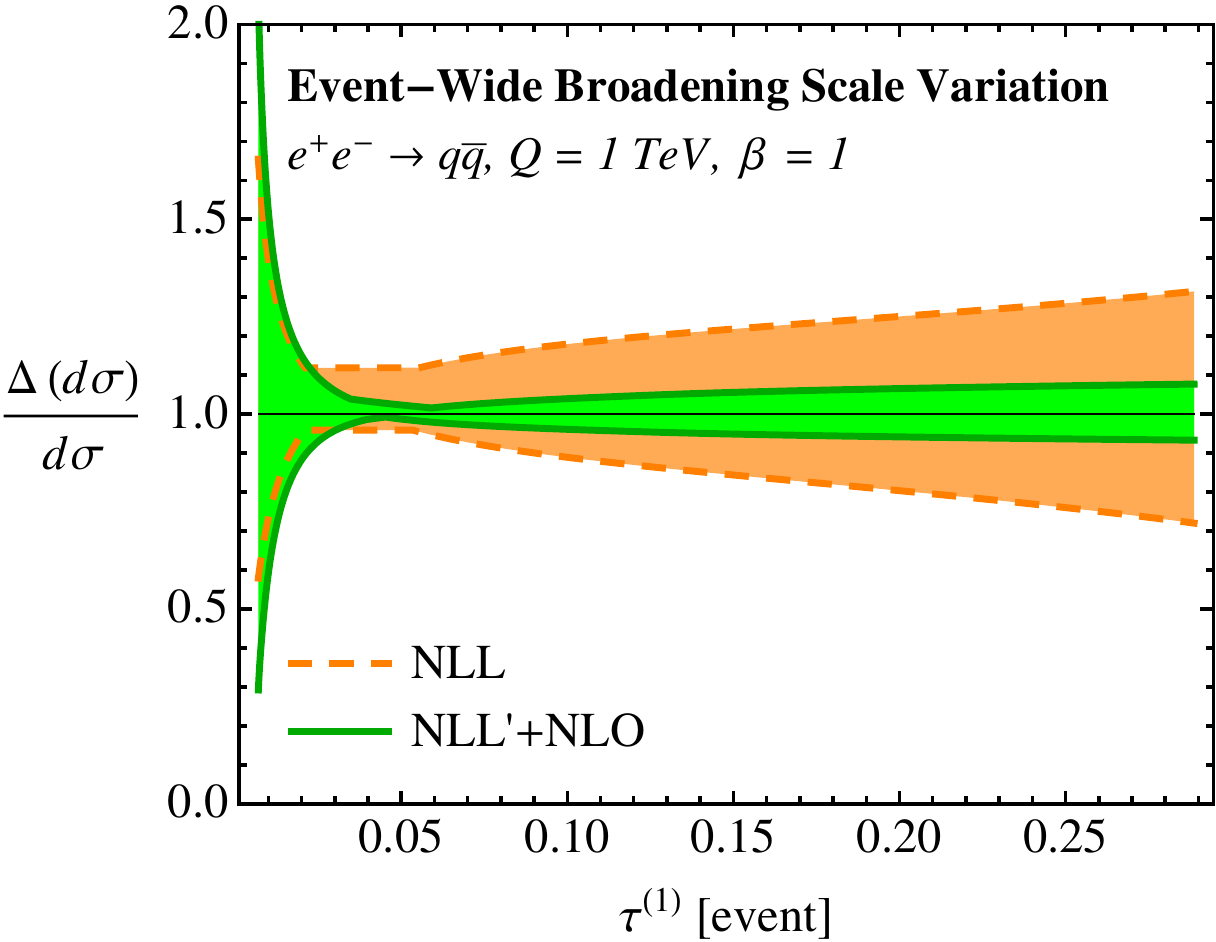}
}\\
\subfloat[]{\label{fig:beta0p5dist}
\includegraphics[height=5.05cm]{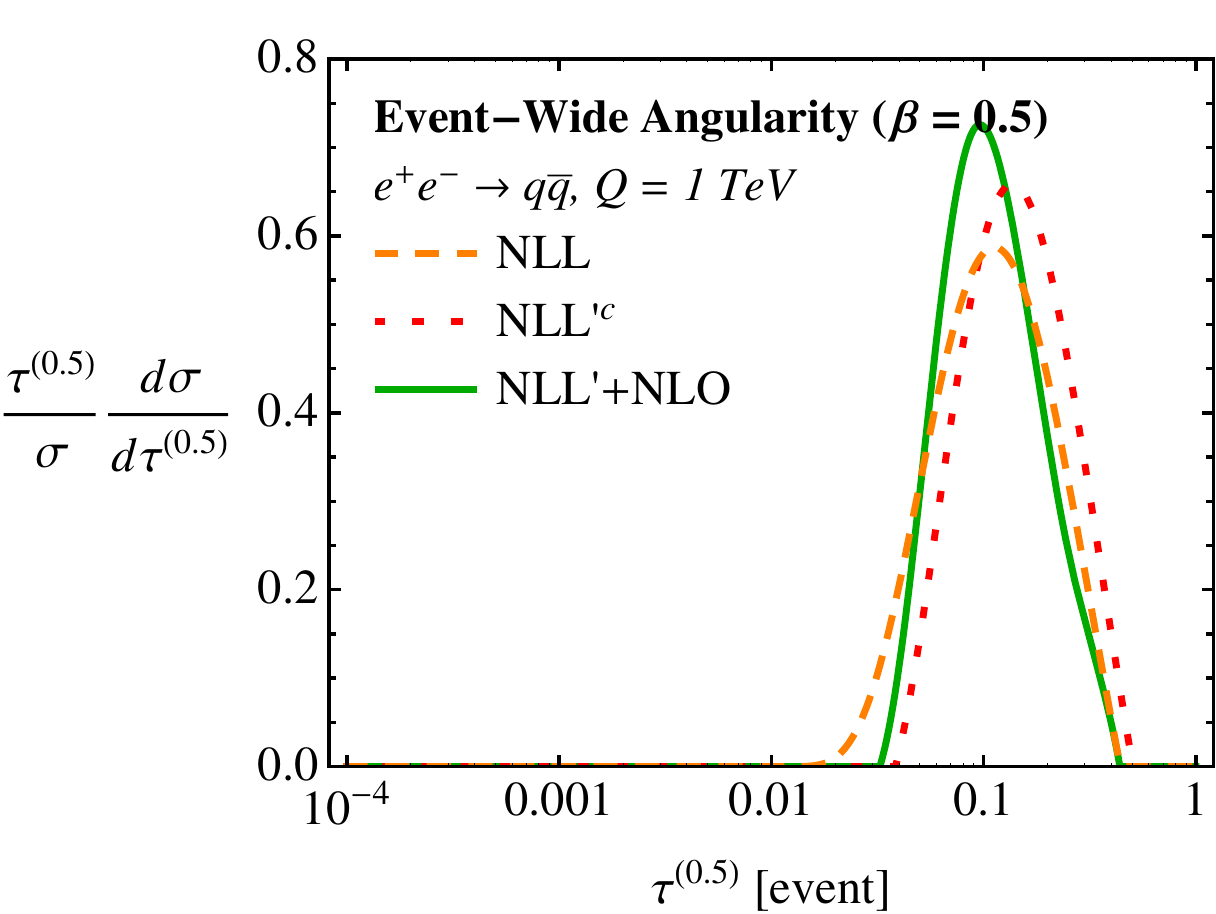}
}
$\quad$
\subfloat[]{
\includegraphics[height=5.05cm]{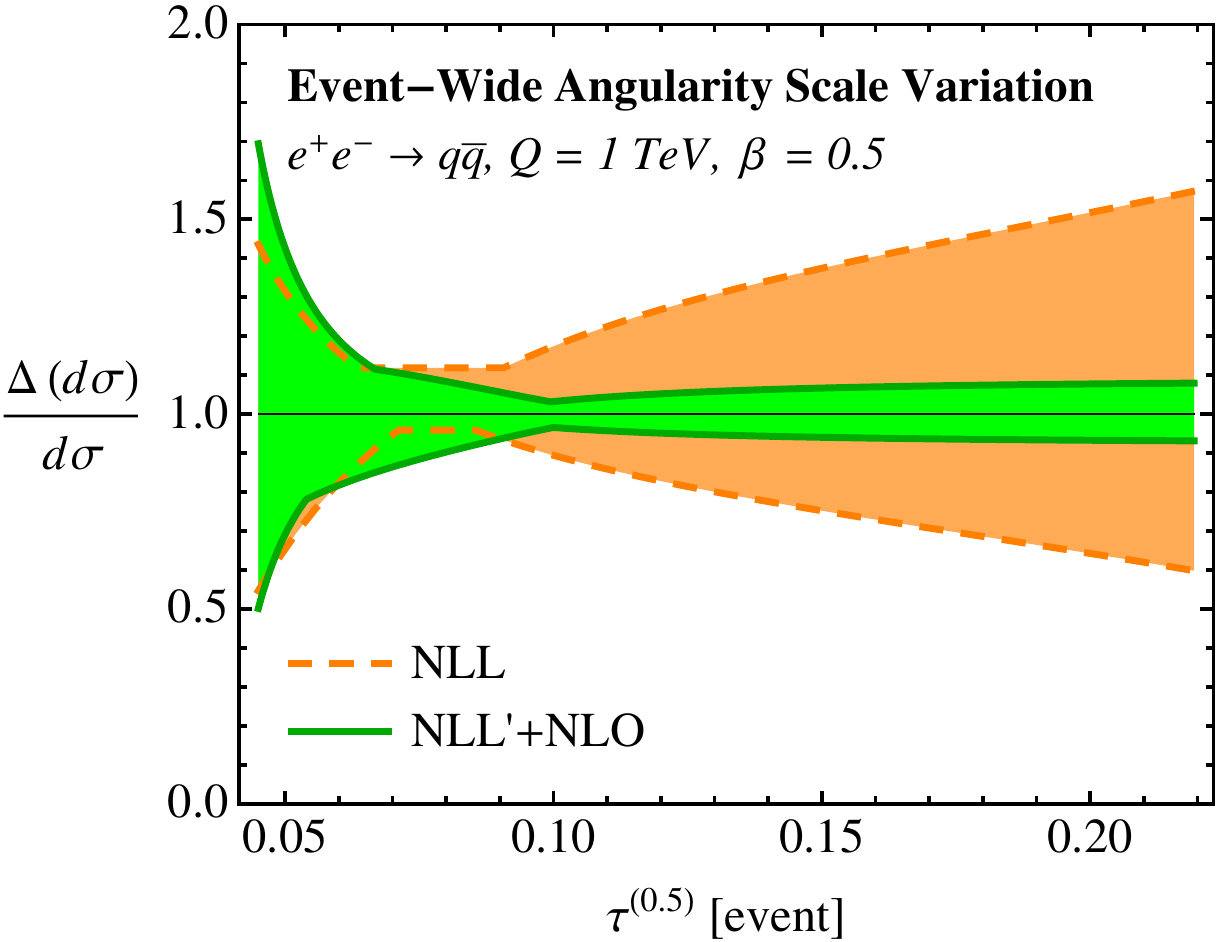}
}
\end{center}
\caption{
Event-wide broadening-axis angularities (left) and fractional uncertainties (right) in $e^+ e^- \to q \bar{q}$ events with $\beta = 2$ (top), $1$ (middle), and $0.5$ (bottom).  Here, we compare the distributions with NLL resummation, \nllp~resummation with the canonical scale choice (\nllp$^c$), and \nllp~with profiled scales matched to fixed-order (\nllp+NLO).  As $\beta$ decreases, the impact of \nllp~resummation and the choice of scales is enhanced, because the corresponding cusp anomalous dimension scales like $1/\beta$.  The uncertainties as defined by scale variation decrease significantly in going from NLL to \nllp\ order.  Note that the plotted range is reduced on the right-hand plots.
}
\label{fig:nll_to_nllprime}
\end{figure}

\begin{figure}
\begin{center}
\subfloat[]{\label{fig:nllprime_angvseec_NLLp}
\includegraphics[width=7cm]{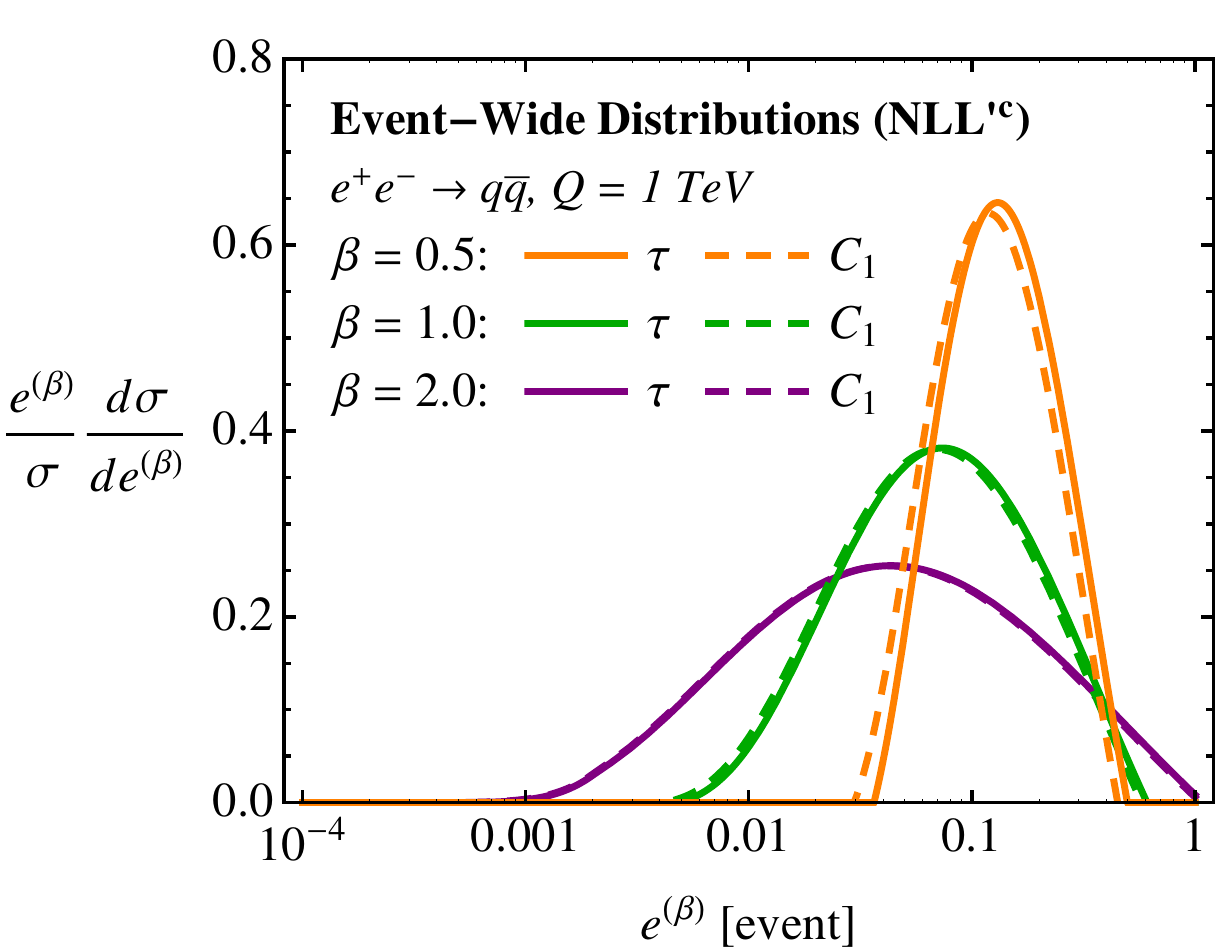}
}
$\quad$
\subfloat[]{\label{fig:nllprime_angvseec_pythia}
\includegraphics[width=7cm]{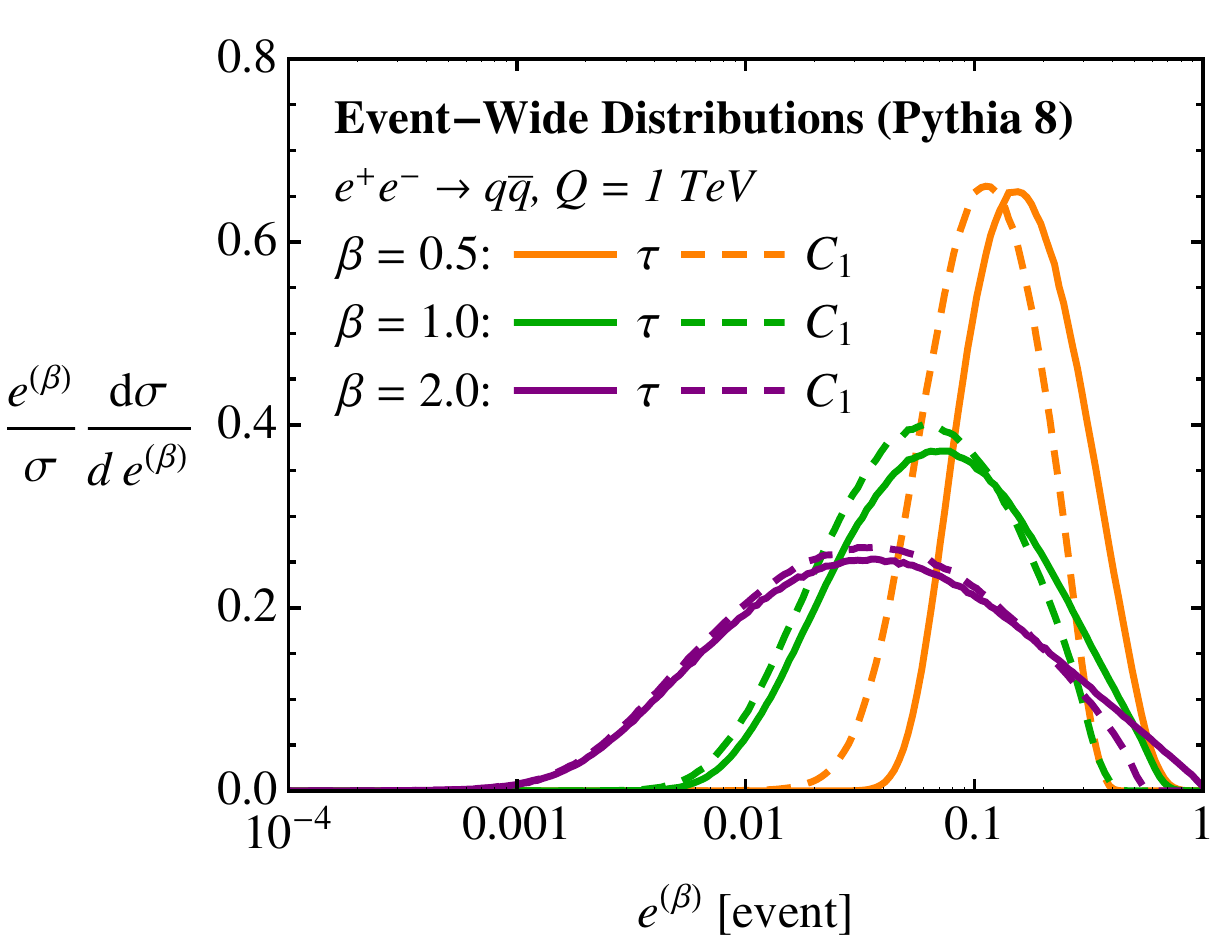}
}
\end{center}
\caption{Comparison of event-wide $\ang{\beta}$ and $\C{1}{\beta}$ in the \nllp$^c$~SCET calculation (left) and in \textsc{Pythia} (right).  Especially at small $\beta$, the difference is larger in the \textsc{Pythia} event sample because of important $\mathcal{O}(\alpha_s^2)$ non-singular effects that are absent from the \nllp\ result.}
\label{fig:nllprime_angvseec}
\end{figure}

\begin{figure}
\begin{center}
\subfloat[]{
\includegraphics[width=7cm]{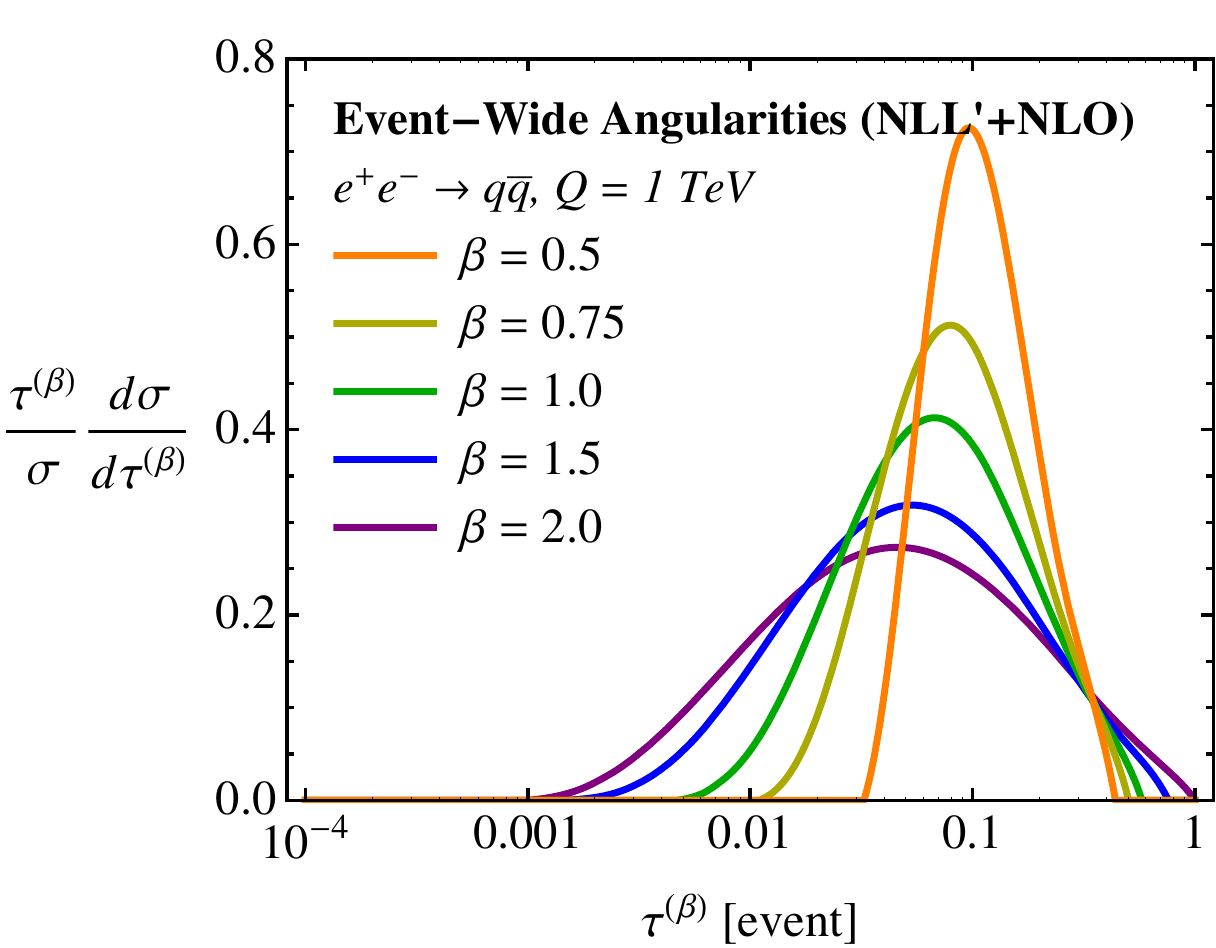}
}
$\quad$
\subfloat[]{
\includegraphics[width=7cm]{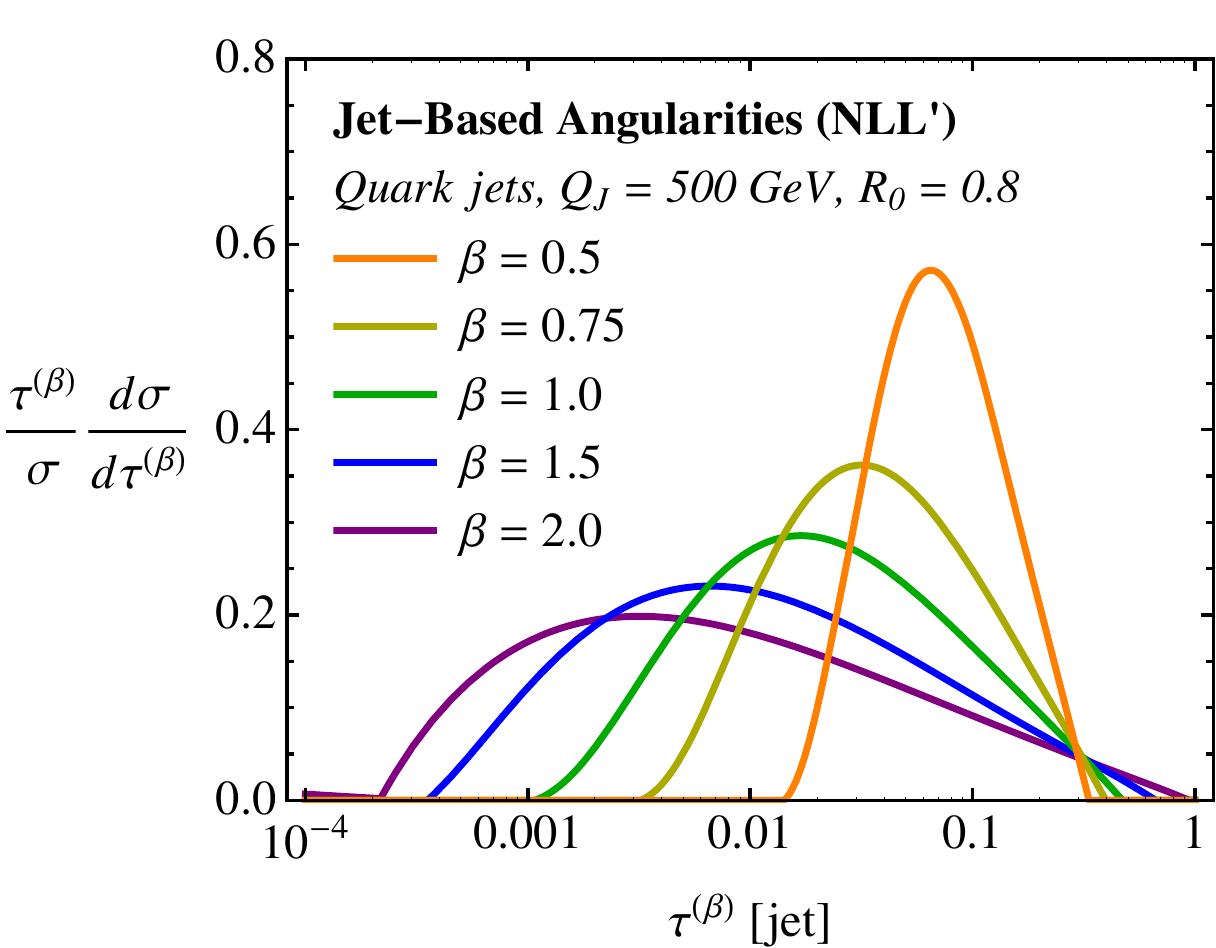}
}
\end{center}
\caption{Event-wide (left) and jet-based (right) angularities at \nllp\ order for $\beta = \{0.5,0.75,1.0,1.5,2.0\}$.  The \nllp\ event-wide distribution is further matched to the ${\cal O}(\alpha_s)$ fixed-order result (\nllp+NLO).  These plots are directly comparable to \Fig{fig:caesar_betasweep}.}
\label{fig:nllprime_betasweep}
\end{figure}

In \Fig{fig:nll_to_nllprime}, we compare the event-wide recoil-free angularities at NLL, \nllp$^{c}$, and \nllp+NLO for $\beta = 2,1,0.5$.  The differences between the distributions is small at large $\beta$, but because the cusp anomalous dimension scales like $1/\beta$, \nllp~corrections grow as $\beta$ decreases.  The difference between canonical scales and profiled scales is large for $\beta = 0.5$, which is indicative of large uncertainties in the \nllp~calculation.  We also compare the dependence of the distributions on the renormalization scale $\mu$ at NLL and \nllp+NLO.  For the NLL distribution, we vary the renormalization scale in the hard and soft functions up and down by a factor of 2 and the scale variation for the \nllp+NLO distribution is discussed in \App{sec:scalechoice}.  The error bands are defined by the envelope of all scale variations.  This nicely illustrates the reduced scale dependence at \nllp+NLO compared to NLL.

As argued in \Sec{subsec:anomdimrelatsion}, the broadening-axis angularities $\ang{\beta}$ and the energy correlation functions $\C{1}{\beta}$ have identical anomalous dimensions and therefore identical NLL distribution.  In going from NLL to \nllp\ order, though, the $\ang{\beta}$ and $\C{1}{\beta}$ distributions are no longer identical, since finite terms in their jet and soft functions differ.  In \Fig{fig:nllprime_angvseec}, we compare event-wide $\ang{\beta}$ and $\C{1}{\beta}$ in both a \nllp$^c$ calculation and in the \textsc{Pythia} event sample from \Sec{subsec:pythia}.  Note that $\C{1}{1}$ is generally smaller than $\ang{1}$, which can be understood because for a hemisphere with two constituents with energy fractions $z$ and $(1-z)$, $\C{1}{1}$ is proportional to $z (1-z)$ while $\ang{1}$ is proportional to just $z$.  For the \nllp$^c$ result in \Fig{fig:nllprime_angvseec_NLLp}, the resulting difference is rather mild.  For the \textsc{Pythia} distributions in \Fig{fig:nllprime_angvseec_pythia}, however, there is a much larger offset between $\ang{\beta}$ and $\C{1}{\beta}$, especially at small $\beta$.  The most dramatic difference is for $\ang{0.5}$, for which the peak region lies at large values of $\ang{0.5}$, where non-singular corrections as important as resummation.  In fact, the \textsc{Pythia} distribution for $\ang{0.5}$ extends well beyond the end point of the $\mathcal{O}(\alpha_s)$ fixed-order cross section (see \Fig{fig:beta0p5dist}), suggesting that this part of the distribution is being dominated by non-singular out-of-plane emissions, which only show up at $\mathcal{O}(\alpha_s^2)$.  For $\C{1}{\beta}$, there is surprisingly good quantitative agreement between \nllp$^c$ and \textsc{Pythia}, suggesting that this observable is less sensitive to higher order corrections than the angularities.

Finally for completeness, we show the event-wide and jet-based angularities for a wider range of $\beta$ values in \Fig{fig:nllprime_betasweep}.  We emphasize that these distributions were all obtained with the same factorization theorem, and the $\beta = 1$ curve was obtained by using $\beta = 1.001$, exploiting the continuity of the cross section through $\beta = 1$.

\section{Historical Perspective}
\label{sec:history}

We have seen that the broadening axis yields very interesting properties with respect to factorization and resummation of event shapes.  It is therefore curious why the broadening axis was not previously defined in the 35+ years of development of QCD observables.  Indeed, the history is quite interesting, and we will give a brief historical perspective on the ``spherocity axis'', the intellectual precursor to the broadening axis.

An early event shape observable introduced to study the jetty nature of QCD was sphericity \cite{Bjorken:1969wi,Ellis:1976uc}, defined as
\begin{equation}
S = \frac{3}{2}\min\frac{\sum_i p_{T i}^2}{\sum_i p_i^2},
\end{equation}
where the sums are over all particles in the event.  However, sphericity is not IRC safe, which led to the development of spherocity by Georgi and Machacek \cite{Georgi:1977sf} in 1977.  Spherocity, which is IRC safe, is defined as
\begin{equation}
S' = \left( \frac{4}{\pi} \right)^2 \left(\frac{\sum_i |p_{\perp i}|}{\sum_i |p_i|}\right)^2,
\end{equation}
where the transverse momentum is measured with respect to the spherocity axis.  The spherocity axis minimizes the scalar sum of momentum transverse to it:
\begin{equation}
\text{Spherocity axis}~\hat{s}: \quad \min_{\hat{s}} \sum_{i \in \text{event}} \left|\hat{s} \times  \vec{p}_i \right|.
\end{equation}
The spherocity axis only differs from the broadening axis by the fact that the broadening axis only considers particles in a single hemisphere (or jet region) of an event.  Like the broadening axis, \Ref{Georgi:1977sf} notes that the spherocity axis ``typically...will be the direction of the largest particle momentum''.  However, this distinction between a global spherocity axis (proposed in \Ref{Georgi:1977sf}) and two kinked broadening axes (defined here in \Sec{sec:broad_def}) had very important consequences for the study of spherocity in the future.

Shortly after spherocity was introduced, Farhi defined thrust  \cite{Farhi:1977sg} and the thrust axis.  Applied to an entire event, the thrust axis maximizes the scalar sum of momentum longitudinal to it:\footnote{This definition is equivalent to \Eq{eq:introthrustaxisdef} applied to each hemisphere of an $e^+e^-$ event.}
\begin{equation}
\text{Thrust axis}~\hat{t}: \quad \max_{\hat{t}} \sum_{i \in \text{event}} |  \hat{t} \cdot \vec{p}_{i}|.
\end{equation}
In the following years, it was realized that while the spherocity and thrust axes are identical at ${\cal O}(\alpha_s)$ in perturbation theory for $e^+e^-$ collisions, the thrust axis was greatly preferred.  The thrust axis nicely partitions the event into two hemispheres of equal and opposite momentum, is very stable to perturbations of the momenta of particles in the event, and can be determined by an exact procedure for events with arbitrary numbers of particles.  By contrast, the spherocity axis does not divide events nicely into two hemispheres, is very sensitive to perturbations of the momenta of particles in the event, and admits no analytic procedure to determine its direction.  For these reasons and others, by the early 1980s, spherocity was no longer seriously considered as a sufficiently interesting QCD observable, both theoretically and experimentally.\footnote{The only experimental measurements of spherocity that we know of are in \Refs{Schmitz:1979fc,Ford:1989iz,Achard:2004zw}.}

In the early 1990s, the resummation of large logarithms in perturbative cross sections was becoming more well understood, with the resummation of thrust to NLL in \Ref{Catani:1991kz} and heavy jet mass to NLL in \Ref{Catani:1991bd}.\footnote{To our knowledge, the spherocity distribution has never been resummed.}  In 1992, event-wide broadening with respect to the thrust axis was studied in \Ref{Catani:1992jc} by Catani, Turnock, and Webber (CTW), and CTW claimed to resum broadening to NLL.  In their calculation, they approximated the recoil of the hard quark off of the jet axis by half of the value of the broadening.  A few years later, Dokshitzer, Lucenti, Marchesini, and Salam (DLMS) realized that the CTW calculation of broadening neglected to properly account for the effect of the vector sum of soft emissions on the direction of the quark, which was necessary for NLL resummation \cite{Dokshitzer:1998kz}.  

At first glance, the observation of these recoil effects by DLMS would seem to 
imply that the CTW calculation was simply obsolete.  As we have seen in this paper, though, ignoring the effect of the vector sum of soft emissions on the recoil in broadening is the same (to leading power) as measuring recoil-free broadening with respect to two kinked broadening axes.  Therefore CTW had actually calculated the recoil-free broadening to NLL, over 20 years prior to this paper!

Going back to the CTW paper, they make a direct comparison between broadening and spherocity  \cite{Catani:1992jc}:
\begin{quote}
Thus [broadening] is similar to the spherocity, except that it is defined with respect to the thrust axis instead of being minimized with respect to the choice of axis.
... As far as we know, there is no convenient procedure for finding the spherocity axis and it does not permit the simple resummation of logarithmic contributions.
\end{quote}
In light of subsequent developments, one has to appreciate some of the irony in this last statement about resummation.  While the thrust axis is convenient for many purposes, it leads to recoil-sensitivity in observables like broadening, making resummation more difficult.  Despite the apparent drawbacks of the spherocity axis, if one allows for a separate spherocity axis (i.e.~broadening axis) in each event hemisphere, then spherocity/broadening becomes a recoil-free observable, greatly simplifying factorization and resummation.  And while there is no convenient procedure for finding the broadening axis, the winner-take-all axis (\Sec{subsec:WTA}) is easily obtained via recursive clustering and has the same recoil-free properties as the broadening axis.  So while the QCD community had good reasons to dismiss the (global) sphericity axis initially, we see with the benefit of hindsight that the (local) broadening axis has a special role to play in defining recoil-free observables.

\section{Conclusions}
\label{sec:conclude}

In this paper, we have (re)introduced the broadening axis.  While at first glance, the jet momentum axis (i.e.\ the thrust axis) would seem like the most natural choice for jet studies, the four-momentum of a jet includes contributions from both collinear and soft modes.  In contrast, the broadening axis is recoil free, meaning that it is independent of soft degrees of freedom and only depends on the kinematics of the collinear modes.  Using the broadening axis dramatically simplifies the structure of the factorization theorem for angularities at all $\beta > 0$, including broadening itself at $\beta = 1$.  While previous studies of thrust-axis angularities required a different treatment of $\beta = 1$ compared to $\beta \ge 2$, our \nllp\ resummation of the broadening-axis observables achieves a smooth interpolation between the \scetii\ and \sceti\ regimes.

Here, we used the broadening axis to define recoil-free jet observables, but one could envision other applications of the broadening axis, particularly at the LHC.  By definition, a recoil-free axis is insensitive to perturbative soft radiation, but this same requirement means that it is insensitive to other kinds of soft jet contamination, including initial state radiation, underlying event, pileup contamination, detector noise, and even the QCD fireball in heavy-ion collisions.  The advantage of defining jets with respect to the broadening axis is that the jet center will always be aligned along the hard collinear modes, though of course the jet energy will still be impacted by soft effects.  In practice, we suspect that the winner-take-all axis (instead of the broadening axis) will become the preferred recoil-free axis for future jet studies, since it can be easily implemented in existing jet clustering algorithms like anti-$k_T$.

In the context of jet substructure studies, the broadening axis can be used to measure the degree of recoil within a jet.  Because a jet from a boosted object with $N$ hard prongs is unlikely to have one of its prongs aligned along the jet momentum axis, one can use the angle between the thrust axis and the broadening axis as a boosted object discriminant.\footnote{A similar observation was made in \Ref{tmb}.  Amusingly, for a single emission within a jet, the thrust-broadening angle is identical to recoil-free broadening, thus the two observables have the same resummation to NLL order, though not the same factorization theorem.}  This kind of logic was exploited in \Ref{Curtin:2012rm} to define ``axis pull'' and ``axis contraction'' variables based on $N$-subjettiness.  Similarly, measuring the thrust-broadening angle may help in discriminating pileup jets from QCD jets \cite{CMS:2013wea}.

Finally, our focus has been on perturbative aspects of the broadening axis, but it would be interesting to study non-perturbative power corrections to $\ang{\beta}$, especially for $0 < \beta < 1$.  The simple form of the broadening-axis factorization theorem suggests that it should be straightforward to include appropriate non-perturbative shape functions \cite{Korchemsky:1999kt,Korchemsky:2000kp}.  We expect that the form of broadening-axis power correction should differ from the thrust-axis case \cite{Dokshitzer:1998qp}, and it may be that the broadening-axis case more resembles the earlier broadening power correction study in \Ref{Dokshitzer:1998pt}.  Of particular interest would be to compare the non-perturbative corrections between the angularities $\ang{\beta}$ and the energy correlation functions $\C{1}{\beta}$ for $0 < \beta < 1$.  While both are recoil-free, $\ang{\beta}$ is sensitive to $\mathcal{O}(\Lambda_{\rm QCD}/Q)$ angles between individual collinear modes and the broadening axis while $\C{1}{\beta}$ is sensitive to $\mathcal{O}(\Lambda_{\rm QCD}/Q)$ angles between all pairs of collinear modes.  Hopefully, such a study would shed light on the comparative advantages of axis-based versus axis-free jet observables.

\begin{acknowledgments}
We thank Steve Ellis, Gavin Salam, and Iain Stewart for helpful discussions.  This work is supported by the U.S. Department of Energy (DOE) under cooperative research agreement DE-FG02-05ER-41360. D.N. is also supported by an MIT Pappalardo Fellowship.  J.T. is also supported by the DOE Early Career research program DE-FG02-11ER-41741 and by a Sloan Research Fellowship from the Alfred P. Sloan Foundation.
\end{acknowledgments}

\appendix

\section{Broadening Axis for Three Coplanar Particles}
\label{sec:3 particle broadening axis}

Unlike for the thrust axis, there is no closed form expression for the broadening axis in a jet with an arbitrary number of particles.  Indeed, no closed form exists even for a jet with only three particles.  Here, we will consider a jet with three constituents in special phase space configurations so as to explore the behavior of the broadening axis.  This example will illustrate how the broadening axis is affected by emissions that are moderately soft as well as explicitly show how the broadening axis is IRC safe.

\begin{figure}
\begin{center}
\subfloat[]{\label{fig:init_config}
\includegraphics[width=6cm]{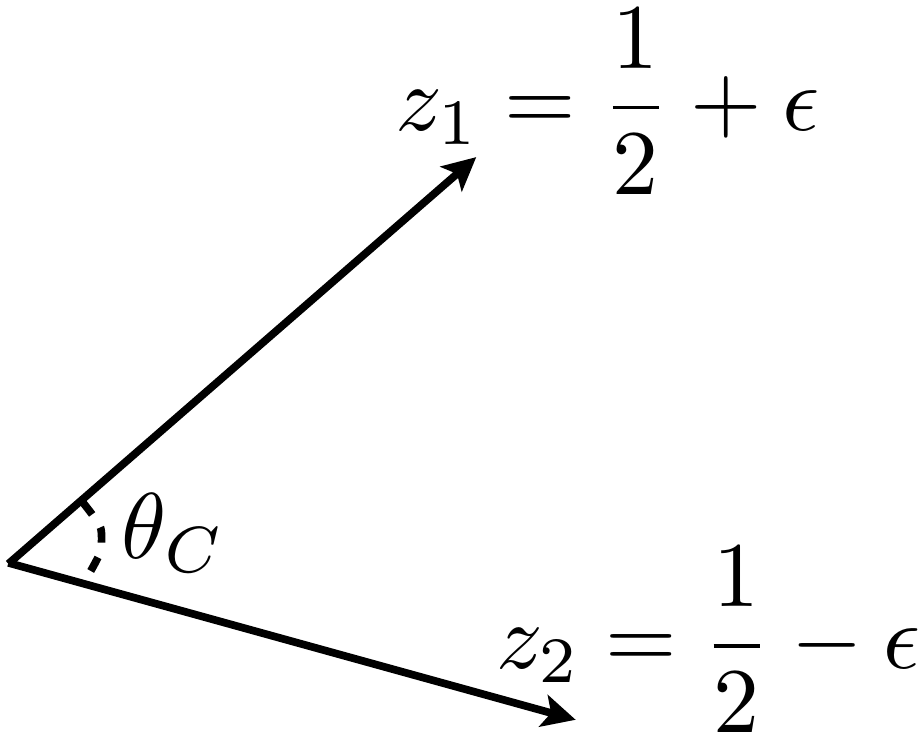}
}
$\qquad$
\subfloat[]{\label{fig:fin_config} 
\includegraphics[width=6cm]{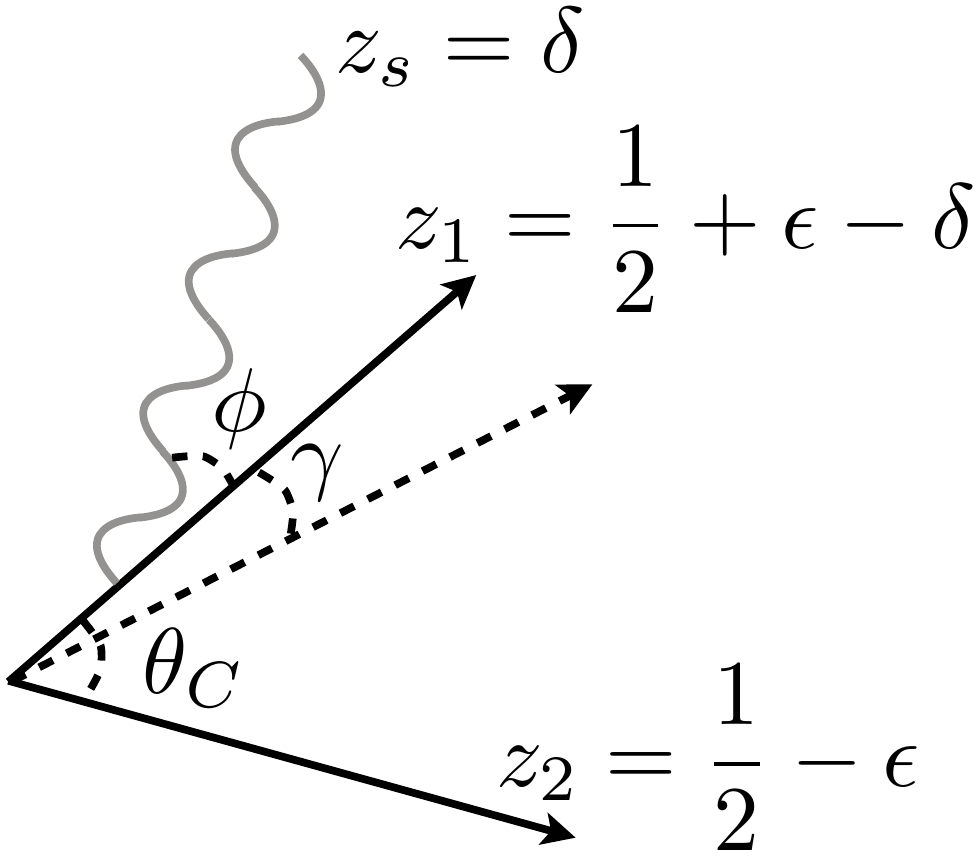}
}
\end{center}
\caption{
Figures for studying the broadening axis for a jet with three coplanar particles. Left: the initial jet with two particles with energy fractions $z_1=\frac{1}{2}+\epsilon$ and $z_2=\frac{1}{2}-\epsilon$ separated by an angle $\theta_C$.  Right: the jet after an emission of energy $\delta$ off of particle $1$.  The angle of the soft emission with respect to particle 1 is $\phi$, and the angle of the broadening axis is $\gamma$.
}
\label{fig:broad_emis}
\end{figure}

As discussed in \Sec{sec:recbroad}, if a single particle carries more than half of the energy of the jet, then the broadening axis will align along it.  Correspondingly, the behavior of the broadening axis is most subtle when there is roughly equal energy sharing among particles.  Consider first a jet with two particles, as illustrated in \Fig{fig:init_config}.  Particle 1 has energy fraction $z_1=\frac{1}{2}+\epsilon$, particle 2 has energy fraction $z_2=\frac{1}{2}-\epsilon$, and the angle of separation between the particles is $\theta_C$.  For $0 < \epsilon < \frac{1}{2}$, particle 1 has larger energy than particle 2, so the broadening axis lies along the direction of particle 1.

We would like to determine the effect of a soft emission off of particle 1 on the location of the broadening axis, as illustrated in \Fig{fig:fin_config}.  The energy fraction of the soft emission is $\delta < \frac{1}{2}$ and the angle of the soft emission with respect to particle 1 is $\phi$.  Particle 1 now has energy fraction $z_1=\frac{1}{2}+\epsilon-\delta$ while particle 2's energy fraction is unchanged ($z_2=\frac{1}{2}-\epsilon$).  Depending on the relationship between $\delta$ and $\epsilon$, particle 1 may have less energy than particle 2 after the emission.  We assume that the energy $\delta$ of the soft particle is sufficiently small such that the angle between particle 1 and 2 is still $\theta_C$.

To determine the location of the broadening axis, we measure the scalar sum of the transverse momenta with respect to an axis and then minimize.  For simplicity, assume that the three particles lie in a plane and that the broadening axis is at an angle $\gamma$ with respect to particle 1.  Here, the direction of positive angles is indicated in \Fig{fig:fin_config}.  In the small angle limit, the scalar sum of the transverse momentum with respect to this axis is
\begin{equation}
\label{eq:scalarsumthreeparticle}
\left| p_{T1} \right|+\left| p_{T2} \right|+\left| p_{Ts} \right| = \left(\frac{1}{2}+\epsilon-\delta \right) |\gamma| + \left(\frac{1}{2}-\epsilon\right)|\theta_C-\gamma|+\delta|\phi+\gamma|.
\end{equation}
The value of $\gamma$ that minimizes this quantity now depends sensitively on the angle $\phi$ and energy fraction $\delta$.

In fact, in this limit (planar, small angles), finding the broadening axis angle $\gamma$ is the same as finding the median of a (weighted) distribution, where the angles are the distribution entries and the energy fractions are the weights.  It is well-known that the median jumps discontinuously as the entries and weights change, and the same will be true for the location of the broadening axis.  

Consider the case where $\phi$ is positive.  If $\delta < \frac{1}{2}$, then the broadening axis has to lie between the hard particles, so $0 < \gamma < \theta_C$. In this case:
\begin{equation}
\phi > 0: \qquad \left| p_{T1} \right|+\left| p_{T2} \right|+\left| p_{Ts} \right| =\left(\frac{1}{2}-\epsilon\right)\theta_C+\delta\phi + 2\epsilon \gamma,
\end{equation}
which is minimized for $\gamma = 0$.  That is, the broadening axis is unchanged under soft, wide angle emission from hard particles.  Because of this, the change in the value of any angularities $\ang{\beta}$ measured about the broadening axis before and after the soft emission is suppressed by the energy $\delta$ of the soft emission.

Now consider the case where the soft emission lies between particles 1 and 2.  In this case, \Eq{eq:scalarsumthreeparticle} becomes
\begin{equation}
-\theta_C < \phi < 0:  \qquad \left| p_{T1} \right|+\left| p_{T2} \right|+\left| p_{Ts} \right| =\left( \frac{1}{2}-\epsilon \right) \theta_C+\delta\phi + 2(\epsilon-\delta) \gamma \ ,
\end{equation}
and the location of the broadening angle $\gamma$ depends on the precise relationship between $\delta$ and $\epsilon$.  For $\delta > \epsilon$, the sum is minimized when the broadening axis lies along the direction of the soft emission.  While this might seem like it results in the broadening axis being IRC unsafe,  the broadening axis discontinuously moves back to particle 1 once $\delta$ decreases below $\epsilon$ (in particular, in the soft limit $\delta \to 0$).  This discontinuous change should not be cause for concern, though, since the phase space for the soft emission to lie in the plane between particles 1 and 2 has zero measure and is not enhanced by any singularities.  
If instead the soft emission lies out of the plane of the two hard particles, then the broadening axis returns to the hardest particle continuously as the energy of the soft emission becomes small.  Of course, for this to be consistent with the behavior when the soft emission is in the plane, the rate of return increases without bound as the soft emission approaches the plane of the hard particles.  Thus, for sufficiently soft emissions, the broadening axis is unchanged, explicitly illustrating its IRC safety.

\section{Calculational Details}
\label{sec:cal_one_loop_details}

In this appendix, we present the details of the calculation of the jet and soft functions for the angularities $\ang{\beta}$ in \Eq{eq:angularity} measured with respect to the broadening axis.  We also give the $\mathcal{O}(\alpha_s)$ fixed-order corrections in \Sec{eq:fixedordercorrection}.   

\subsection{Jet Function Calculation}

Our calculations are evaluated at one-loop, which means that the jets contain two particles.  We choose to work in a frame where the two particles have light-cone momenta
\begin{equation}
(q^+,q^-,\vec{q}_\perp) \ , \qquad (l^+-q^+,Q-q^-,-\vec{q}_\perp) \ ,
\end{equation}
where the jet has total momentum $(l^+,Q,\vec{0})$ in the $+$, $-$ and $\perp$ components, respectively.  The broadening axis will be aligned with the particle that has larger $-$ component and therefore the softer particle will determine the value of the angularity.  To leading power, the angle between the particles in the jet is
\begin{equation}
\tan \frac{\theta}{2}\simeq \sin \frac{\theta}{2} \simeq \frac{Q q_\perp}{q^- (Q-q^-)} \ ,
\end{equation}
where $\tan \theta$ and $\sin \theta$ are the same to leading power in the jet function.
The energy of the softer particle that contributes to the angularity is 
\begin{equation}
\frac{E}{Q}   = \frac{\min[Q-q^-,q^-]}{2} \ .
\end{equation}
Therefore, the broadening axis angularities at one-loop in the jet function are
\begin{equation}
\ang{\beta} = 2^{\beta-1}Q^{\beta-1} q_\perp^\beta (q^-)^{1-\beta}(Q-q^-)^{1-\beta}\min[(Q-q^-)^{-1},(q^-)^{-1}] \ .
\end{equation}

For all $\beta>0$, the quark jet function can be computed in this frame from
\begin{align}
J^{(1)}_q(\ang{\beta})&=g^2  \bar{\mu}^{2\epsilon} \nu^\eta C_F \int \frac{dl^+}{2\pi} \frac{1}{(l^+)^2}\int \frac{d^d q}{(2\pi)^d}\, (q^-)^{-\eta} \, \left( 4 \frac{l^+}{q^-}+(d-2)\frac{l^+-q^+}{Q-q^-}  \right)\,  \nonumber \\
&\qquad \times  2\pi\delta(q^+q^--q_\perp^2)\Theta(q^+)\Theta(q^-)\nonumber \\
&\qquad \times \Theta(Q-q^-) \Theta(l^+-q^+)  2\pi\delta\left( l^+-q^+ -\frac{q_\perp^2}{Q-q^-}  \right) \nonumber \\
&\qquad \times \left\{   
\Theta\left( \frac{Q}{2}-q^-  \right)
 \delta\left( \ang{\beta} - (2Q)^{\beta-1}(Q-q^-)^{-\beta} (q^-)^{1-\beta}q_\perp^\beta   \right) \right. \nonumber\\
&\left. \qquad \qquad+\ \Theta\left( q^- -\frac{Q}{2} \right)\delta\left( \ang{\beta} - (2Q)^{\beta-1}(Q-q^-)^{1-\beta} (q^-)^{-\beta}q_\perp^\beta   \right)
\right\} \ ,
\end{align}
in $d=4-2\epsilon$ dimensions with $\overline{\text{MS}}$ scale $\bar{\mu}$ defined as 
\begin{equation}
\bar{\mu}^2 = \mu^2 \frac{e^{\gamma_E}}{4\pi} \ ,
\end{equation}
where $\gamma_E$ is the Euler-Maschroni constant.  We have also introduced $\eta$ which regulates rapidity divergences for $\beta=1$ and has a corresponding scale $\nu$.  Similarly, the gluon jet function can be computed from 
\begin{align}
J^{(1)}_g(\ang{\beta})&=2g^2  \bar{\mu}^{2\epsilon}\nu^\eta \int \frac{dl^+}{2\pi} \frac{1}{l^+}\int \frac{d^d q}{(2\pi)^d}\, (q^-)^{-\eta} \,\frac{1}{Q-q^-} \, 2\pi\delta(q^+q^--q_\perp^2)\nonumber \\
&\qquad \times\Theta(q^+)\Theta(q^-) \Theta(Q-q^-) \Theta(l^+-q^+)  
2\pi\delta\left( l^+-q^+ -\frac{q_\perp^2}{Q-q^-}  \right) \nonumber \\
&\qquad \times  
\left[
n_f T_R \left(1-\frac{2}{1-\epsilon}\frac{q^+q^-}{Q l^+}\right) - C_A \left( 2-\frac{Q}{q^-} - \frac{Q}{Q-q^-}-\frac{q^+q^-}{Q l^+}  \right)
 \right]
 \nonumber \\
&\qquad \times \left\{   
\Theta\left( \frac{Q}{2}-q^-  \right)
 \delta\left( \ang{\beta} - (2Q)^{\beta-1}(Q-q^-)^{-\beta} (q^-)^{1-\beta}q_\perp^\beta   \right) \right. \nonumber\\
&\left. \qquad\qquad+\ \Theta\left( q^- -\frac{Q}{2} \right)\delta\left( \ang{\beta} - (2Q)^{\beta-1}(Q-q^-)^{1-\beta} (q^-)^{-\beta}q_\perp^\beta   \right)
\right\} \ .
\end{align}
Performing the integrals, the quark jet function is
\begin{align}\label{bare_quark_jet_function}
J_q^{(1)}(\ang{\beta})&=\frac{\alpha_s}{\pi}\frac{C_F}{\beta}4^{\frac{\beta-1}{\beta}\epsilon} \frac{e^{\gamma_E \epsilon}}{\Gamma(1-\epsilon)}{\ang{\beta}}^{-1-\frac{2\epsilon}{\beta}}\left(\frac{\mu^2}{Q^2}\right)^\epsilon\left(\frac{\nu}{Q}\right)^\eta\left\{\frac{2\beta}{2(\beta-1)\epsilon+\beta\eta}+\frac{3}{2}\right.\nonumber\\
&\qquad\qquad\qquad\qquad\left.-\ \epsilon\left(-\frac{13}{2}+\frac{2\pi^2}{3}+\frac{3}{\beta}-\frac{\pi^2}{3\beta}+\frac{3\log2}{\beta}\right)+{\cal O}(\epsilon^2)\right\} \ ,
\end{align} 
and the gluon jet function is
\begin{align}\label{bare_gluon_jet_function}
J_g^{(1)}(\ang{\beta})&=\frac{\alpha_s}{\pi}\frac{C_A}{\beta}4^{\frac{\beta-1}{\beta}\epsilon}\frac{e^{\gamma_E \epsilon}}{\Gamma(1-\epsilon)}{\ang{\beta}}^{-1-\frac{2\epsilon}{\beta}}\left(\frac{\mu^2}{Q^2}\right)^\epsilon\left(\frac{\nu}{Q}\right)^\eta\left\{\frac{2\beta}{2(\beta-1)\epsilon+\beta\eta}+\frac{11}{6}-\frac{1}{3}\frac{n_f T_F}{C_A}\right.\nonumber\\
&\qquad\left.- \ \epsilon\left(-\frac{67}{9}+\frac{2\pi^2}{3}+\frac{137}{36\beta}-\frac{\pi^2}{3\beta}+\frac{11\log2}{3\beta}+\frac{n_fT_R}{C_A}\left(\frac{23}{18}-\frac{29}{36\beta}-\frac{2\log2}{3\beta}\right)\right)\right. \nonumber \\
&\left.\qquad +\ {\cal O}(\epsilon^2)
\vphantom{\frac12}
\right\} \ .
\end{align} 

For consistency of the factorization, the limit $\eta\to 0$ must be taken first, and then the limit $\epsilon\to 0$ can be taken.  The rapidity regulator $\eta$ is therefore only active if $\beta=1$ and does not appear in the jet function if $\beta \neq 1$.  Expanding in $\eta$ and $\epsilon$ isolates the singularities of the jet function which allows for renormalization.

\subsubsection{Renormalization of the Jet Function for $\beta\neq 1$}

We now present the calculation of the renormalized jet function from which the anomalous dimension of the jet function can be defined.  For $\beta\neq1$, there are no rapidity divergences, and so we consider that case first.  The renormalized jet function $J^{(R)}$ is defined from the bare jet function $J^{(B)}$ calculated above as
\begin{equation}\label{eq:renorm_def}
J^{(B)}(\ang{\beta})=\int de_{\beta}' \, Z_{J}(e_{\beta}')J^{(R)}(\ang{\beta}-{\ang{\beta}}') \ ,
\end{equation}
with the $Z_J$ factor containing all divergences of the bare jet function.  Note that $Z_J$ depends on the value of the angular exponent $\beta$.

To determine the $Z_J$ factor, and therefore the renormalized jet function, requires expanding \Eqs{bare_quark_jet_function}{bare_gluon_jet_function} in $\epsilon$.  This can be accomplished using $+$-distributions defined in \Ref{Ligeti:2008ac} where
\begin{equation}\label{eq:p_dist}
{\cal L}_{n}\left[\ang{\beta},\left(\frac{\mu}{ 2^{\frac{1-\beta}{\beta}}Q}\right)^\beta\right]=\frac{\mu^\beta}{2^{1-\beta}Q^\beta}\left[\frac{2^{1-\beta}Q^\beta}{\ang{\beta}\mu^\beta}\log^n\left(\frac{2^{1-\beta}Q^\beta}{e\mu^\beta}\right)\right]_{+} \ .
\end{equation}
The $+$-distribution integrates to zero on $\ang{\beta}\in[0,1]$ and only has support on $\ang{\beta}\in(0,1]$.  For $\beta\neq1$, the expansion that is needed is
\begin{align}
{\ang{\beta}}^{-1-\frac{2\epsilon}{\beta}}4^{\frac{\beta-1}{\beta}\epsilon} \left(\frac{\mu^2}{Q^2}\right)^\epsilon&=-\frac{\beta}{2\epsilon}\delta(\ang{\beta})+{\mathcal L}_{0}\left[\ang{\beta},\left(\frac{\mu}{2^{\frac{1-\beta}{\beta}}Q}\right)^\beta\right] \nonumber \\
&\qquad +\frac{2\epsilon}{\beta}{\mathcal L}_{1}\left[\ang{\beta},\left(\frac{\mu}{2^{\frac{1-\beta}{\beta}}Q}\right)^\beta\right]+ {\cal O}(\epsilon^2) \ .
\end{align}
Using this in \Eqs{bare_quark_jet_function}{bare_gluon_jet_function} and setting $\eta = 0$, the divergent terms of the quark jet function are
\begin{equation}
Z_{J_q}(\ang{\beta})=\delta(\ang{\beta})+\frac{\alpha_s}{2\pi} C_F\Biggl\{\left(\frac{\beta}{1-\beta}\frac{1}{\epsilon^2}+\frac{3}{2}\frac{1}{\epsilon}\right)\delta(\ang{\beta})+\frac{2}{1-\beta}\frac{1}{\epsilon}{\mathcal L}_{0}\Biggl[\ang{\beta},\Biggl(\frac{\mu}{2^{\frac{1-\beta}{\beta}}Q}\Biggr)^{\! \beta} \Biggr]\Biggr\} ,
\end{equation}
and for the gluon jet function
\begin{equation}
Z_{J_g}(\ang{\beta})=\delta(\ang{\beta})+\frac{\alpha_s}{2\pi} C_A\Biggl\{\left(\frac{\beta}{1-\beta}\frac{1}{\epsilon^2}+\frac{\beta_0}{C_A}\frac{1}{\epsilon}\right)\delta(\ang{\beta})+\frac{2}{1-\beta}\frac{1}{\epsilon}{\mathcal L}_{0}\Biggl[\ang{\beta},\Biggl(\frac{\mu}{2^{\frac{1-\beta}{\beta}}Q}\Biggr)^{\! \beta} \Biggr]\Biggr\} .
\end{equation}
Here, $\beta_0$ is the one-loop $\beta$-function coefficient
\begin{equation}
\beta_0=\frac{11}{6}C_A-\frac{2}{3}n_fT_R \ .
\end{equation}
The one-loop anomalous dimensions as defined in \Eq{eq:anom_dim_def} are therefore
\begin{align}
\label{eq:janom}
\Gamma_{J_q} &= 2 \frac{\alpha_s}{\pi} C_F\frac{\beta}{\beta-1} \ , \qquad \Gamma_{J_g} = 2 \frac{\alpha_s}{\pi} C_A\frac{\beta}{\beta-1} \ , \nonumber \\
\gamma_{J_q} &= \frac{3}{2}\frac{\alpha_s}{\pi} C_F \ , \qquad\qquad \  \gamma_{J_g} = \frac{\alpha_s}{\pi}\beta_0 \ ,
\end{align}  
for the cusp part ($\Gamma$) and the non-cusp part ($\gamma$).

The renormalized quark jet function is then
\begin{align}
\label{eq:renorm_jet_function}
J_{q}^{(R)}(\ang{\beta})&=\delta(\ang{\beta})+\frac{\alpha_s C_F}{\pi}\left\{
\left[
\frac{13}{4}-\frac{3}{8}\pi^2-\frac{3}{2\beta}(1+\log2)+\frac{\pi^2}{8\beta}-\frac{\pi^2}{8(1-\beta)\beta}
\right]\delta(\ang{\beta}) 
 \right.\nonumber \\
&\left. \qquad+\ \frac{3}{2\beta}{\mathcal L}_{0}\left[\ang{\beta},\left(\frac{\mu}{2^{\frac{1-\beta}{\beta}}Q}\right)^\beta\right]+\frac{2}{(1-\beta)\beta}{\mathcal L}_{1}\left[\ang{\beta},\left(\frac{\mu}{2^{\frac{1-\beta}{\beta}}Q}\right)^\beta\right]\right\} \ ,
\end{align}
and the renormalized gluon jet function is
\begin{align}
J_{g}^{(R)}(\ang{\beta})&=\delta(\ang{\beta})+\frac{\alpha_s C_A}{\pi}\left\{
\left[ 
\frac{67}{18}-\frac{3}{8}\pi^2-\frac{5}{72\beta}-\frac{\beta_0}{C_A\beta}(1+\log 2)\right.\right. \nonumber \\
&\left.\qquad\qquad\qquad\qquad\qquad
+\ \frac{\pi^2}{8\beta}-\frac{\pi^2}{8(1-\beta)\beta}+\frac{n_fT_R}{C_A}\left(\frac{53}{36\beta}-\frac{23}{18}\right)
\right]\delta(\ang{\beta})
\nonumber \\
&\left.\qquad+\ \frac{\beta_0}{C_A\beta}{\mathcal L}_{0}\left[\ang{\beta},\left(\frac{\mu}{2^{\frac{1-\beta}{\beta}}Q}\right)^\beta\right]+\frac{2}{(1-\beta)\beta}{\mathcal L}_{1}\left[\ang{\beta},\left(\frac{\mu}{2^{\frac{1-\beta}{\beta}}Q}\right)^\beta\right]\right\}.
\end{align}

\subsubsection{Renormalization of the Jet Function for $\beta=1$}

When $\beta=1$, there are rapidity divergences in the jet function that are regulated by the parameter $\eta$.  To determine the divergences of the jet function we first expand the jet function for $\eta\to 0$ and then for $\epsilon\to 0$.  This ordering is vital for extracting the correct singularities.  The $Z$ factor for quark jets with $\beta = 1$ is
\begin{equation}
Z_{q}(\ang{1})=\delta(\ang{1})+\frac{\alpha_s}{\pi}C_F\left\{\frac{2}{\eta}\frac{e^{\epsilon\gamma_E}}{\Gamma(1-\epsilon)}(\ang{1})^{-1-2\epsilon}\left(\frac{\mu^2}{Q^2}\right)^{\epsilon}-\frac{1}{\epsilon}\left[\frac{3}{4}+\log\left(\frac{\nu}{Q}\right)\right]\delta(\ang{1})\right\} \ ,
\end{equation}
while for gluon jets
\begin{equation}
Z_{g}(\ang{1})=\delta(\ang{1})+\frac{\alpha_s}{\pi}C_A\left\{\frac{2}{\eta}\frac{e^{\epsilon\gamma_E}}{\Gamma(1-\epsilon)}(\ang{1})^{-1-2\epsilon}\left(\frac{\mu^2}{Q^2}\right)^{\epsilon}-\frac{1}{\epsilon}\left[\frac{\beta_0}{2C_A}+\log\left(\frac{\nu}{Q}\right)\right]\delta(\ang{1})\right\} \ .
\end{equation}
The renormalized jet function for quark jets is then
\begin{equation}\label{eq:renorm_qjet_beta_1}
J^{(R)}_q(\ang{1})=\delta(\ang{1})+\frac{\alpha_s}{\pi}C_F\left\{\left[\frac{7}{4}-\frac{\pi^2}{6}-\frac{3}{2}\log2\right]\delta(\ang{1})+\left(\frac{3}{2}+2\log\left(\frac{\nu}{Q}\right)\right){\mathcal L}_0\left[\ang{1},\frac{\mu}{Q}\right]\right\} \ ,
\end{equation}
and for gluon jets, we have
\begin{align}\label{eq:renorm_gjet_beta_1}
J^{(R)}_g(\ang{1})&=\delta(\ang{1})+\frac{\alpha_s}{\pi}C_A\left\{\left[\frac{131}{72}-\frac{\pi^2}{6}-\frac{17n_fT_F}{36C_A}+\frac{\beta_0}{C_A}\log2\right]\delta(\ang{1})\right.\nonumber \\
&\left.\qquad\qquad\qquad\qquad+\left(\frac{\beta_0}{C_A}+2\log\left(\frac{\nu}{Q}\right)\right){\mathcal L}_0\left[\ang{1},\frac{\mu}{Q}\right]\right\} \ .
\end{align}

\subsection{Soft Function Calculation}

At one-loop, the soft function consists of a single soft emission off of the hard jet.  The momentum of the soft emission is
$$
(k^+,k^-,\vec{k}_\perp) \ ,
$$
in the $+$, $-$ and $\perp$ light-cone coordinates measured with respect to the thrust axis of the event.  While the observables that we consider are measured with respect to the broadening axis, and not the thrust axis, the angle between the thrust and broadening axes is power-suppressed in the soft function.  Therefore, to leading power, we can consider the soft modes as measured with respect to the thrust axis of the jet.

  For dijet events in $e^+e^-$ collisions, the soft function consists of two contributions, depending on which jet in the event the soft emission is closer to in angle.  If $k^->k^+$, then the soft emission is closer to the $\bar{n}$ axis and its angle with respect to the $\bar{n}$ axis is
\begin{equation}
\sin^2\frac{\theta}{2} = \frac{k^+}{k^- + k^+} \ .
\end{equation}
The energy of a soft emission is
\begin{equation}
E=\frac{k^- + k^+}{2} \ ,
\end{equation}
and so the value of the angularity when $k^->k^+$ is
\begin{equation}
\ang{\beta} = 2^{\beta-1}Q^{-1}(k^+)^{\beta/2}(k^- + k^+)^{1-\beta/2}\ .
\end{equation}
When $k^+>k^-$, the expression for the angularity is the same, with $k^+\leftrightarrow k^-$. 

Then, at one-loop for dijet events, the soft function can be computed from
\begin{align}
S^{(1)}(\ang{\beta})&=4g^2\bar{\mu}^{2\epsilon}\nu^\eta C_i\int\frac{d^dk}{(2\pi)^d}  
\frac{\left|k^-- k^+\right|^{-\eta} }{k^+ k^-}\,2\pi \delta(k^2)\,\Theta(k^0)  \nonumber \\
& \qquad ~ \times \left\{
\Theta(k^- - k^+) \delta\left( \ang{\beta}-2^{\beta-1}Q^{-1}(k^+)^{\beta/2}(k^- + k^+)^{1-\beta/2}   \right)\right.\nonumber \\
&\qquad\qquad\left.+\ \Theta(k^+-k^-) \delta\left( \ang{\beta}-2^{\beta-1}Q^{-1}(k^-)^{\beta/2}(k^++k^-)^{1-\beta/2}   \right)
\right\} \ .
\end{align}
$C_i$ is the total color of the jets in each hemisphere and $\eta$ is the rapidity regulator.  Evaluating this expression, we find 
\begin{align}\label{eq:softfunc}
S^{(1)}(\ang{\beta})&= \frac{\alpha_s}{\pi} C_i \frac{2^{1-(1-\beta)(\eta+2\epsilon)}e^{\gamma_E\epsilon}}{\Gamma(1-\epsilon)}{\ang{\beta}}^{-1-2\epsilon-\eta}\left(\frac{ \mu^2}{Q^2}  \right)^\epsilon \left( \frac{\nu}{Q} \right)^\eta \nonumber \\
& \qquad ~ \times \frac{\Gamma(1-\eta)\Gamma\left(\frac{\beta}{2}\eta+(1-\beta)\epsilon\right)}{\Gamma\left(1-(1-\beta)\epsilon-\frac{1}{2}(2-\beta)\eta\right)}\nonumber\\
&\qquad ~ \times \,_{2}F_{1}\left(-(1-\beta)\epsilon+\frac{\beta}{2}\eta,\frac{1}{2}(\beta-2)(2\epsilon+\eta);1-(1-\beta)\epsilon+\frac{1}{2}(\beta-2)\eta;-1\right),
\end{align}
where $_2F_1(a,b;c;z)$ is the hypergeometric function.  As for the jet function, we must first take $\eta\to0$ and then $\epsilon\to 0$ and so the rapidity regulator is only relevant for $\beta = 1$.

\subsubsection{Renormalization of the Soft Function for $\beta \neq 1$}
As with the jet function, the soft function can be renormalized by isolating the divergences.  We first consider the soft function for $\beta\neq1$ so that there are no rapidity divergences.  The renormalization $Z$ factor defined in \Eq{eq:renorm_def} for the soft function computed above is
\begin{equation}
Z_{S}(\ang{\beta})=\delta(\ang{\beta})-\frac{\alpha_s}{2\pi}\frac{C_i}{1-\beta}\left\{\frac{1}{\epsilon^2}\delta(\ang{\beta})-\frac{4}{\epsilon}{\mathcal L}_0\left[\ang{\beta},\frac{\mu}{2^{1-\beta}Q}\right]\right\} \ ,
\end{equation}
where we have set $\eta=0$ in \Eq{eq:softfunc}.  The soft anomalous dimension is therefore
\begin{equation}
\label{eq:sanom}
\Gamma_{S_i} = -2\frac{\alpha_s}{\pi}C_i \frac{1}{\beta-1} \ .
\end{equation}
The renormalized soft function is then
\begin{equation}
\label{eq:renorm_soft_function}
S^{(R)}(\ang{\beta})=\delta(\ang{\beta})+\frac{\alpha_s}{\pi}\frac{C_i}{1-\beta}\left\{\left(\frac{\pi^2}{4}+\frac{\pi^2}{12}(1-\beta)(\beta-2)\right)\delta(\ang{\beta})+4{\mathcal L}_1\left[\ang{\beta},\frac{\mu}{2^{1-\beta}Q}\right]\right\} \ ,
\end{equation}
where the $+$-distributions are defined in \Eq{eq:p_dist}.

\subsubsection{Renormalization of the Soft Function for $\beta = 1$}
For $\beta=1$, there are rapidity divergences in the soft function that are regulated by the parameter $\eta$.  Thus, the renormalization factor $Z_S$ for $\beta=1$ will have dependence on both the dimensional regularization parameter $\epsilon$ as well as $\eta$.  The $Z$ factor is then
\begin{equation}
Z_{S}(\ang{1})=\delta(\ang{1})-\frac{\alpha_s}{\pi}C_i\left\{\frac{8}{\eta}\frac{e^{\gamma_E\epsilon}}{\Gamma(1-\epsilon)}(\ang{1})^{-1-2\epsilon}\left(\frac{\mu^2}{Q^2}\right)^{\epsilon}+\left[\frac{1}{\epsilon^2}-\frac{2}{\epsilon}\log\left(\frac{\nu}{\mu}\right)\right]\delta(\ang{1})\right\} \ .
\end{equation}
The renormalized soft function for $\beta=1$ follows:
\begin{equation}\label{eq:renorm_soft_beta_1}
S^{(R)}(\ang{1})=\delta(\ang{1})+\frac{\alpha_s}{\pi}C_i\left\{\frac{\pi^2}{2}\delta(\ang{1})-4\log\left(\frac{\nu}{\mu}\right){\mathcal L}_0\left[\ang{1},\frac{\mu}{Q}\right]+4{\mathcal L}_1\left[\ang{1},\frac{\mu}{Q}\right]\right\} \ .
\end{equation}

\subsection{Fixed-Order Corrections}
\label{eq:fixedordercorrection}

For the $e^+e^-$ event shape, we give the necessary expressions to reproduce the fixed-order non-singular corrections in \Eq{eq:fixedorderpiece}. At ${\cal O}(\alpha_s)$ in the differential cross section, there are three partons in the final state: $q,\bar{q},g$. The two broadening axes coincide with the momenta of the two most energetic partons, and the least energetic parton sets the value of the angularity.  Because we use thrust axes to partition the event into hemispheres, the angularity is set by the angle between the least energetic and the second-most energetic parton.  This effectively partitions the phase space into six sectors, depending on the ordering of particle energies. 

For each of the six sectors, the phase space is identical, so after a trivial relabeling of variables in the squared matrix element, the six sectors can be summed into a single integral.  The differential cross section for $\ang{\beta}>0$ is then given by:
\begin{align}\label{eq:fo_exp}
\frac{1}{\sigma_0}\frac{d\sigma}{d\ang{\beta}}=&\ \frac{\alpha_s C_F}{\pi}\int_0^{1}dx\int_x^{1}dy\, \Theta(1-x-2y)\, \delta\left(\ang{\beta}-2^{\beta-1}x^{\frac{\beta}{2}}(1-y)^{-\frac{\beta}{2}}(x+y)^{1-\frac{\beta}{2}}\right)\nonumber\\
&\qquad\times\left\{\frac{2-3(1+x)(1-y)+3(1+x^2)(1-y)^2+3x(1+x)(1-y)^3}{x (1-y)(1-(1+x)(1-y))}\right\},
\end{align}
where within a sector, before relabeling, $x$ is the invariant mass between second-most and least energetic partons, normalized to $Q^2$. These particles form the broadening axis and the measured parton. Then $y$ is the invariant mass between the least energetic parton and the most energetic parton.

If $\beta=2$, all integrals can be performed analytically, and the result is:
\begin{align}
\frac{1}{\sigma_0}\frac{d\sigma}{d\ang{2}} &=\frac{\alpha_s C_F}{\pi}\Bigg\{6\frac{(1-\ang{2})(-32-4\ang{2}+6(\ang{2})^2-4(\ang{2})^3+(\ang{2})^4)}{\ang{2}(2-\ang{2})(4-\ang{2})^2(2+\ang{2})}\nonumber\\
&\qquad-\frac{4}{\ang{2}(2-\ang{2})}\text{ln}\Big[\frac{\ang{2}}{2-\ang{2}}\Big]+\frac{3}{2}\text{ln}\Bigg[\frac{\ang{2}(4-\ang{2})}{4-(\ang{2})^2}\Bigg]\Bigg\}
\end{align}
When $\beta\neq 2$, one of the integrals can be done via the delta function constraint after the change of variables:
\begin{align}
y&\rightarrow u-(1-u)v \ ,\nonumber \\
x&\rightarrow (1-u)v \ .
\end{align} 
The remaining integral can be computed numerically.  

We can split the fixed-order cross section at ${\cal O}(\alpha_s)$ into two contributions.  First, there is the singular contribution that can be computed in the effective theory.  This is defined as the part of the cross section in \Eq{eq:totangexp} that remains when $\ang{\beta}\to 0$:
\begin{align}\label{eq:sing_distribution_event}
\frac{\ang{\beta}}{\sigma_0} \frac{d\sigma^\text{sing}}{d\ang{\beta}} &= \lim_{\ang{\beta}\to 0}\frac{\ang{\beta}}{\sigma_0} \frac{d\sigma}{d\ang{\beta}}\nonumber \\
&=-\frac{\alpha_s C_F}{\pi\beta}\left[4\log \left(2^{1-\beta}\ang{\beta}\right)+3\right] \ .
\end{align}
After subtracting off the singular contribution, the rest defines the non-singular contribution.  The non-singular contribution is beyond the scope of the effective theory and is necessary for matching the resummed cross section to achieve formal next-to-leading order (NLO) accuracy. 

There are two non-trivial checks of the fixed-order result  in \Eq{eq:fo_exp}.  The first is that it must reproduce the singular behavior in \Eq{eq:sing_distribution_event}.  The second is that the cross section must vanish at
\begin{align}\label{eq:max_ang}
\ang{\beta}_{\max}&=3^{\frac{\beta}{2}-1}.
\end{align}
This is because the maximum value of the angularity at $\mathcal{O}(\alpha_s)$ is given by the ``Mercedes-Benz'' configuration when all three partons have equal energy.  In \Eq{eq:fo_exp}, this corresponds to $x$ and $y$ in the integrand both equaling $1/3$, at which point $\ang{\beta} = \ang{\beta}_{\max}$.

\begin{figure}
\begin{center}
\includegraphics[width=10cm]{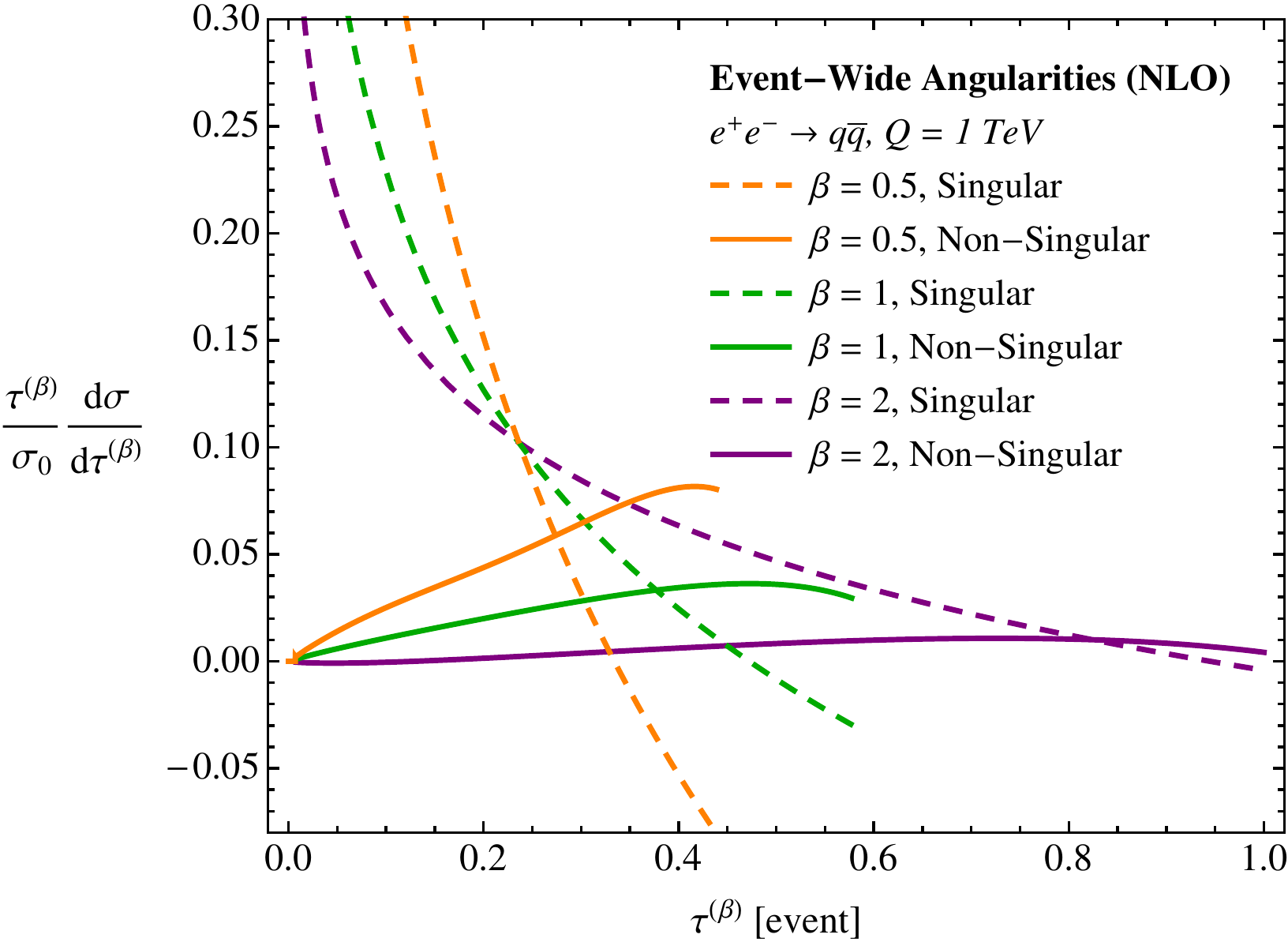}
\end{center}
\caption{Comparison of the contributions of the singular and non-singular components of the cross section at ${\cal O}(\alpha_s)$ as a function of $\ang{\beta}$. Where the dashed and solid curves of like color cross corresponds to when the resummation should be turned off.
 }
\label{fig:non-sing}
\end{figure}

In \Fig{fig:non-sing}, we plot singular and non-singular contributions to the cross section at ${\cal O}(\alpha_s)$ for $\beta = 0.5,1,2$.  The two above cross checks are satisfied because the non-singular piece goes to zero as $\ang{\beta} \to 0$ and $\sigma^\text{sing} + \sigma^\text{non-sing} = 0$ at $\ang{\beta} = \ang{\beta}_{\max}$.  The resummation is important in the region where the singular contribution to the cross section dominates and the fixed order result is important where the non-singular contribution dominates.  As we discuss in \App{sec:scalechoice}, we will turn off the resummation when $\sigma^\text{sing}$ and $\sigma^\text{non-sing}$ are equal.

\section{Check of the $\beta  = 1$ Limit}
\label{app:betaonecheck}

When $\beta=1$, the bare jet and soft functions develop new divergences associated with rapidity integrals. These divergences can be renormalized like the standard ultraviolet divergences by including two renormalization parameters $\mu$ and $\nu$ in each sector:
\begin{align}
F^B(\ang{1})&=\int d\ang{1}\,'\,Z_F(\ang{1}\,',\mu,\nu)F^R(\ang{1}-\ang{1}\,',\mu,\nu) \ .
\end{align}
The corresponding RG equations for evolution in $\mu$ and $\nu$ are then 
\begin{align}
\mu\frac{d}{d\mu}F^R(\ang{1},\mu,\nu)&=\gamma_{F}^{\text{UV}}(\nu_F/\nu)F^R(\ang{1},\mu,\nu) \ ,\label{eq:UVanomdimconv}\\
\nu\frac{d}{d\nu}F^R(\ang{1},\mu,\nu)&=\int d\ang{1}\,'\,\gamma_{F}^{\text{Rap}}(\ang{1}\,',\mu)F^R(\ang{1}-\ang{1}\,',\mu,\nu) \ .
\end{align}
The scale $\nu_F$ is the intrinsic rapidity scale of the function $F^R$, and each anomalous dimension is labeled according to its physical origin.

The convolution structure of these objects (or lack thereof) can be understood as follows. Each function has a cusp double logarithmic structure, where one of these logarithms is tied the invariant mass of the sector, and the other its rapidity scale. Thus, in general one expects the anomalous dimension to depend on the observable, whose values is set by the invariant mass. However, since the invariant mass of the sector is controlled by the RG parameter $\mu$, the $\mu$-derivative in the standard UV anomalous dimension removes the observable dependence, so the structure of \Eq{eq:UVanomdimconv} must be multiplicative. By contrast, in the rapidity anomalous dimension, one is intrinsically sensitive to the observable in the anomalous dimension, since the $\nu$ derivative does not remove the observable dependence in the cusp double logarithmic structure. 

We wish to compare with the $\beta\rightarrow 1$ limit of \Sec{sec:anom_dim_beta_1_limit}. To make a direct comparison, we Laplace transform the one-loop matrix elements for the jet and soft functions, \Eq{eq:renorm_qjet_beta_1} and \Eq{eq:renorm_soft_beta_1}, and consider $e^+e^-\to q\bar{q}$ events, yielding the UV anomalous dimensions:
\begin{align}
\label{eq:br_anom}
\gamma_J^{\text{UV}}(Q/\nu)&=\frac{\alpha_s }{\pi}C_F\left[\frac{3}{2}+2\log\left(\frac{\nu}{Q}\right)\right] \ ,\nonumber\\
\gamma_{S}^{\text{UV}}(\mu/\nu)&=-4\frac{\alpha_s}{\pi}C_F\log\frac{\nu}{\mu} \ ,
\end{align}
for the jet and soft functions, respectively. From the same matrix elements, we find the rapidity anomalous dimensions to be:
\begin{align}
\label{eq:br_anom_rap}
\gamma_J^{\text{Rap}}(s_1,\mu)&=\frac{2\alpha_s C_F}{\pi}\log\left(s_1e^{\gamma_E}\frac{\mu}{Q}\right) \ , \nonumber\\
\gamma_{S}^{\text{Rap}}(s_1,\mu)&=-\frac{4\alpha_s C_F}{\pi}\log\left(s_1e^{\gamma_E}\frac{\mu}{Q}\right) \ .
\end{align}
Therefore, under the rapidity renormalization group, the product of jet and soft functions in Laplace space becomes
\begin{align}
\tilde{J}^R_n\left(s_{1},\frac{\mu}{Q},\frac{\nu}{Q}\right)&\tilde{J}^R_{\bar{n}}\left(s_{1},\frac{\mu}{Q},\frac{\nu}{Q}\right)\tilde{S}^R\left(s_{1},\frac{\mu}{Q},\frac{\nu}{\mu}\right)
\nonumber \\
&=\text{exp}\left[-\frac{4\alpha_s C_F}{\pi}\log\left(\frac{\nu_J}{\nu_S}\right)\log\left(s_1e^{\gamma_E}\frac{\mu}{Q}\right)\right]\nonumber\\
&\qquad \times \tilde{J}^R_n\left(s_{1},\frac{\mu}{Q},\frac{\nu_J}{Q}\right)\tilde{J}^R_{\bar{n}}\left(s_{1},\frac{\mu}{Q},\frac{\nu_J}{Q}\right)\tilde{S}^R\left(s_{1},\frac{\mu}{Q},\frac{\nu_S}{\mu}\right)
\end{align}
If we take $\mu=\mu_J=\mu_S$ and $\nu=\nu_S$ and $\nu_J=Q$, we recover the $\beta\rightarrow1$ limit of the soft evolution factor in \Eq{eq:betaonelimitSRG}.

\section{Profile Scales for General $\beta$}
\label{sec:scalechoice}

The canonical RG scales in \Eq{eq:natural_scales} are appropriate in the region $\ang{\beta}_{L,R}\ll 1$ where resummation is needed, but at larger values of $\ang{\beta}_{L,R}$, fixed-order corrections dominate the cross section.  In order to smoothly interpolate between the resummed and fixed-order results, we can use profiling of the resummation scales \cite{Abbate:2010xh}, where the RG scales depend on the measured value of $\ang{\beta}_{L,R}$.

\begin{figure}
\begin{center}
\includegraphics[scale=0.7]{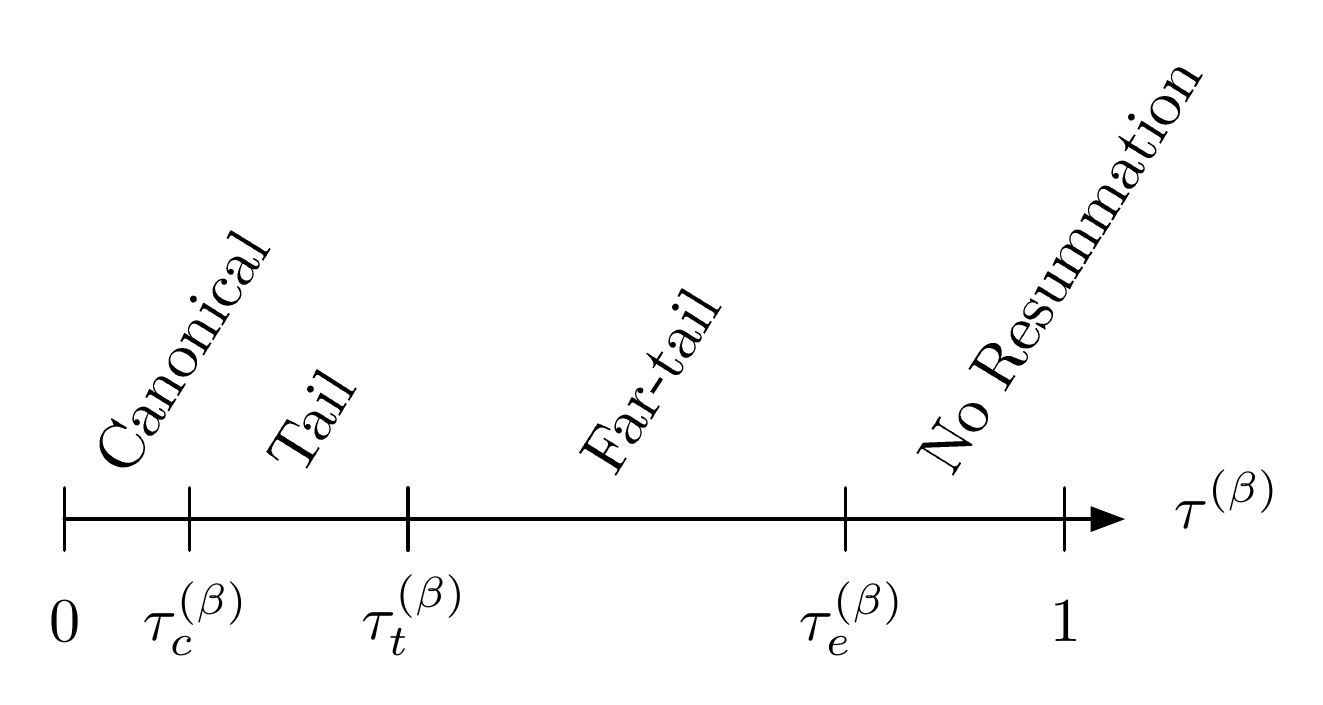}
\end{center}
\caption{Definition of the profile regions and transition points.}
\label{fig:profileboundaries}
\end{figure}

Since we are ignoring the inclusion of non-perturbative effects, we have three dominant regions as shown in \Fig{fig:profileboundaries}: a canonical (peak) region, a tail region, and a far-tail region.  In the canonical region, we set scales to minimize the logarithms in the low-scale matrix elements, while in the tail and far-tail regions, we gradually turn off resummation to return to the fixed-order cross section.   We also include small transition regions to make sure that the scales are everywhere differentiable. 

To set the functional form of the profile functions, we first have to determine the location of the canonical/tail boundary ($\ang{\beta}_c$), the tail/far-tail boundary ($\ang{\beta}_t$), and the point at which resummation is completely turned off ($\ang{\beta}_e$).  The canonical/tail boundary occurs when resummation of logarithms starts to become less relevant.  For thrust-like observables as considered in \Ref{Abbate:2010xh}, the sector of lowest virtuality is the soft sector, so the canonical/tail boundary is set by the condition $\alpha_s\log\ang{\beta}\sim \mathcal{O}(1)$ (i.e.\ when the soft logarithms are no longer large). However, if $\beta<1$, the collinear sectors have the lowest virtuality, and collinear logarithms of the form $\frac{\alpha_s}{\beta}\log \ang{\beta}$ can still be relevant  up to larger value of $\ang{\beta}$ (and therefore must still be resummed).  Hence, we adopt the prescription for the canonical/tail boundary $\ang{\beta}_c$ to be set by the sector of lowest virtuality: 
\begin{align}
\label{def_can_region_soft}\text{Soft:  } \alpha_s\left(\ang{\beta}_cQ\right)C_i\log\ang{\beta}_c \sim 1\,\text{if } \beta>1,\\
\label{def_can_region_jet}\text{Jet:   }\frac{\alpha_s \left(\ang{\beta}_cQ\right)}{\beta}C_i\log \ang{\beta}_c\sim 1\,\text{if } \beta<1,
\end{align}
where $C_i$ is the appropriate color factor for the jet. The resummation should be completely turned off when the perturbative non-singular corrections become as important as the singular pieces, so
\begin{align}
\frac{d\sigma^\text{sing}}{d\ang{\beta}}\left(\ang{\beta}_e\right)\sim\frac{d\sigma^\text{non-sing}}{d\ang{\beta}}\left(\ang{\beta}_e\right).
\end{align}
Together, the tail and far-tail regions is defined by $\ang{\beta}_c<\ang{\beta}<\ang{\beta}_e$, and the boundary between them is somewhat ambiguous.  For concreteness, we set the tail/far-tail boundary by $\ang{\beta}_t\sim\ang{\beta}_e/3$.

A peculiar feature of these profiles is that the $\ang{\beta}_c$ and $\ang{\beta}_e$ boundaries start to coincide as $\beta \to 0$.  This can be understood because $\ang{\beta}_c$ grows as $\beta\rightarrow 0$ but the position where the singular corrections vanish remains fixed.  Hence the tail plus far-tail region necessarily shrinks.  Thus, unlike the case for thrust, as $\beta \to 0$ one can expect both singular and non-singular corrections to be substantial in the peak/tail regions of the cross section.  We saw this in the plots in \Sec{subsec:NLLprime}, where there are large differences between the NLL and \nllp+NLO calculations even in the peak region.

\begin{figure}
\begin{center}
\subfloat[]{\label{fig:prof_2}
\includegraphics[width=4.4cm]{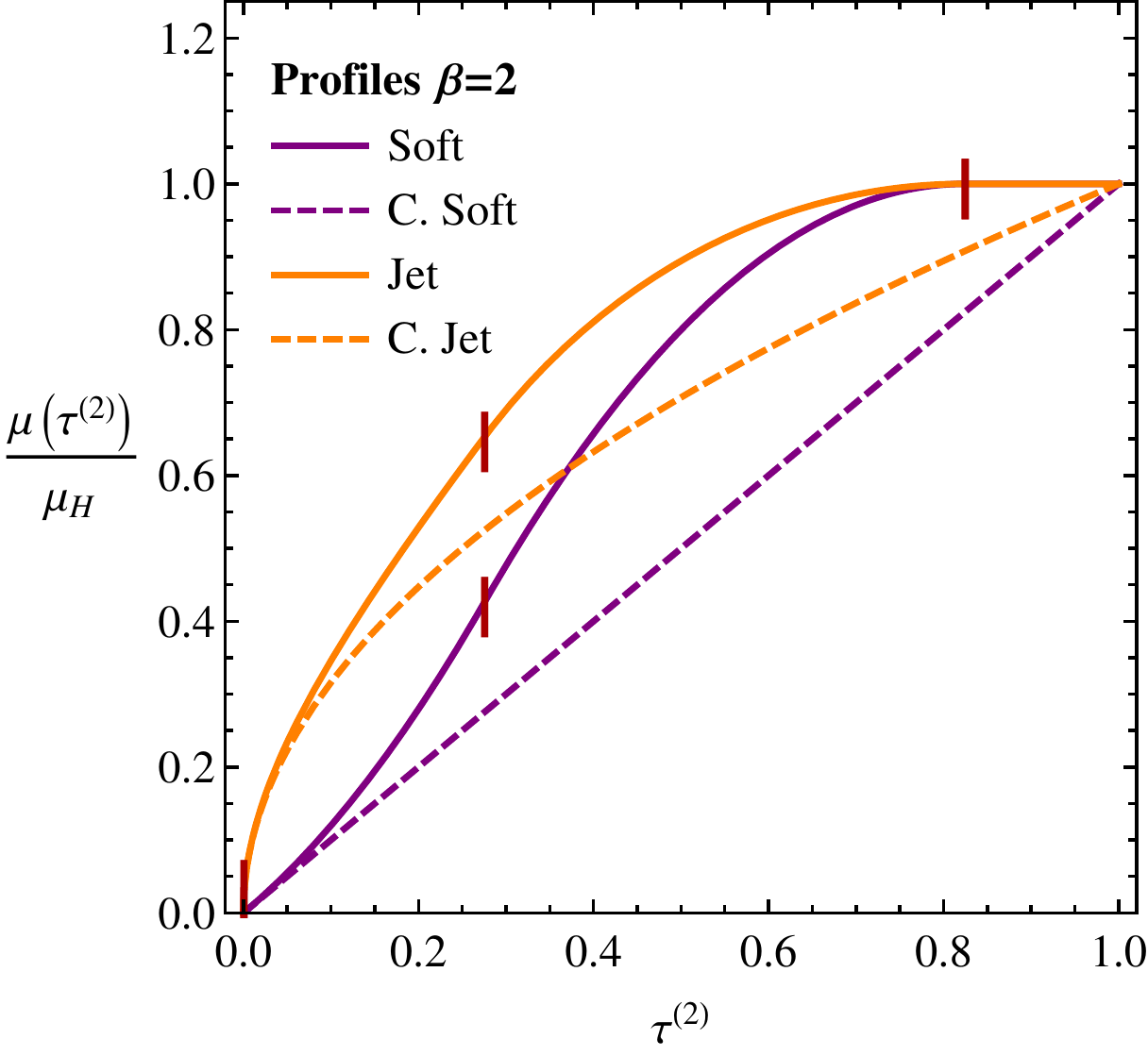}
}
$\quad$
\subfloat[]{\label{fig:prof_1}
\includegraphics[width=4.4cm]{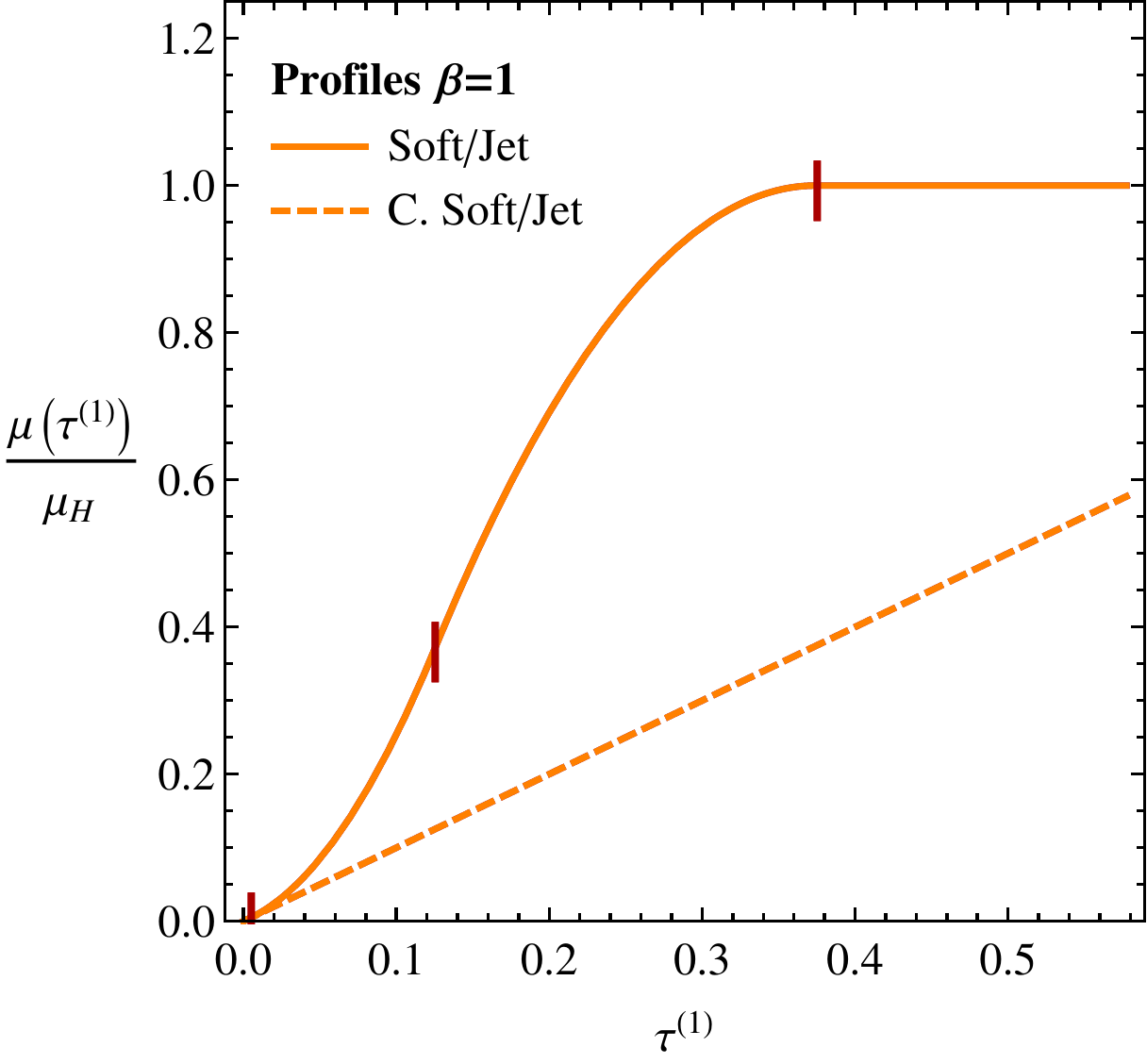}
}
$\quad$
\subfloat[]{\label{fig:prof_05}
\includegraphics[width=4.4cm]{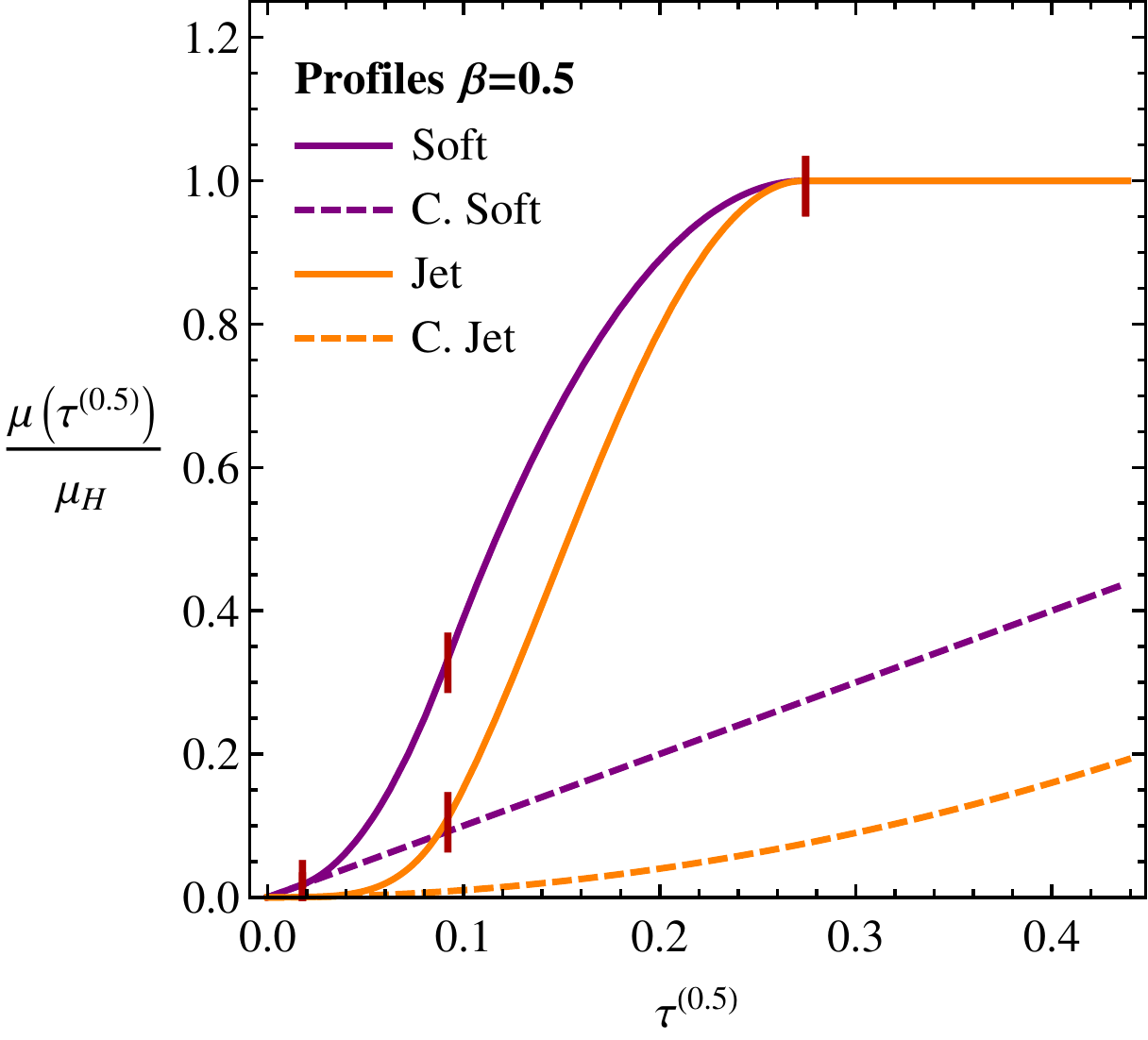}
}
\end{center}
\caption{Profile functions for $\beta = 2$ (left), $1$ (middle), and $0.5$ (right).  The solid curves are the profile scales used in this paper while the dashed curves are the canonical scale choices from \Eq{eq:natural_scales}, and the red lines mark the transitions between regions in the profiles.  Notice that the turning off of resummation and the onset of the canonical scaling changes as a function of $\beta$. In particular, as $\beta\rightarrow 0$, the distance between the peak region and the far tail region shrinks.   }
\label{fig:profile}
\end{figure}

Having set the boundaries of the regions, we can now set the profiles for arbitrary $\beta$.  The simplest profile to fix is the soft profile, since the canonical soft scale in \Eq{eq:natural_scales} is a linear function of the observable.   We therefore make the profile for $\mu_S$ to be a linear function of $\ang{\beta}_{L,R}$ near the origin, and then use quadratic splines in the vicinity of $\ang{\beta}_c$ and $\ang{\beta}_e$ to move the soft scale $\mu_S$ to the hard scale $\mu_H$, turning off resummation.  At each transition, the splines are continuous through the first derivative, so the cross section exhibits no kinks.  The final functional form is shown in \Fig{fig:profile}.  Specifically we use:
\begin{align}
\mu_S(\ang{\beta};a,\mu_H)=
\begin{cases}
a\ang{\beta},&\quad \ang{\beta}<\tau_{c},\\
b_1+b_2\ang{\beta}+b_3\left(\ang{\beta}\right)^2,&\quad\tau_{c}<\ang{\beta}<\tau_{t},\\
c_1+c_2\ang{\beta}+c_3\left(\ang{\beta}\right)^2,&\quad\tau_{t}<\ang{\beta}<\tau_{e},\\
\mu_H,&\quad \tau_{e}<\ang{\beta}.
\end{cases}
\end{align}
The soft profile has as free parameters the slope of the canonical region $a$, as well as the boundary points $\tau_{c},\tau_{t},\tau_{e}$. The constants $b_i$ and $c_i$ are set by demanding continuity through the first derivative at each boundary. 

Once we have fixed the soft profile for $\mu_S$, the jet profile for $\mu_J$ is fixed by the relative scalings between the all modes in the theory.  Motivated by \Eq{eq:natural_scales}, we take
\begin{align}
\mu_J(\ang{\beta};a,\mu_H,e_J)&=\left(\mu_H+e_J\left(\mu_H-\mu_S(\ang{\beta};a,\mu_H)\right)\right)^{\frac{1-\beta}{\beta}} \left(\mu_S(\ang{\beta};a,\mu_H)\right)^{\frac{1}{\beta}},
\end{align}
The jet profile has another free parameter $e_J$, whose default value we take to be $0$.  As $\beta\rightarrow 1$, the parameter $e_J$ controls the effective rapidity scale variation, as described in \Sec{sec:anom_dim_beta_1_limit}. 

To gauge the $\mu$-variation of the profiled \nllp\ result, we vary $e_J$ from $-\frac{1}{2}$ to $1$, $a$ from $\frac{1}{2}$ to $2$, $\mu_H$ from $\frac{Q}{2}$ to $2Q$, and $\tau_t^{(\beta)}$ from half to twice its default value.  The bands in \Fig{fig:nll_to_nllprime} correspond to the envelope of these variations. 

\bibliography{broadening}

\end{document}